\documentclass[conference]{IEEEtran}

\usepackage{graphicx}
\usepackage{booktabs}
\usepackage{array}   %
\usepackage{amsmath,amssymb}
\usepackage{siunitx}
\usepackage{enumitem}
\usepackage[numbers]{natbib}
\usepackage[hidelinks]{hyperref}

\setlist{nosep, leftmargin=1.4em, topsep=2pt, itemsep=1pt, parsep=0pt}

\newcommand{\lead}[1]{\par\smallskip\noindent\textbf{#1}\ \ignorespaces}

\newcommand{\astar}{a^{*}}                 %
\newcommand{\edge}{\varepsilon}            %
\newcommand{\defrate}{\rho_{d}}            %
\newcommand{\lam}{\lambda}                 %
\newcommand{\advers}{\alpha}               %
\newcommand{\escfrac}{\phi}                %
\newcommand{\opasym}{\kappa}               %
\newcommand{\pin}{\textsc{pin}}
\newcommand{\router}{\textsc{router}}

\begin{document}

\title{False Floors: LLM Safety Routing\\
       Evaluations Break Under Distribution Shift}

\author{\IEEEauthorblockN{Amit Singh Bhatti and Vishal Vaddina}
        \IEEEauthorblockA{Quantiphi Analytics\\
        amit.bhatti@quantiphi.com}}

\maketitle

\begin{abstract}
Safety routers classify each request and send it to one of several models, and are judged against
the best single model. That judgement fails where safety routing is meant to work. A major routing
benchmark picks the comparator on the evaluation data. In that benchmark's own setting the practice
is harmless, its Best Single costing nothing at its full model pool. Under distribution shift it is
not. On HELM Safety the cost is
\numrange{0.003}{0.030} of harm under random splits and \numrange{0.045}{0.113} under held-out
categories, its direction holding under either published judge alone, comparable to the whole deficit attributed to routing, and it rises seven- to ninefold on
AgentDojo when suites are held out. On seven safety corpora chosen by rules fixed in advance three meet a registered interval test and four beat a permutation null added later, and three of the four interval misses are corpora where some models have zero observed harm. Prior work proves the direction of this bias. We size it on harm
and accuracy, show that it is larger under the held-out splits we measure, and bound it by optimism plus a shift-dependent regret.
Scored honestly under shift, routing buys little on these benchmarks. In most pool cells the nested router simply serves the honest
baseline's model, and on the nearly saturated AgentDojo corpus a perfect router that commits before
the injection arrives is worth at most two points of harm. Pre-dispatch headroom on these agentic grids is small, and we separately find the model's expressed recognition of a late injection steerable. On held-out reruns an attacker who
knows which model it faces lowers GPT-5.4's judged recognition by \num{19.6} points, the only flag in
the pool with room to fall, confirmed by an independent label. In an offline counterfactual that assumes a flag before any action and replaces silent targets by fresh fallback runs, the same attack raises or lowers estimated harm depending on the fallback model. Four action-level policy settings
record zero judged successes on one shared set of episodes and their traces are consistent with the same recognition. Safety routing should be evaluated under shift, against a baseline chosen without the
test labels, and defences that rely on recognition should be scored on harm against an attacker who
chooses what the model sees.

\end{abstract}

\section{Introduction}
\label{sec:intro}

\begin{figure*}[!t]\centering
  \includegraphics[width=.8593\textwidth]{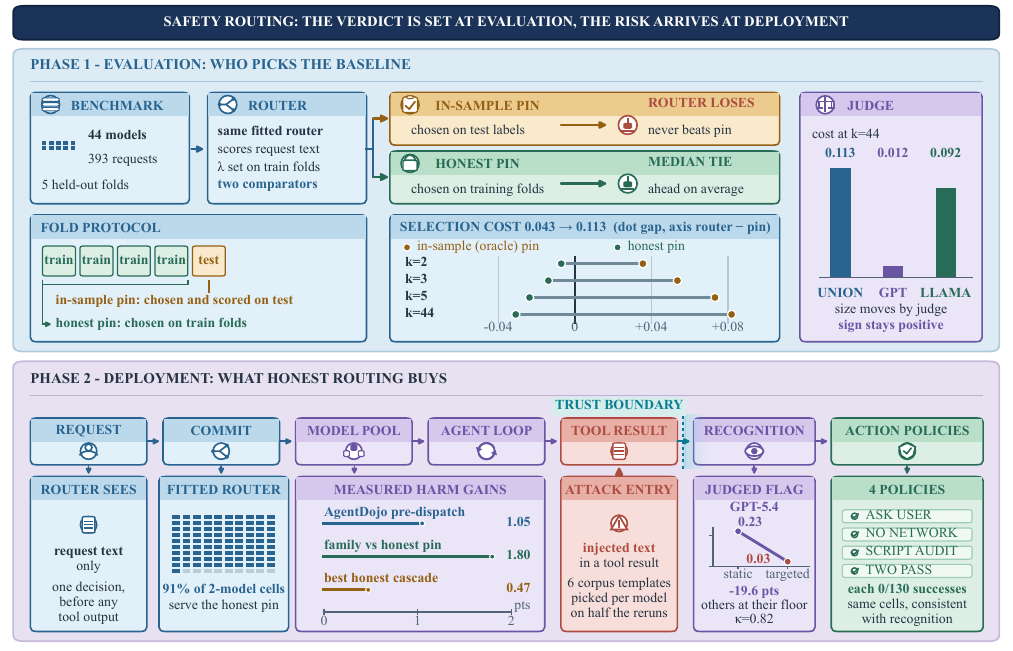}
  \caption{\textbf{The baseline, not the router, decides the verdict.} \emph{Phase 1, evaluation.}
  One fitted router is scored against two fixed-model comparators on identical folds (E24, union
  label). The fold protocol shows where each comparator is chosen, the honest pin on the four
  training folds, the in-sample pin on the held-out fold it is then scored on. Against the
  in-sample pin the router never beats it. Against the honest pin it ties in the median and is ahead
  on average. The gap is the
  selection cost, \numrange{0.043}{0.113}, whose sign is guaranteed (Observation~1) and whose size
  depends on the judge (E24j). \emph{Phase 2, deployment.} The router sees only the request text and decides once, before any
  tool output exists. What routing can buy, each on its own surface, is small. The fitted router serves the honest pin's own model on every held-out request in \SI{91}{\percent} of two-model pool cells, a perfect pre-dispatch
  router on AgentDojo stays within two points of harm (E43), and the best honest cascade is one cheap model (E34). After the router commits, an
  injection arrives in a tool result and defence rests on the model's recognition. On held-out
  reruns a targeted template lowers \texttt{gpt-5.4}'s judged recognition by \num{19.6} points,
  confirmed by an independent label (E54b, E55). Four
  action-level policy settings record zero judged successes on one shared set of episodes.}
  \label{fig:overview}
\end{figure*}

Production serving stacks already route on safety. A router classifies each request and selects
which of a mixed pool of proprietary and open models answers it, blocking or steering flagged
requests before dispatch \citep{vllmsemanticrouter2026}. The case for doing so rests on an
\emph{oracle gap}. On a benchmark of attacks, the per-request minimum over models sits well
below the best single model's rate. Whether a router reaches that headroom is judged against a
fixed-model comparator, and in one major routing benchmark that comparator is picked from the
evaluation data while the router is held to out-of-sample discipline. \Citet{commercial2026} say so
outright. Their Best Single baseline selects ``a single model with the highest average accuracy
across all datasets in hindsight'' and is scored on the same test split as the routers.
Classifier-selection work, by contrast, tunes it on validation \citep{lassance2026dcs}.
\Citet{shihab2026opportunity} show that choosing the best fixed model on the same examples biases
the comparison in the comparator's favour, and prove its direction on accuracy under random splits.
How large that bias is on harm, and what drives it, is our question. Safety is where it matters,
because a safety router exists for requests unlike the ones it was tuned on.

\begin{figure}[t]\centering
  \includegraphics[width=\linewidth]{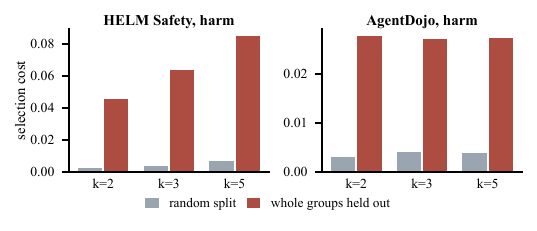}
  \caption{\textbf{The comparator's cost appears under shift, on both safety corpora.} Selection
  cost in harm under a random split and with whole HELM categories or AgentDojo suites held out
  (E25, E25a).}
  \label{fig:shift}
\end{figure}

On the protocols we measure, the answer is distribution shift. On the chat corpus the in-sample comparator costs almost nothing
under random splits, \numrange{0.003}{0.030} of harm, and there a nested router beats even the in-sample
comparator (Sections~\ref{sec:baseline} and~\ref{sec:surface}). When test requests come from
categories held out of training, the cost rises to \numrange{0.045}{0.113} (E25, and under the benchmark's own
full-sample convention, Appendix~\ref{app:comparators}), comparable to the
deficit it is used to establish. The same contrast appears on AgentDojo, seven- to ninefold when whole suites are held out
(Figure~\ref{fig:shift}). The in-sample comparator's harm is then a false floor, a level set by a choice no
deployer can make, because making it needs the test labels, and routers are judged against it. The same router then never wins a single chat pool cell against the
in-sample comparator, while against a comparator chosen on training data the median difference is
zero (Section~\ref{sec:baseline}). The direction is guaranteed, since an in-sample comparator can
only flatter a fixed model, and its bound carries a regret term with a shift component, while the measured held-out protocols show larger costs on the named corpora (Observation~1, Appendix~\ref{app:comparators}). An oracle cost frontier makes the
same error, hiding configurations that beat an honest frontier at a matched planned budget. For the tested fold-shared policy under category-held-out evaluation, almost all of its gain is fixed-selector gain (E66), and it ends up serving the honest pin's own model on every held-out request in \SIrange{82}{91}{\percent}
of pool cells (E24), so under this protocol the comparison is mostly between two ways of choosing one model, and which model is best is itself unstable across scoring
rules \citep{lee2026routinggap}.

Scored honestly under shift, routing buys little on these benchmarks. The accuracy a router needs is a surface rather than a
number (Section~\ref{sec:surface}). On AgentDojo a perfect router that commits before execution is
worth at most two points of harm, and the cheaper configurations an oracle frontier hides are
mostly one cheap model rather than a router (Sections~\ref{sec:two-settings}
and~\ref{sec:mechanisms-die}). That headroom is small on these grids, so we turn to the model's own recognition of an injection that arrives after dispatch. On a skill-injection corpus that signal is steerable. An attacker that
knows which model it faces picks its template on some reruns and, on held-out reruns of the same
configurations, lowers one model's judged recognition while no other model's flag moves. The
drop reproduces under a lexicon label the judge did not write, within each of the three run
configurations that contain both templates.
We call the gap between how often a model notices and whether it can be relied on to notice the
\emph{vigilance illusion}, and scope it to what we measured, expressed recognition on one model
(Section~\ref{sec:mechanisms-die}). Composed offline into a flag-gated controller from logged runs, the same attacker raises or
lowers harm depending on which fallback serves the escalation, so a flag-level measurement cannot
certify a recognition-based defence.

\lead{Contributions.} One thesis runs through them. A safety evaluation that scores a component on
what a deployer cannot use, a comparator chosen on the answers or a flag read without the controller
that acts on it, misstates what the deployment gets. (1) The size of the first error under shift.
Its direction is the optimizer's curse \citep{smithwinkler2006}, which
\citet{shihab2026opportunity} prove for routing accuracy, and what is new is its size on safety
corpora, on identical folds across two corpora, on seven corpora chosen by rules fixed in advance,
split into how often and how far the choice moves, and on LLMRouterBench's own release
(Section~\ref{sec:baseline}, Appendix~\ref{app:comparators}). (2) The four-axis
threshold surface a router must clear, and the price of the alternative. The lowest-harm model in a
\num{44}-model pool, selected in sample, refuses harmless requests at \num{1.62} times the median
rate, a cost that harm-only evaluation cannot see (Section~\ref{sec:surface},
Appendix~\ref{app:axes}). (3) A bound on pre-execution routing in agents and an honest cost
frontier (Sections~\ref{sec:two-settings} and~\ref{sec:mechanisms-die}). (4) An adaptive
template-selection attack on the recognition signal, independently annotated, and its effect
through flag-gated controllers. (5) Four action-level
policy settings in the class of \citet{debenedetti2025camel} and \citet{beurerkellner2025patterns},
which record zero judged successes on one shared set of cells and whose traces are consistent with recognition
(Section~\ref{sec:what-works}). We also correct three instruments. Rank-inversion counts sit below
their own null, full-pool headroom is combinatorially vacuous, and AUROC is an insufficient summary
(Appendix~\ref{sec:instruments}).

\section{Problem Setup and Threat Model}
\label{sec:setup}

A pool $\mathcal{M}=\{1,\dots,M\}$ of models. For request $x$ a router emits per-model scores
$s_i(x)$ and serves $\arg\min_i s_i(x)$. Write $h_i(x)\in\{0,1\}$ for whether model $i$ produces
harmful output on $x$, and $u_i(x)$ for whether it completes the user's task.

\lead{Baselines.}
Alongside random selection and the oracle gap, the routing comparisons we examined score against a
\textbf{best-fixed pin}, $\pin = \arg\min_i \mathbb{E}[h_i]$, the single best-on-average model. It
is a harder comparison and the policy a deployer already has. We also report a \emph{signal-free}
router (predictions shuffled within model, marginals preserved), isolating $\arg\min$ geometry from
signal. Four terms are fixed throughout. The \emph{oracle} is the per-request minimum over models,
as in the literature's ``oracle gap'', and an \emph{oracle frontier} is a cost--harm hull fitted with
hindsight. The \emph{in-sample pin} ($\pin_{\text{oracle}}$, ``oracle pin''
in figures and tables) is the fixed model with the lowest harm on the evaluation data, chosen with hindsight,
either over all evaluation items (\emph{full-sample}) or inside each held-out fold
(\emph{within-fold}). The two differ at $k=M$ by more than several of the effects reported, so every table names
which it uses. The \emph{honest pin} ($\pin_{\text{honest}}$) is chosen on training folds only and is
what a deployer can run. The \emph{predicted-marginal pin} is the fixed model a router collapses to at
$\lam=0$ (below).

A fixed \emph{mixing} baseline degenerates to the pin under a single objective, since expected harm is linear in the weights. A router conditions on $x$ and \emph{can} leave that hull, though it need not (Appendix~\ref{app:comparators}).

Full-strength routing is one point in a family indexed by shrinkage toward the pool marginal:
\begin{equation}
  s_i^{(\lam)}(x) \;=\; \lam\, p_i(x) \;+\; (1-\lam)\, \overline{p_i},
  \qquad \overline{p_i} = \mathbb{E}_x[p_i(x)] ,
  \label{eq:shrinkage}
\end{equation}
where $\lam=1$ recovers the near-universal per-request router. At $\lam=0$ the router becomes the
predicted-marginal pin, the model with the lowest mean \emph{predicted} harm. It matches the honest
pin only when predicted and realised training rankings agree. On HELM \texttt{harm\_bench} the two rankings
correlate at Spearman \num{0.96}, yet the $\lam{=}0$ choice issues the honest pin's decisions in only
three folds of five at $k{=}44$ (Section~\ref{sec:mechanism}). Treating $\lam=1$ as ``the router'' is a modelling choice, and an expensive one
(Section~\ref{sec:surface}).

Three quantities carry the analysis. The \emph{deferral rate} $\defrate$ is the fraction of requests
routed away from the pin. The \emph{conditional tail edge} is
\begin{equation}
  \edge \;=\; \mathbb{E}\!\left[h_{\pin} \mid \text{deferred}\right] \;-\;
              \mathbb{E}\!\left[h_{\text{chosen}} \mid \text{deferred}\right],
  \label{eq:edge}
\end{equation}
positive when deferrals genuinely beat the pin. It exists only where $\defrate>0$, so every $\edge$
and $\defrate$ below is marked conditional on deferring or unconditional. At small $k$ the two differ
by nearly an order of magnitude. Excess harm decomposes exactly as
\begin{equation}
  \underbrace{\mathbb{E}[h_{\router}] - \mathbb{E}[h_{\pin}]}_{\text{deficit}}
  \;=\; \defrate \cdot (-\edge),
  \label{eq:decomp}
\end{equation}
an identity under either baseline convention, with $\defrate$ and $\edge$ recomputed against
whichever pin is used. The identity closes per fold and per pool rather than after averaging (Appendix~\ref{app:comparators}).

We write $\astar$ for the signal AUROC at which the deficit crosses zero. It is a functional of the
whole score-generating process, including cross-model calibration and the correlation of scoring
errors, neither fixed by AUROC. Hence no $\astar$ is quoted here without the pool, estimator and
baseline convention producing it. Section~\ref{sec:surface} shows it is not a single number.

\subsection{Threat Model}
\label{sec:threat}
Four adversaries appear here. A result against one does not transfer to another, and none observes
weights or activations. A1 is an assumed ability to influence routing before dispatch. A2--A4 select templates for an
indirect injection arriving in a tool result after the model is chosen, the chronology of
Figure~\ref{fig:overview}, phase 2 (Section~\ref{sec:mechanisms-die}). 
\begin{itemize}
  \item \textbf{A1, routing-steering ($\advers$).} Forces a fraction $\advers$ of requests onto a
  cascade's cheap branch, and observes the routing policy but not the payload. $\advers$ means this and
  nothing else. It is motivated by, not drawn from, \citet{zhang2026rerouteguard}, who manipulate
  routing decisions by prepending triggers to a query before the router acts, which they report
  can bypass safety guardrails.
  \item \textbf{A2, post-hoc template selector.} Sees all measured outcomes and picks the
  lowest-flagging template after the fact. This is selection rather than adaptation, so it bounds A3 and A4 above.
  \item \textbf{A3, generic adaptive attacker.} Picks one template on attacker-train reruns,
  scored on held-out ones, and must serve the whole pool with it, since it cannot see which model answers.
  \item \textbf{A4, targeted adaptive attacker.} As A3, but per model, requiring \emph{target
  observability}, meaning knowledge of which model answers. That is a precondition on the deployment
  rather than a value of $\advers$. A1 and A4 are different adversaries that no single
  experiment varies together.
\end{itemize}

\section{Data and Evaluation Protocol}
\label{sec:data}

\begin{table}[t]\centering\footnotesize\setlength{\tabcolsep}{3pt}
\caption{The four corpora and the analysis set every number is computed on. XSTest enters as HELM Safety's over-refusal scenario, and its human-labelled release also supplies one row of Table~\ref{tab:e58}. Analysis sets are
smaller than the raw releases because a pool comparison is defined only where every model is scored
on every item; that complete-case filter is what removes the difference.}
\label{tab:corpora}
\begin{tabular}{@{}>{\raggedright\arraybackslash}p{0.30\linewidth}
                  >{\raggedright\arraybackslash}p{0.30\linewidth}
                  >{\raggedright\arraybackslash}p{0.32\linewidth}@{}}\toprule
corpus & raw release & analysis set \\ \midrule
HELM Safety \citep{liang2023helm} &
  \num{44} models $\times$ \num{393} behaviours, \num{2} judges &
  \num{393} behaviours, \num{7} categories \\
XSTest \citep{rottger2024xstest} &
  \num{450} prompts & \num{250} safe, \num{200} unsafe \\
HarmBench \citep{mazeika2024harmbench} &
  \num{54581} cells, \num{16}$\times$\num{29} &
  complete \num{19}$\times$\num{16} subgrid \\
AgentDojo \citep{debenedetti2024agentdojo} &
  \num{33119} runs, \num{28} configs, \num{15} attacks &
  \num{478} scenarios; \num{949} on the \num{5} complete \\
skill injection \citep{cua2026handcrafted} &
  \num{1862} episodes &
  \num{1342} undefended; \num{130} defended per defence \\
\bottomrule\end{tabular}
\end{table}

Table~\ref{tab:corpora} gives the four corpora and the analysis set each number here is computed on.
In the primary complete-case pools no model has zero observed harm over the full analysis set. The safest reaches harm \num{0.0331} on chat, \num{0.0021} on agentic and \num{0.2633} on skill injection (E53).
Three points do not fit the table. The HELM \num{393} are the complete-case intersection over both
published annotators (\texttt{safety\_gpt\_score}, \texttt{safety\_llama\_score}) and all
\num{44} models, so the label we route on is the union of two judges and is neither judge's own. AgentDojo's \num{5} complete configurations are only \num{4} distinct
base models, since the two safest are one injection-hardened model with and without a prompt defence
(Section~\ref{sec:two-settings}). The skill-injection corpus is narrower than its episode count suggests, covering a single harm
category, exfiltration, and \SI{91}{\percent} of its undefended episodes one domain. It was released in \num{2026} and
carries
the adaptive-attack result, so the newest surface here is also the one the adaptive-attack result is
measured on. Retrieval URLs are in Appendix~\ref{app:experiments}.

\lead{Instruments.}
Two common statistics do not survive their own nulls. Rank inversions across attacks, read as
evidence of routable heterogeneity, are produced by a zero-heterogeneity null more often than they are observed, putting the observed rate below its own null. Interaction is therefore
measured against a purely additive logit model resampled at the true per-cell counts. Under it,
model$\times$attack interaction on HarmBench is real, while \SI{84}{\percent} of a widely quoted
share on a second corpus is sampling noise (Appendix~\ref{sec:instruments}).

\lead{Evaluation discipline, and its departures.}
Cross-validation is group-aware by default (by attack family, semantic category, repository or user
task) and never by random splits over templates. AUROC is computed within each fold then
averaged, never pooled. Matched nulls or permutation tests accompany the claims that a signal exists
(predictability, interaction, headroom, rank inversion, geometry/signal split). The later comparative
results on the cost frontier, the per-family router and the adaptive attacker report effect sizes
with pool-level dispersion instead, marked as such in Appendix~\ref{app:experiments}. A non-semantic
baseline runs on identical input, and over-refusal or task utility is reported alongside the core harm
comparisons. Each rule followed a specific failure, recorded in the artifact's \texttt{ERRATA.md}.

The departures belong here rather than in a footnote. E25 adds a stationary random-split protocol as a deliberate contrast. Intervals are exact Clopper--Pearson counts or item-level bootstraps over behaviours, which condition on the observed groups (resampling unit recorded in the artifacts). E67 separately describes group-resampling sensitivity without a population coverage claim. The agentic per-family AUROC of \num{0.8839} has no interval
at all and an effective sample far below its \num{949} scenarios, which hold \num{99} distinct
feature vectors in \num{40} user-task groups. The base-rate axis of Section~\ref{sec:surface} selects
its honest pin on a \emph{random} within-band split and carries no null. Routers in E24--E27, E31 and E34 are fitted on outer-training columns through
\texttt{safety\_routing.nested}, a module tested by perturbing outer-test labels. A label-free
vocabulary fitted over all requests remains in the base-rate axis (E21, E39), featuriser ablation
(E23), encoder TF-IDF arm (E41) and the AUROCs quoted from E25 and E31. Refitting it on training requests
moves E25's two AUROCs to \num{0.6418} and \num{0.8549}. E30's anchor averages all
columns, and E2, E7--E9, E11, E14, E20 and E28 predate the module, so none of these is nested.
In-sample pins appear only as labelled controls, including the E4b sweep of Figure~\ref{fig:heatmap}.

\section{Baseline Selection Decides the Verdict}
\label{sec:baseline}

\begin{figure}[t]\centering
  \includegraphics[width=\linewidth]{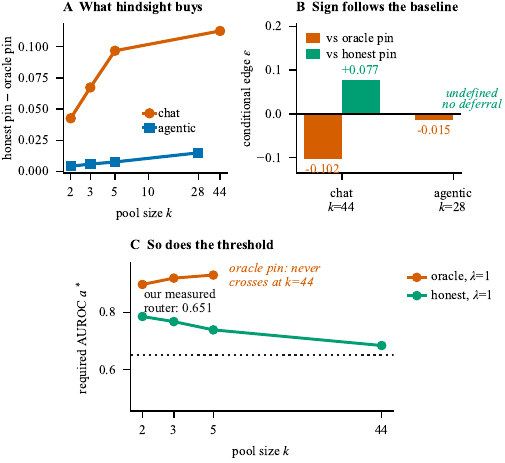}
  \caption{\textbf{Baseline choice changes the verdict.} All panels use within-fold in-sample pins,
panel A from E24 (chat) and E27 (agentic, the lower line), panels B and C from E26 on its own grid. (A) The harm advantage a pin gains purely from being selected on the
evaluation data rather than on training folds. (B) The conditional tail edge changes sign
once the pin is selected honestly, which dissolves an apparent chat-versus-agentic reversal. (C) The AUROC a router must reach at full strength ($\lam=1$), by pin protocol. The shrunken arm is reported in the text rather than drawn.}
  \label{fig:baseline}
\end{figure}

Every routing comparison we examined, ours included, scores a router against the best fixed model
in the pool. How that model is identified is rarely stated, yet it determines the answer.

Write $\pin_{\text{oracle}} = \arg\min_i \widehat{h}_i(\text{eval})$ for the model with lowest mean
harm on the evaluation data and $\pin_{\text{honest}} = \arg\min_i \widehat{h}_i(\text{train})$ for
the model selected on training folds and frozen. The first is the quantity \citet{commercial2026} report. We call it an \emph{in-sample} pin and keep the subscript for continuity (Section~\ref{sec:setup}).
Throughout this section it is minimised \emph{within} each held-out fold, the convention
Table~\ref{tab:pin-protocol} uses. The selection procedure cannot be run prospectively, since it needs the evaluation labels. The second is what a deployer has.

The asymmetry matters because the router is already held to out-of-sample discipline. Scoring an
out-of-fold router against an in-sample baseline charges the router for its uncertainty and forgives
the baseline its own. Selecting the router's own shrinkage on the test set likewise makes it ``win'' at every pool
size, so the two disciplines are one problem.

\begin{table}[!t]\centering\footnotesize\setlength{\tabcolsep}{2.8pt}
\caption{HELM harm on identical category-held-out folds (seed \num{24}).
Means over \num{200} pools $\times$ \num{5} folds for $k\leq5$; $k=44$ is one pool
$\times$ \num{5} folds. The in-sample pin is selected on each evaluation fold, the
honest pin on training folds; $\lam$ is tuned on training data. Table~\ref{tab:router-vs-pin}
instead uses a full-sample pin and $\lam=1$.}
\label{tab:pin-protocol}
\begin{tabular}{lcccc}
\toprule
$k$ & 2 & 3 & 5 & 44 \\
\midrule
pin (oracle) & 0.1873 & 0.1278 & 0.0654 & 0.0000 \\
pin (honest) & 0.2299 & 0.1951 & 0.1622 & 0.1127 \\
winner's curse & $+0.0426$ & $+0.0674$ & $+0.0968$ & $+0.1127$ \\
router $-$ pin (oracle) & $+0.0355$ & $+0.0536$ & $+0.0732$ & $+0.0819$ \\
router $-$ pin (honest) & $\mathbf{-0.0071}$ & $\mathbf{-0.0138}$ & $\mathbf{-0.0236}$ & $\mathbf{-0.0308}$ \\
\bottomrule
\end{tabular}
\end{table}

Table~\ref{tab:pin-protocol} and Figure~\ref{fig:baseline}A size the effect on HELM
\texttt{harm\_bench}. The gap to the in-sample pin,
$\widehat{h}(\pin_{\text{honest}}) -
\widehat{h}(\pin_{\text{oracle}})$, is nonnegative by construction, as the in-sample pin minimises over a set containing the honest one. It grows with pool size (Table~\ref{tab:pin-protocol}).
This growth is measured rather than entailed, since enlarging a pool can
lower either pin. It reaches the same order as the entire deficit we had attributed to routing.
It is not purely a winner's curse. Under a stationary split, with no category shift, the same
estimator gives \num{0.0026} at $k=2$ and \num{0.0305} at $k=\num{44}$ against \num{0.0454} and
\num{0.1127} on disjoint categories (E25, Figure~\ref{fig:surface}E). The contrast says most of the
gap is absent under the stationary split. It is a protocol contrast, not an additive split, and pools reuse the
same labels, so their count is not a count of independent deployments.

\lead{The headline loss does not survive the correction.}
Against the in-sample pin the router wins no pool cell at any pool size (E24). The cost is uneven across held-out folds, \numrange{0.013}{0.118} at $k=2$, and without its largest fold the mean
is \num{0.024} (E24j). On AgentDojo, whose held-out user tasks shift less, the router edges past
even the in-sample pin at $k=2$ and $k=3$ (E27). Under a random split
of the same requests the selection cost is only \numrange{0.003}{0.030}, and the router beats even
the in-sample pin at every pool size (E25), so the cost is a price of shift. A request-level
bootstrap within categories puts the held-out-minus-random difference above zero at every pool
size, \numrange{0.032}{0.049} at $k=2$ (E25b). Under the benchmark's own full-sample
convention, the comparator chosen once on all evaluation data, the cost is \numrange{0.001}{0.018}
under random splits and \numrange{0.022}{0.084} with categories held out (E25f). Against
$\pin_{\text{honest}}$ the mean difference reverses sign at every pool size
(Table~\ref{tab:pin-protocol}, E24, $-\num{0.0071}$ at $k=2$). The median over pool cells is
exactly zero at every pool size, not because routing works but because nested selection chooses
small $\lam$ and pulls the router toward one fixed choice. E26, a second run on its own grid, shows why. Its router never defers in \SI{91.5}{\percent} of
\num{750} cells at $k=2$, and its mean of $-\num{0.0155}$ comes from the \SI{8.5}{\percent} that
defer (Appendix~\ref{app:supporting}).

\lead{The pattern holds on two further benchmarks.} On AgentDojo, for $k=2$, $3$ and $5$, the cost is \numrange{0.003}{0.004}
under random splits and \numrange{0.027}{0.028} when whole suites are held out (E25a). On
LLMRouterBench's own released results \citep{commercial2026}, the benchmark's Best Single costs
nothing at its full \num{20}-model pool, under its random split or with whole datasets held out
(E57), so its convention is sound in the setting it is used for, and our registered fivefold
shift test there failed (Appendix~\ref{app:comparators}). On all six safety corpora a
rule fixed in advance admits, the held-out cost beats a group-permutation null on three, and on
AIR-Bench \citep{zeng2024airbench}, registered separately, too (E61). Against a perfect per-item router it is
\SI{57}{\percent} of the possible gain on HELM and \SI{40}{\percent} on SORRY-Bench with groups held out,
but \SI{3}{\percent} on HarmBench, whose fold assignment gives a lower cost than each of \num{200} random ones
(E58, E59, E64, Table~\ref{tab:e58}).
Three of seven meet the registered interval criterion, and four reach $p=\num{0.005}$ against the permutation null added after verification, HarmBench passing that test and not the interval one. The other three interval failures are the corpora where some models have zero observed harm, which leaves a selection cost little room.

\begin{table*}[t]\centering\scriptsize\setlength{\tabcolsep}{4pt}
\caption{Selection cost at $k=2$ on every eligible corpus (E58). The registered criterion,
an interval above zero, is met on the first three rows. The group-permutation null, added after
verification, is beaten at $p=\num{0.005}$ on the first four, at \num{0.010} on HELM \texttt{xstest},
where it depends on the pool draw, and not on the last two. AIR-Bench was registered separately
and its row enumerates every pair of models (E62), except its permutation $p$, which comes from E61's sampled pairs, while the other rows average \num{200} sampled pairs per fold. The null shuffles group labels at the same
group and fold sizes. Headroom is the held-out harm of the honest pin minus that of a perfect per-item
router over the pool, and $R$ is the selection cost over headroom, both with groups held out (E59).}
\label{tab:e58}
\begin{tabular}{lcccccccc}\toprule
corpus & models $\times$ items & groups & held out & random & difference, \SI{95}{\percent} & null $p$ & headroom & $R$ \\ \midrule
HELM \texttt{harm\_bench} & $44\times393$ & 7 & \num{0.0454} & \num{0.0026} & $[\num{0.0329}, \num{0.0491}]$ & \num{0.005} & \num{0.0797} & \num{0.57} \\
SORRY-Bench & $43\times44$ & 4 & \num{0.0076} & \num{0.0036} & $[\num{0.0007}, \num{0.0098}]$ & \num{0.005} & \num{0.0192} & \num{0.40} \\
AIR-Bench 2024 (E61) & $87\times5694$ & 16 & \num{0.0053} & \num{0.0002} & $[\num{0.0048}, \num{0.0055}]$ & \num{0.005} & \num{0.0431} & \num{0.12} \\
HarmBench & $19\times2210$ & 16 & \num{0.0019} & \num{0.0006} & $[\num{-0.0002}, \num{0.0023}]$ & \num{0.005} & \num{0.0617} & \num{0.03} \\
HELM \texttt{xstest} & $44\times200$ & 8 & \num{0.0044} & \num{0.0022} & $[\num{-0.0013}, \num{0.0035}]$ & \num{0.010} & \num{0.0084} & \num{0.52} \\
HELM \texttt{simple\_safety\_tests} & $44\times99$ & 5 & \num{0.0064} & \num{0.0074} & $[\num{-0.0041}, \num{0.0047}]$ & \num{0.343} & \num{0.0106} & \num{0.60} \\
XSTest & $5\times200$ & 8 & \num{0.0003} & \num{0.0005} & $[\num{-0.0027}, \num{0.0008}]$ & \num{0.930} & \num{0.0018} & \num{0.17} \\
\bottomrule\end{tabular}
\end{table*}

\lead{The direction holds under each judge, and the size does not.} HELM's label is the union of two
published judges. Scored under each judge alone, with categories held out in E25's folds or one at a
time, the held-out cost exceeds the random-split cost at $k=2$, $3$ and $5$ under both, \num{0.029}
against \num{0.002} at $k=2$ under the GPT judge and \num{0.043} against \num{0.002} under the Llama
judge (E65). At the full pool the size depends on the judge, \num{0.113} under the union against
\num{0.012} under the GPT judge, which is why we lead with small pools. The safety-routing evaluations
we found \citep{saferoute2025, kassem2025routerrobust, luo2026cascade} release no per-query outcomes, so
none can be re-scored under both conventions (search scope in Appendix~\ref{app:related-extended}).

\lead{Under shift the router chooses a fixed model.} On the category-held-out protocol, a fixed
selector based on training-only predicted mean risk accounts for \SI{98.2}{\percent} of the tuned
router's mean gain over the honest pin at $k=2$ and all of it at $k=3$, $5$ and $44$. The tuned policy
serves more than one model across a fold's held-out requests in only \SI{0.5}{\percent} of $k=2$ pool
draws and in none of the larger ones. Under random splits it does so in \SIrange{59.1}{100}{\percent}
of draws and beats the in-sample fixed model on average (E66). Allowing $\lam$ to vary by pool yields
occasional dispatch and slightly lower harm under category holdout, so the collapse characterises the
fold-shared tuned policy rather than every routing rule.

\lead{The contrast survives dropping any one group.} Exact-pair held-out-minus-random costs are
\num{0.0409} on HELM and \num{0.0224} on AgentDojo's \num{28}-configuration, \num{478}-scenario common
grid. Deleting one group at a time leaves them positive, \numrange{0.0107}{0.0490} and
\numrange{0.0238}{0.0412}. Copyright is influential on HELM. Its \num{100} items have a
leave-one-category-out cost of \num{0.119}, and removing it cuts the contrast to \num{0.0107} (E67).
A duplicate-safe group-resampling sensitivity analysis is positive in every valid sampled replicate,
which is not a population confidence guarantee, since exact enumeration finds rare negative HELM
resamples and AgentDojo's three suites allow only seven distinct values.

\subsection{What the Deferrals Are Worth}
\label{sec:mechanism}

On chat the conditional tail edge changes sign with the baseline, $+\num{0.0770}$ at $k=44$ against
$-\num{0.1023}$ under an in-sample pin. On agentic it is undefined, as the router never defers from the
honest pin at the full pool (Section~\ref{sec:two-settings}). The reversal is not a mechanism either, because at $k=M$
the deferral rate is \num{1.0} on the \num{2} folds of \num{5} that contribute, so the edge is the
whole-sample difference sign-flipped rather than an independent quantity
(Appendix~\ref{app:supporting}). Equation~(\ref{eq:decomp}) holds under either convention.

What the data do support is that correct nested tuning shrinks the router toward a fixed
choice and, at the full pool, collapses it onto one. At $k=44$ selection picks $\lam=0$ in all five
folds, and that fixed policy issues the pin's own decision on every request in \SI{60.0}{\percent} of
them. Selection now picks $\lam=0$ in every fold at $k=3$ and $k=5$ as well. Only at $k=2$ does
any fold take a positive weight, one of five at $\lam=\num{0.2}$. The collapse is the rule across
pool sizes rather than an artefact of the full pool. A $\lam=0$ policy is fixed-model selection, so
its wins over the pin concern choosing the fixed model, not per-request dispatch. Shrinkage acts through the
deferral rate rather than the edge. That leaves \num{2} of \num{5} folds at $k=44$ with any deferral at
all to estimate one from (Appendix~\ref{app:supporting}).
Appendix~\ref{app:supporting} gives the full decomposition, the conditioning that must be held fixed,
and the geometry-versus-signal control.

\lead{What a pre-execution router can buy.} A router choosing before execution selects among the
models' complete behaviours, so the per-scenario oracle bounds it. On complete subgrids of the
AgentDojo traces that gap is at most \num{0.0105}, on grids running from \num{1735} scenarios at
\num{5} configurations down to \num{477} at \num{28} (E43, reconciled with
Section~\ref{sec:two-settings}'s \num{478} in Appendix~\ref{app:supporting}). Taken against the
honest pin, the convention this section argues for, the family pool's headroom is \num{0.0180}.
Executed \emph{perfectly}, pre-execution routing is worth one to two points of harm on this
surface. The bound covers routing that selects among complete runs. A mid-run switch changes the
run itself, and scoring it by replay misstates the outcome at a rate of the same order as the
headroom \citep{replaygap2026}, so it is a separate estimand (Appendix~\ref{app:supporting}). Inside
the loop a harm signal is available (Table~\ref{tab:in-loop}, E18c, E18d). What we can test on the model's own recognition is
whether an attacker can move it, and template choice does (Section~\ref{sec:mechanisms-die}). Concurrent work finds a
hidden-state signal survives such variation \citep{ipiexposure2026}, which leaves the gap between
recognising and acting as the binding constraint.

The required-accuracy threshold $\astar$, where the deficit against the pin crosses zero, inherits
the convention. Against $\pin_{\text{oracle}}$ at $k=44$ no such accuracy exists, through degeneracy
rather than difficulty. That pin attains \emph{zero} harm on every evaluation fold
(Table~\ref{tab:pin-protocol}), so the deficit is bounded below by zero for any router at any accuracy,
AUROC \num{1.0} included. This is a property of the baseline. Against $\pin_{\text{honest}}$ on identical folds a full-strength router crosses at
\num{0.6842} at $k=44$ and \num{0.7857} at $k=2$, against \num{0.8976} at $k=2$ under the in-sample
convention (Figure~\ref{fig:baseline}C). The shrunken arm is identified too, crossing at
\num{0.6165} at $k=44$ and \num{0.6313} at $k=2$. An anchor computed over the evaluation
columns would censor it at the grid edge (Appendix~\ref{app:axes}). A router at the accuracy measured above therefore sits below the
full-strength threshold wherever it exists, and no single number here can be read without its
baseline convention.

\section{What Bounds a Safety Router}
\label{sec:surface}

\begin{figure*}[t]\centering
  \includegraphics[width=\textwidth]{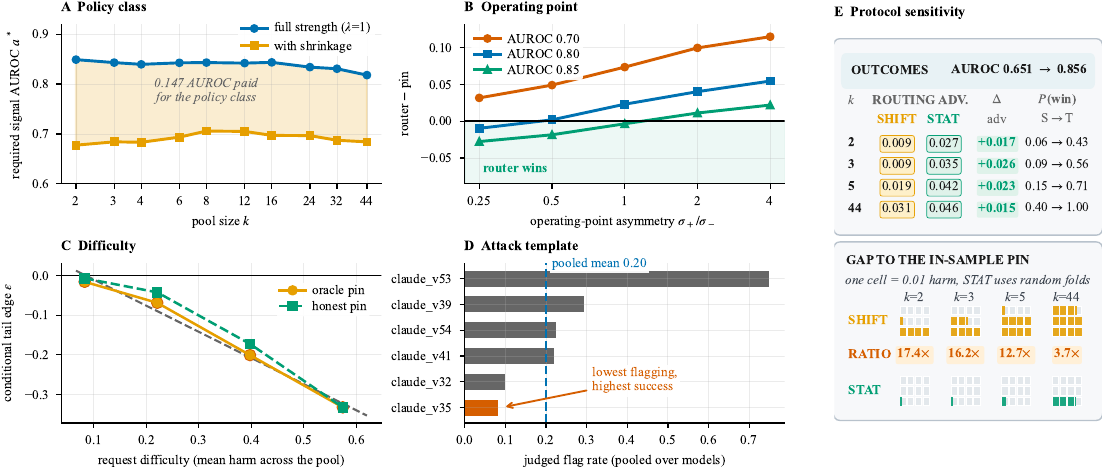}
  \caption{\textbf{Four axes of routing fragility, none of them accuracy.}
(A) \emph{Policy class.} Shrinking a per-request router toward the pool marginal lowers the AUROC it
needs by \num{0.147} under an in-sample pin, which is not itself a deployable threshold.
(B) \emph{Operating point.} At one fixed AUROC, changing only the shape of the score distribution
flips the sign of excess harm. (C) \emph{Difficulty.} Across bands of empirical difficulty the conditional tail edge falls,
under either pin, so deferral is worst on the requests most models fail.
(D) \emph{Attack template.} Pooled flagging varies by an order of magnitude across templates.
Templates are not scenario matched, so template and domain are confounded here.
(E) \emph{Protocol sensitivity.} The upper scorecard compares protocols. Each lower waffle cell is
one percentage point of the gap to the in-sample pin. STAT is the stationary protocol and SHIFT
the category-held-out one, and labels give the ratio. The STAT/SHIFT comparison is a protocol contrast rather than an additive decomposition.}
  \label{fig:surface}
\end{figure*}

The accuracy a router needs is not a scalar. Four axes move it, at least two by more than an accuracy
change of the size usually argued over, even before the baseline convention of
Section~\ref{sec:baseline}. Figure~\ref{fig:surface3d} in Appendix~\ref{app:axes} shows the operating-point slice,
Figure~\ref{fig:surface} gives all four axes, and Figure~\ref{fig:heatmap} gives the break-even
contour against pool size, an in-sample control. Appendix~\ref{app:axes} gives the protocols.

\subsection{Signal Exists, and the Router Still Loses}
On HELM \texttt{harm\_bench} (\num{44} models, \num{393} behaviours), per-model harm is predictable
from request text at AUROC \num{0.6509}, above a label-permutation null of \num{0.5049}, a sanity floor
from which we report no $p$-value (Appendix~\ref{app:supporting}). The signal generalises to
held-out semantic categories. On out-of-fold predictions the router still loses at
every pool size, beating a full-sample in-sample pin in at most \SI{4.0}{\percent} of \num{200}
sampled pools and losing outright at $k=\num{44}$ (Table~\ref{tab:router-vs-pin}). We retain that baseline only as the convention the
literature reports, and Section~\ref{sec:baseline} gives the deployable comparison.

\begin{table}[!t]\centering\footnotesize\setlength{\tabcolsep}{3pt}
\caption{HELM harm under out-of-fold, full-strength routing ($\lam=1$).
The pin is selected on all \num{393} evaluation behaviours; the oracle is the
per-request minimum. ``Beats pin'' covers \num{200} pools for $k\leq5$ and
one full pool at $k=44$. Table~\ref{tab:pin-protocol} uses a different pin protocol.}
\label{tab:router-vs-pin}
\begin{tabular}{lcccccc}\toprule
$k$ & router & pin & random & oracle & r.\ $-$ pin & beats pin \\ \midrule
2 & 0.2241 & \textbf{0.2157} & 0.3317 & 0.1572 & $+0.0084$ & 4.0\% \\
3 & 0.1904 & \textbf{0.1662} & 0.3437 & 0.0906 & $+0.0243$ & 3.5\% \\
5 & 0.1545 & \textbf{0.1062} & 0.3268 & 0.0293 & $+0.0483$ & 2.5\% \\
44 & 0.0840 & \textbf{0.0331} & 0.3363 & 0.0000 & $+0.0509$ & 0 of 1 \\
\bottomrule\end{tabular}\end{table}

Small encoders do not rescue the router. MiniLM scores \num{0.6060}, \num{0.0449} below TF-IDF,
against a non-semantic length baseline of \num{0.5264} that still beats TF-IDF on \num{12} of the
\num{44} models. Encoder scale, however, does move the signal. Across a \num{360}$\times$ span from MiniLM to Qwen3-Embedding-8B on identical folds,
AUROC rises to \num{0.6831} at \num{4}B and \num{0.6901} at \num{8}B against an honest-pin bar
of \num{0.6842} (Table~\ref{tab:featurizer}). Only the \num{8}B arm clears it. That bar comes from an
injected-signal sweep, so clearing it sizes the signal rather than showing that a router on this
encoder beats the pin. Cost does not rule out the large arms. An embedding router reads the request without generating and the \num{8}B arm adds about
\SI{1.2}{\percent} to the cheapest routing decision in our pool (Appendix~\ref{app:router-config}). Against it, the crossing is thin at
\num{0.0059}, per-model dispersion is
wide, and precision is not matched across the sweep, and the threshold $\astar$ moves with both the
baseline convention and the operating point. Defined here at $\opasym=1$, it shifts by
\numrange{0.09}{0.16} AUROC across the $\opasym$ sweep, many times the crossing
(Appendix~\ref{app:axes}). One further quantity moves this more than any encoder does, and it is
not a property of the router at all. Scored on a stationary split rather than a category-grouped
one, the same router on the same features reaches \num{0.8561} against \num{0.6509}. When test attacks come
from categories seen in training, the request's semantic category becomes learnable
(Appendix~\ref{app:supporting}). In a genuinely stationary deployment, this is legitimate information rather than leakage. There the router beats even the in-sample pin on average at every pool size, by \num{0.0154} to
\num{0.0357} (\num{0.0241} at $k=2$). Under grouped splits it loses to it by \num{0.0361}
to \num{0.0819} (E25).
Section~\ref{sec:limitations} states the scope condition that implies.

Scoring behaviour $n$ for model $i$ by the difficulty of $n$ estimated from the other \num{43}
models' outcomes, using no text at all, attains AUROC \num{0.7993}, above every text arm at any
scale we swept. We report it as a diagnostic rather than a bound in either direction
(Appendix~\ref{sec:instruments}).

\subsection{The Accuracy a Router Needs Is a Surface}
The threshold a router must clear moves with four quantities, each swept in
Appendix~\ref{app:axes}. With synthetic signal injected at a target accuracy, a full-strength router
needs $\astar\approx\num{0.84}$ against an in-sample pin and $\approx\num{0.69}$ under optimal
shrinkage. At a fixed AUROC of \num{0.80} the operating point alone decides whether the same router
wins or loses. The deferral edge falls from $-\num{0.016}$ to $-\num{0.332}$ as requests get riskier,
so it is worst where most models fail. The fourth quantity is the baseline convention of
Section~\ref{sec:baseline}. These sweeps inject signal rather than build routers on real features.

\section{The Agentic Setting}
\label{sec:two-settings}

Against an in-sample pin the mechanism appears to reverse between the chat and agentic settings, a
negative conditional edge on HELM against a positive one on AgentDojo. The correction of
Section~\ref{sec:baseline}, applied to both, removes the contrast.

\begin{table}[!t]\centering\footnotesize\setlength{\tabcolsep}{3pt}
\caption{Conditional tail edge, measured only when the router
defers. The chat comparison retains \num{2}/\num{5} folds (\num{146}/\num{393}
behaviours). The agentic comparison retains \num{0}/\num{5}: at the full pool the
router issues the honest pin's decision on every held-out scenario, so nothing is
deferred and the edge is undefined rather than zero. Unconditionally the agentic
figure is \num{0.0000}, the router being the pin by construction in all five folds.}
\label{tab:edge-common}
\begin{tabular}{lcccc}
\toprule
setting & $k$ & \multicolumn{2}{c}{conditional edge vs.} & median, deferring \\
\cmidrule(lr){3-4}
 & & in-sample pin & honest pin & router $-$ pin \\
\midrule
    chat (HELM) & 44 & $-0.1023$ & $\mathbf{+0.0770}$ & $-0.0770$ \\
    agentic & 28 & $-0.0149$ & n/a & n/a \\
\bottomrule
\end{tabular}
\end{table}

Chosen honestly, the pin removes the router's apparent disadvantage. Against the honest pin on
identical held-out folds, the conditional edge is positive on chat, \num{0.0770}, and does not exist on
agentic (Table~\ref{tab:edge-common}). The edge is conditional
on the router's choice differing from the pin's. On chat this retains \num{2} of \num{5} folds and
\num{146} of \num{393} behaviours, and the unconditional edge is $+\num{0.0308}$. On agentic it
retains none, because the router issues the pin's own decision on every held-out scenario, so there is no
deferred set and no edge. The reversal was an artefact of the in-sample pin biting harder on the more heterogeneous corpus. In both settings, how badly an out-of-fold router appears to fail depends on how much hindsight the
comparator is given.

The correction is small here because the corpus is saturated, leaving hindsight little to exploit.
One configuration, \texttt{claude-3-5-sonnet}, carries it, at harm
\num{0.0021} against \num{0.0105} for the next (E46). In the \num{5}-configuration family
pool, the two safest are the same injection-hardened weights with and without a prompt defence
(Section~\ref{sec:data}).

\lead{The agentic numbers cover one attack family, not fifteen.} A pool comparison is well defined only when every configuration is scored on every scenario. Enforcing this complete-case filter collapses
AgentDojo from \num{7735} scenarios across \num{15} attack families to \num{478} across a single
family, \texttt{important\_instructions}, and every agentic quantity reported above is that family
alone. Sparse coverage is the cause
(Appendix~\ref{app:agentic-detail}).

\lead{A strong per-request signal, and a policy that declines to condition on the request.} Dropping
the cross-family complete-case requirement and rebuilding \texttt{important\_instructions} on its
own \num{5} configurations raises the usable scenario count to \num{949} and the per-request
predictor to AUROC \num{0.8839}. The policy we then score declines to condition on the request. On
this family nested selection chose $\lam=0$ in all \num{5} folds, so the rule issues one pick for
all held-out items and never conditions on the input (E31). The predictor enters only through a row
mean, which severs AUROC \num{0.8839} from the harm figures below. We call the rule the
\emph{predicted-marginal pin}. Fitted without sight of the evaluation scenarios and read in canonical configuration order, it is
the honest pin, issuing that pin's choice on every held-out item in all \num{5} folds. Averaged over
the pool's permutations it is behind the pin by \num{0.0005} harm and ahead by \num{0.0021}
utility, and beats it in no sampled pool (Table~\ref{tab:family-router}). The harm margin is not routing. One fold has an exact training-harm tie that permutations split between two configurations,
and the margin sits inside that fold's swing (E48). Of the \num{3} families with
enough coverage to analyse, it is
also the only one whose headroom clears our \num{0.01} threshold for attempting a router
(Appendix~\ref{app:agentic-detail}).

\begin{table}[!t]\centering\footnotesize\setlength{\tabcolsep}{3pt}
\caption{AgentDojo by attack family, without cross-family
complete-case filtering. Headroom is fixed-pin harm minus per-request oracle
harm. In the lower block nested tuning chooses $\lam=0$ in all five folds:
the marginal pin is a fixed model selected from mean predicted scores,
not a request-level router; its per-request AUROC \num{0.8839} is not the
performance of the scored fixed policy. Fitted without sight of the evaluation
scenarios and read in canonical order it selects the honest pin's own configuration in every
fold, and the figures average permutations, one fold tying (E48).}
\label{tab:family-router}
\begin{tabular}{lccccc}
\toprule
attack family & cfg & scen. & fixed & oracle & headroom \\
\midrule
    \texttt{important\_instr.} & 5 & 949 & $0.0211$ & $0.0053$ & $\mathbf{0.0158}$ \\
    \texttt{direct} & 4 & 949 & $0.0126$ & $0.0074$ & $0.0053$ \\
    \texttt{ignore\_previous} & 4 & 949 & $0.0137$ & $0.0095$ & $0.0042$ \\
\midrule
 & \multicolumn{2}{c}{harm} & \multicolumn{3}{c}{utility} \\
\cmidrule(lr){2-3}\cmidrule(lr){4-6}
    marginal pin ($\lam{=}0$) & \multicolumn{2}{c}{$0.0238$} &
      \multicolumn{3}{c}{$0.8115$} \\
    honest pin  & \multicolumn{2}{c}{$0.0232$} &
      \multicolumn{3}{c}{$0.8093$} \\
    difference  & \multicolumn{2}{c}{$\mathbf{+0.0005}$} &
      \multicolumn{3}{c}{$\mathbf{+0.0021}$} \\
\bottomrule
\end{tabular}

\smallskip
\footnotesize The marginal pin's choice differs from the honest pin's in
\num{0} of the \num{5} held-out folds; median difference
\num{0.0000}.
\end{table}

The corpus cannot settle the case the proposal is meant to address. With one configuration already
near-perfect, AgentDojo shows routing to be unnecessary when a dominant model exists, not what it
does when several models have genuine and complementary vulnerabilities. No corpus we have exhibits
that naturally, and our attempt on a live skill-injection surface returned AUROC \num{0.5000} under
complete-case filtering and is void (Appendix~\ref{sec:experiments}). This is the principal gap in
the evidence, and we do not treat the agentic result as closing the question.

\lead{An artificially de-saturated pool.} Dropping the dominant configurations does let a router
beat the honest pin on most held-out folds, at a cost in task utility. We do not report this as
evidence that routing works, because the dropped configurations were chosen by their measured harm on the
same data, so the pool embeds the very hindsight Section~\ref{sec:baseline} corrects
(Appendix~\ref{app:agentic-detail}).

\lead{Structural blindness of pre-execution routing.} Harm appears to vary far more by attack
type than by user task, but \num{12} of the \num{15} families ran on one configuration only.
Holding the configuration fixed, the two spreads are \num{0.1512} and \num{0.1261}, so most of that
gap is composition (E46). A prompt-only router reaches AUROC \num{0.647} here, which is one
featuriser's measured value rather than a ceiling we computed
(Appendix~\ref{app:agentic-detail}).

The released corpora differ greatly across surfaces, under different attacks and evaluation designs. Browser-borne injection recorded no success on
two frontier models in a handcrafted \num{140}-attack sample \citep{cua2026handcrafted} while the skill-loading path stays open
(Appendix~\ref{app:attack-surface}). Defences move harm further than model choice does. On
\texttt{gpt-4o}, matched to the single attack family the defended runs cover, harm falls from
\num{0.4769} undefended to \num{0.0684} under the strongest defence (E45). A pooled comparison appears to show
one defence raising harm to \num{0.417}, but it sets defended runs on one family against undefended
runs on all fifteen. Matched by family, every defence lowers harm. Pool
elements must therefore be (model, defence) pairs.

\section{Defences Scored on a Signal}
\label{sec:mechanisms-die}
\label{sec:what-works}

Section~\ref{sec:baseline} scored a baseline on something no deployer has. The defences we examined next each look better on a signal than on harm through the system that reads it. A cascade's saving is mostly its cheap model, a triage idea did not survive pricing, and a flag-gated controller's outcome is set by its fallback. Appendix~\ref{app:defences-detail} gives each in full.

\lead{A cheaper cascade is mostly a cheaper model.} With pin, hull and budget fixed on training
data, \num{16} of \num{96} genuine cascades have lower held-out harm than the honest hull at the same
planned budget with no adversary ($\advers=0$), configurations an oracle frontier had hidden, by at
most \num{0.47} points, and \num{6} of them cost more at realised cost (E34). For \num{12} of the
\num{13} that hold utility the saving is the cheap model, not escalation, and for the thirteenth
escalation cuts harm by \num{5.68} points (E47). An adversary able to force cheap-model routing, an
assumed steering ability and not an attack we ran, gets whatever that model costs.

\lead{Recognition can be steered.} On the skill-injection corpus harm occurs in \num{0.0037} of
undefended episodes where the agent flags the injected content and \num{0.6309} where it does not
(E44), so a defence gated on the flag looks near-perfect. The flag and the success label come from
one judge call, so the rubric may account for part of that gap. An attacker that knows which model
it faces, choosing its template on half of the reruns and scored on the other half, lowers
\texttt{gpt-5.4}'s matched judged flag from \num{0.226} to \num{0.030}, \num{19.6} points (E54b),
and every other model is already at its lowest-flag released template under the static arm. A fixed
lexicon that reads no judge field agrees with the judge on \texttt{gpt-5.4} ($\kappa=\num{0.82}$)
and reproduces the drop within each of the three run configurations that contain both templates
(permutation $p\leq 10^{-4}$, E55). The attacker chooses among the released templates each model was
run on, three to six, so this measures selection, not optimisation.

\lead{Through a controller, the fallback decides.} In an offline composition of the logged reruns into controllers, which assumes the flag is seen
before the target acts and replaces a silent episode by a fresh draw of the fallback,
escalate-on-silence covers \SIrange{94}{98}{\percent} of tasks because two models never flag and a
third flags \num{2} of \num{420}, so the system is effectively its fallback (E56, E56b). The
flag-lowering attacker cuts system harm by \numrange{4.9}{10.6} points only because one fallback,
\texttt{claude-sonnet-4-6}, resists every template but the static one, and without that model the
sign reverses (\numrange{2.3}{8.9}). Here the fallback's vulnerabilities, not the flag, set the
outcome, so a flag measurement alone did not certify the defence.

\lead{Action-level policies, as run, are consistent with recognition.} Four action-level policy
settings, user confirmation, denied network egress, code audit and a second pass over the plan,
record zero judged successes on the same \num{130} episodes from \num{30} rerun indices
(Table~\ref{tab:defences}). Matched on model, template, rerun index and sandbox mode, only \num{10}
episodes are informative, with a Clopper--Pearson bound of \num{0.308} on residual success (E40).
Every configuration sets \texttt{docker\_sandbox: false} and the release has no harness enforcing
the policy. No defended episode contains a mechanical-block marker, while \num{86} of \num{520} carry
policy language against \num{0} of \num{1342} undefended. The zeros are consistent with the agent
declining, the recognition an attacker's template choice moved for \texttt{gpt-5.4}. The design
these results point to, untested here, enforces policy outside the agent's context
\citep{debenedetti2025camel}.

\section{Related Work}
\label{sec:related}

\lead{What we add.}
\Citet{shihab2026opportunity} name same-sample selection of the best fixed model as a flaw in routing
evaluation and prove the direction of its bias on accuracy. We size it on harm, contrast random with
category-held-out splits on identical folds, and decompose it, with the sign of the tail edge under
each convention. \Citet{lassance2026dcs} count how often deferral beats the single
best overall, choosing single-best and threshold on a validation split, so that sign is not a quantity their
construction surfaces (Appendix~\ref{app:related-extended}). To it we add the
geometry-versus-signal decomposition and a held-out adaptive attack on the flag the defence is
gated on. Adaptive attacks defeat many injection detectors \citep{zhan2025adaptive, nasr2025attacker}, one built on the agent's own backbone, with strings optimised per test case. Detectors designed against an attacker who targets them are evaluated against one \citep{datasentinel2025, pishield2025}, and DataSentinel reports a failure when the injected and target tasks share a type. All of these score a separate detection call. We measure the acting model's own expressed recognition and move it with templates chosen on training reruns and scored on held-out ones.

\lead{Safety routing as deployed, and attacks on it.}
Safety routing exists as deployable open-source infrastructure \citep{vllmsemanticrouter2026}. Agent routers already fix one model per task at admission, for cost and quality rather than safety \citep{tracerouter2026}, and Section~\ref{sec:baseline} bounds that design's safety value.
Attacks on the routing decision are established. \citet{zhang2026rerouteguard} prepend triggers
that manipulate which model is chosen and report that this can bypass safety guardrails, which is the mechanism our $\advers$ adversary abstracts. Where routing is applied to safety it selects a
\emph{detector} rather than the model that answers \citep{saferoute2025, mo2026triaging}. Closest to
our agentic half, \citet{replaygap2026} show that scoring a per-step model switch by replay
misstates the outcome, which is why we bound the pre-execution case and decline the in-loop one
(Section~\ref{sec:limitations}).

\lead{Collapse toward the pool marginal.}
\Citet{lai2026routing}, \citet{routingplateau2026} and \citet{garg2026router} each document routers
converging on a single model, from small-margin selection failure, a narrow accuracy plateau,
and a degenerate likelihood ratio under a strong label prior. What we add is a measurement of one nested rule's behaviour under each baseline convention, compared with a signal-free null. The winner's curse
\citep{smithwinkler2006, bastani2025winners} and validation-tuned deferral from a single best
model \citep{lassance2026dcs} are established, so Eq.~\ref{eq:decomp} is not novel.

\lead{Constraints on actions.}
Indirect prompt injection is due to \citet{greshake2023}. CaMeL \citep{debenedetti2025camel}
enforces the principle we argue for, keeping untrusted data from changing program flow or acquiring
authority while its policies still inspect that data's provenance.
\Citet{beurerkellner2025patterns} catalogue the pattern class. We measure four instances of it on one live surface and find them consistent with recognition
(Section~\ref{sec:what-works}).

\lead{Concurrent and adjacent.}
\citet{chen2026gap} and \citet{chen2026cofailure} bound the headroom a router could recover and
\citet{dekoninck2024router} characterise optimal routing, whereas we come from the other side.
\Citet{manczak2024primeguard} route within one model, so there is no pin to beat. Our
grader swap lacks the ordering preservation \citet{benchmarkless2026} report, and template fragility
is scoped against \citet{ipiexposure2026} to adversaries that optimise against the gate.
Appendix~\ref{app:related-extended} gives all of these in full, with the AI-control comparison
\citep{greenblatt2023control, adaptive2024untrusted} and the judge-instability results
\citep{strongreject2024, chouldechova2026comparison, cyclicjudge2026}.

\section{Limitations}
\label{sec:limitations}

\lead{Where the results apply.} The four primary corpora cover two chat, one agentic and one
skill-injection setting, in English and largely single-turn on the chat side. The shift we study is
a held-out request category. Of seven safety corpora its size is large only on HELM
\texttt{harm\_bench}, and nothing we registered predicts it, including the size we predicted for
AIR-Bench (E58, E61). Its size also depends on which groups share a fold. Over \num{200} random
assignments the HELM $k=2$ cost is \num{0.037} on average, $[\num{0.026}, \num{0.045}]$, our fixed
assignment sitting high, against \num{0.0026} under random splits (E64). Under a stationary mix the in-sample comparator is nearly harmless and
routing pays, beating even the in-sample pin at every pool size (Section~\ref{sec:surface}), so a deployment that genuinely
never meets new categories sits outside our warning. No agentic surface we have is naturally
non-saturated. The one pool with real headroom is de-saturated by dropping configurations on their
measured harm (Section~\ref{sec:two-settings}).

\lead{How harm and recognition are measured.} Every harm number except the human-labelled XSTest row is read through an automated judge,
which on the skill corpus is also the target in some of the defended episodes it grades
(Section~\ref{sec:what-works}), and the same judge call writes the flag, so flag and success share
one annotator. Grading its own output it records attack success \num{0.263} against \num{0.575} for
the other four models (E36), which cannot separate being the safest model from being lenient on
itself. Grader recall $\hat{\beta}$ is unmeasured. If a grader misses a fraction $\beta$ of true
harms, attack-success rates are deflated by roughly $1/(1-\beta)$, and the zeros of
Section~\ref{sec:what-works} and the $[0, \num{0.028}]$ bound are the most exposed. Which chat model
is safest, and the size of the selection cost, depend on the judge, while the collapse onto a fixed
choice holds under each (Appendices~\ref{sec:instruments} and~\ref{app:comparators}). An independent lexicon label reproduces the recognition
result, measuring expressed rather than internal recognition.

\lead{What the attackers can do.} $\advers$ forces cheap-model routing on a random subset rather than
a targeted one, so only $\advers=0$ and $\advers=1$ are trustworthy (Appendix~\ref{app:supporting}).
The template-choosing attacker picks among six corpus templates, whose names suggest tuning against
one model family, so the result is about choosing a template, not optimising one. The two adversaries have different preconditions and are joined by
argument rather than a joint experiment. The controller results assume a fallback's re-run is a
fresh draw from its own episodes on the same payload (Appendix~\ref{app:controller}).

\lead{The next measurements.} An in-loop \emph{router} needs episodes resumable under a different
model from a chosen step, and the corpora here store one complete run per configuration. The
literature is arranged the same way (Appendix~\ref{app:related-extended}), and we found no evaluation
of mid-trajectory model switching against an attack. It needs branching rollouts on a surface whose
models fail in complementary ways, which ours do not (E43). The structural constraints are costed on
harm only, and whether the four policy settings are enforced rather than recognised needs a harness
we could run. Our recall is bounded by what a web and arXiv search surfaced, and two references
cited in Appendix~\ref{app:experiments} are unread in full text, with no claim resting on either.

\section{Conclusion}
\label{sec:conclusion}

Safety routing is being graded against a false floor. A comparator chosen on the test labels costs almost nothing when tomorrow's requests look like today's, and there the router genuinely wins, on the stationary protocols we measured. Once
the categories shift, which is the regime safety routing exists for, the same comparator absorbs a
cost as large as the deficit blamed on routing, and a router that never beats it ties an honest
baseline in the median. The pattern holds on both corpora we study in depth and on three of seven by a registered interval test and four by a later permutation null, while in the routing
benchmark's own setting the convention is harmless. Its direction is guaranteed, so it recurs in any evaluation that picks its baseline with the answers in hand. Report both conventions, on the same folds, under shift.

Scored that way, the case for routing is thin. The accuracy a router needs is a surface rather than a
number, and the deferral edge is worst on the requests most models fail. A request-text encoder
reaches that surface only at \num{8}B and thinly, as a signal size rather than a routing policy we
ran (Section~\ref{sec:surface}). Under grouped splits a tuned router mostly serves the honest
baseline's own model (Section~\ref{sec:baseline}), so pinning is defensible on simplicity, though not
free.

\lead{What to do instead.}
A router that commits before the injection arrives is worth at most two points of harm on these AgentDojo grids, and the model's expressed recognition, measured on GPT-5.4, bends to an attacker's template choice. Composed offline into a controller, the same attacker raises or lowers harm
depending on which fallback serves the escalation, and no flag measurement reveals which
(Section~\ref{sec:mechanisms-die}). The four policy settings recorded zero judged successes, and their traces are consistent with that same recognition, not proof of it (Section~\ref{sec:what-works}). The defence worth building is
enforcement over actions in the harness \citep{debenedetti2025camel}, outside the agent's context,
scored end to end on harm, under shift, against an attacker who chooses what the model sees.

\lead{An evaluation protocol.} Declare the deployment split, naming which groups are held out. Choose
and freeze the baseline on training data only. Report the in-sample comparator separately, as a
hindsight reference and not a baseline. Publish the paired per-request losses and the grouping
units, so that a reader can re-score both conventions.

\section*{Ethical Considerations}
\addcontentsline{toc}{section}{Ethical Considerations}

The paper's security-relevant result is a signal-evasion measurement. Section~\ref{sec:mechanisms-die} shows that an attacker
who knows which model it faces, choosing its template on some reruns and scored on held-out
reruns, lowers the pooled judged flag rate by about four points on matched configurations, entirely
on one of five models, while its success gain depends on how configurations are weighted. Composed offline into flag-gated controllers from logged episodes, the attack's effect on harm is set by the fallback
model rather than by the flag, so we claim a steerable input signal, not a defeated defence, and a
deployer relying on a model's own recognition should know both before trusting it.

We disclose no new attack technique. Every template used is already published in the corpus we draw
on \citep{cua2026handcrafted}, and the contribution is the measurement that per-model selection
transfers to held-out reruns, not the construction of a payload. No model was jailbroken to produce novel harmful content, and every retained harm label comes from
previously released evaluations scored by previously published judges. The one exception is a
withdrawn pilot (E19) that generated agent actions on \num{100} sampled cases, and its outputs are kept
in the artifact and enter no result.

\lead{Vendor disclosure.} The evading templates are already public. The post-hoc evasion is on the
commercial model named in Table~\ref{tab:flag-suppression}, and the held-out flag reduction is on
\texttt{gpt-5.4} (Table~\ref{tab:adaptive}). What is new is the measurement, that selecting a
template against a known model transfers to reruns held out from that selection.
We treat that measurement, and not the template, as the disclosable object. We have notified no
vendor as of this version. We intend to notify both affected model vendors through their published
reporting channels and will record the date here. We
judge this sequence proportionate under the Menlo Report's beneficence and respect-for-persons criteria,
since the marginal capability this paper adds is a selection procedure over already-published
payloads rather than a new attack, and the population whose risk changes is whoever would deploy a defence
that treats a model's flag as evidence of safety, who are also the audience this result is addressed to.

Two further considerations. The corpora contain harmful requests and successful attacks, and we
release analysis code and derived statistics but no generated harmful outputs. And our strongest
positive result, that an honest frontier reveals a configuration cheaper than the pin at no more
harm (Section~\ref{sec:mechanisms-die}), holds only where the attacker cannot influence routing, or where
the model it can force is itself safe. Reporting it without that condition would invite deployment
without checking the forced destination, so we state the condition wherever the result appears.

\section*{Open Science}
\addcontentsline{toc}{section}{Open Science}

We release the full analysis artifact: every experiment script, the JSON output each one writes, the
table generator, and the figure generators. Each statistic is the output of a script in that
artifact, and Appendix~\ref{app:experiments} registers every
experiment against the file that produces it, including the experiments we killed. Of the \num{17} table sources the shared generator emits \num{16}, and every one of
those reproduces each printed number, differing at most in layout or in captions corrected during
review. The generator asserts that the experiment register covers every result file on disk, so a
run that is added and not registered fails the build rather than shipping a register that
under-reports. Where we call a criterion or prediction registered, it was written in our append-only
decision log before the computation. The artifact's \texttt{PREREGISTRATION.md} copies those entries
verbatim. They are not independently time-stamped, and each departure from them is reported as a
deviation. Tables written inline in the section files, including Table~\ref{tab:e58} and the corpus summary, are not generated, and their numbers are checked against the result files by the claim map instead. We claim less than
bit-exact recovery of every printed numeral, and we say where. Pool-sampling summaries are averages
over randomly drawn pools, and where a value in the text was read off a larger draw than the released
script fixes by default, re-running the released script reproduces it only up to sampling noise,
and those quantities are written to the JSON of the script that owns them, at that script's own seed and
pool count, so a reader can see both the released value and the value we quote. A small number of
diagnostics are recorded in the script's log rather than in its JSON. A few tables are computed by the table generator
from the released episode records rather than read from a JSON field, and the artifact's claim map
lists every mapped number with its source and every unmapped one with a reason.

\lead{Artifact availability.} The artifact, with every experiment script, its committed result file, the
claim map and the reproduction commands, is available from the authors on request.

\lead{Data.} Every corpus is public and was retrieved, not generated. The four that carry the
routing results are: HELM Safety
\citep{liang2023helm} (its \texttt{harm\_bench} and \texttt{xstest} scenarios, drawing on
HarmBench \citep{mazeika2024harmbench} and XSTest \citep{rottger2024xstest}, \num{44} models, two
published judges), HarmBench \citep{mazeika2024harmbench}, AgentDojo
\citep{debenedetti2024agentdojo}, and a released skill-injection corpus
\citep{cua2026handcrafted}. The instrument checks of
Appendix~\ref{sec:instruments} additionally use a released SORRY-Bench result grid, which also supplies one row of the comparator audit in Table~\ref{tab:e58}. Retrieval scripts are included. We redistribute no
model outputs containing harmful content.

\lead{Protocol.} The core routing comparisons use group-aware cross-validation, grouped by attack
family, semantic category, repository or user task and never randomly over templates, and fit
policies and honest baselines on outer-training data. Section~\ref{sec:data} identifies the stationary splits, transductive diagnostics, in-sample
controls and legacy non-nested analyses that depart from this. AUROC is
computed within each fold and then averaged, never pooled across folds. In those comparisons the baseline policies (the pin,
the cost-harm frontier, the safety floor, the cheap model and the escalation threshold) are selected
on training folds only and evaluated on held-out folds. Section~\ref{sec:baseline} explains why this
distinction decides the verdict, and the artifact's \texttt{ERRATA.md} records the experiments in which we
originally got it wrong. Over-refusal or task utility is reported alongside the core harm comparisons.
Permutation nulls accompany the existence claims and not every number, while the comparative results
report effect sizes with pool-level dispersion instead.

\lead{Known limits on exact reproduction.} Three quantities are not bit-reproducible. Cost
figures depend on a list-price table fixed at the time each model was current, though the table is
included and the qualitative verdict is reported under a sensitivity sweep over the price ratio.
Pool-sampling results average over randomly drawn pools at a fixed seed, so counts may shift by one
or two under a different seed. Fine-tuned encoder results depend on hardware. We train one fold per
process after finding that folds sharing a process on Apple Silicon produced non-finite losses from
the second fold onward.

\section*{LLM usage considerations}
\addcontentsline{toc}{section}{LLM usage considerations}

LLMs were used for editorial purposes in this manuscript, and all outputs were inspected by the
authors to ensure accuracy and originality. Large language models also assisted with literature search, with writing and debugging analysis code,
and with drafting and editing text, including typographical and grammatical corrections. The
authors directed the research, verified every result and are responsible for all content. Every
reference was checked at its primary source, every script was run and its output inspected, and
the numbers in the manuscript are mapped to their artifact sources by an automated check, with any
unmapped number listed in the artifact's register with a reason. All harm, flag and
over-refusal labels come from released evaluations scored by previously published judges. The only
hosted model calls embed request texts for the encoder arms (E41, E41b) and generate actions for a
withdrawn pilot (E19), whose outputs are kept in the artifact.

\bibliographystyle{IEEEtranN}
\bibliography{bib/references}

\appendices
\section{Instruments}
\label{sec:instruments}

\begin{figure}[t]\centering
  \includegraphics[width=.85\linewidth]{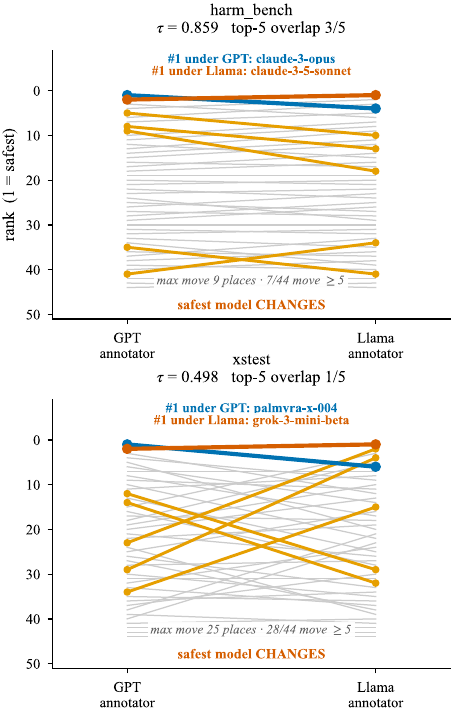}
  \caption{\textbf{Swapping only the judge changes which model is safest.} Two independently
  published annotators on identical instances. Kendall $\tau=0.859$ on \texttt{harm\_bench}, where the
  safest model still changes and the judges share 3 of the top 5, and $0.498$ on the over-refusal
  axis, where they share 1 of 5.}
  \label{fig:grader-swap}
\end{figure}

Four statistics in common use do not support the weight placed on them. We give the replacements and
the evidence, because the rest of the paper depends on measuring correctly.

\lead{Rank-inversion counts.}
Observing that model rankings invert across attacks is routinely offered as evidence of routable
heterogeneity. Under a zero-heterogeneity null (one true success rate per attack, binomial sampling
noise, $n=\num{139}$), \SI{99.2}{\percent} of model pairs invert on at least one of \num{16} attacks
(\SI{95}{\percent} interval \numrange{97.1}{100.0}). The observed rate is \SI{43.8}{\percent}, which is
below its own null. Rank inversion is what sampling noise looks like.

\lead{Use an additive null for interaction.}
The right instrument is a two-way variance decomposition against a purely additive (zero-interaction)
logit model, resampled binomially at the true per-cell counts. Under it, the model$\times$attack
interaction on HarmBench is real (\num{0.185} against a null of \num{0.065}), but \SI{84}{\percent} of
a widely quoted \num{0.2025} interaction share on SORRY-Bench is sampling noise
(Table~\ref{tab:instruments}).

\lead{Full-pool headroom.}
``Refused by some models but not all'' saturates as the pool grows. At $M=\num{44}$ it reads
\num{0.945}, while a one-dimensional pool calibrated to the same marginals gives
\numrange{0.918}{0.942}. The deployable quantity is measured at $k\in\{2,3,5\}$ against a matched
one-dimensional (Rasch) null. Across the deployable pool sizes and both detectors, \num{6} of \num{6}
cells show no excess over a one-dimensional pool.

\lead{Insufficiency of AUROC.}
Holding AUROC fixed and varying only the shape of the class-conditional score distributions moves the
routing outcome more than a \num{0.05} change in AUROC does, and the argmin router has no threshold. At
AUROC \num{0.80} the same router either wins or loses depending on that shape alone (Appendix~\ref{app:axes}). This confirms, for model selection,
an argument made for model evaluation by \citet{chouldechova2026comparison}.

\lead{Reliability of the label itself.}
Two independently published judges scoring identical instances (Figure~\ref{fig:grader-swap}) agree
on \SI{82.8}{\percent} of cells, yet a perfect predictor of one judge's label scores only AUROC
\num{0.746} against the other's. We report this as target ambiguity and not as an information bound,
since a predictor is free to fit either label exactly. It bounds nothing, and in particular it is not
refuted by any predictor that scores above \num{0.746} against a different target: the \num{0.746}
is one judge's label scored against the other's, and it can only be compared with numbers measured
against that same pair. What it does mean is that an objective defined by one judge and audited by
another is effectively specified to within \num{0.746}, and any safety comparison should state which
judge fixed the target.

One consequence deserves stating once, because it applies to every harm number we report on this
corpus. The label we route on is the \emph{union} of the two judges, so that an instance counts as
harmful if either judge scores it harmful and the target is neither judge's label. Quantities defined with
respect to one judge and quantities defined with respect to the union are not interchangeable, and we
do not compare them.

\begin{table}[!t]\centering\footnotesize\setlength{\tabcolsep}{3pt}
\caption{Interaction variance against a \emph{purely additive} (zero-interaction) null. The
interaction is real on one corpus and largely sampling noise on the other.}
\label{tab:instruments}
\begin{tabular}{lccc}\toprule
corpus & grid & \shortstack{observed\\interaction} & \shortstack{no-interaction\\null (90\%)} \\ \midrule
HarmBench & $19\times16$ & 0.1854 &
  [0.0580, 0.0715] \\
SORRY-Bench & $43\times44$ & 0.2025 &
  [0.1624, 0.1759] \\ \bottomrule
\end{tabular}
\end{table}

\lead{A reference point for what item identity carries.} Scoring behaviour $n$ for model $i$ by the
difficulty of $n$ estimated from the other \num{43} models' outcomes on it, using no text at all,
attains AUROC \num{0.7993}. It needs labels for the very item it scores, so it cannot be deployed.
Regressing text embeddings onto that same target recovers \num{0.6505}: as excess over chance, text
carries $(\num{0.6505}-\num{0.5})/(\num{0.7993}-\num{0.5})=\SI{50}{\percent}$ of what those outcomes
carry.

Fine-tuning end to end instead, using a shared transformer with one sigmoid head per model retrained
inside every fold, reaches \num{0.6674}, \num{0.0120} above the best frozen arm and \num{0.1319}
short of \num{0.7993}, so learning the representation closes about a tenth of that distance. On the
mean the remainder is not the vectoriser. Per-model TF-IDF AUROC nevertheless spans
\numrange{0.499}{0.797}, as wide as the distance discussed, so this is a statement about averages.

\lead{What this does and does not bound.} We report \num{0.7993} as a diagnostic, not a ceiling. It
bounds no router conditioning on item identity, a class that contains the map from an item to its
own label and attains AUROC \num{1.0}. Nor, in either direction, routers conditioning on shared item
difficulty, absent a cross-model conditional-independence assumption that we do not make and that
our own interaction result argues against (\num{0.185} against a null of \num{0.065},
Table~\ref{tab:instruments}), since the \num{43}-draw difficulty estimate attenuates while residual
cross-model dependence lets the other models' realised outcomes carry what no shared difficulty
parameter holds. It is a reference point for how far text-only features sit from a non-deployable
label-based one, and no necessity claim rests on it. The fine-tuned arm uses
\texttt{distilroberta-base} rather than a larger encoder, a hardware constraint and not a design
choice.

The label our routers are trained against is the union of the two judges, so the \num{0.746} of
the reliability paragraph above applies to neither of them directly.

\subsection{Router training configuration}
\label{app:router-config}

Every router in this paper is a per-model harm predictor over a frozen featurisation, with one
exception noted below. The configuration is identical across featuriser arms so that a difference
between arms is a difference in the features and not in the fit.

\lead{Shared predictor.} \texttt{LogisticRegression}, $C=\num{1.0}$, \texttt{max\_iter}
$=\num{2000}$, fixed random state, fitted once per model per fold. Cross-validation is
\texttt{GroupKFold} with $\min(\num{5}, n_{\text{groups}})$ splits, grouped by semantic category on
HELM and by user task on AgentDojo. E25's stationary protocol instead uses random five-fold
\texttt{KFold}. AUROC is computed \emph{within}
each fold and then averaged, and out-of-fold probabilities are never pooled before scoring, a
measured error recorded in the artifact's \texttt{ERRATA.md}. A fold whose
training split is single-class falls back to the training mean and contributes no AUROC.

\lead{Featuriser arms.} TF-IDF is $1$--$2$-gram, \texttt{sublinear\_tf}, \texttt{min\_df}
$=\num{2}$, English stop words. Sentence encoders are used frozen with mean pooling and
$L_2$-normalised outputs. The six frozen arms are TF-IDF, MiniLM, MPNet, BGE, the character-length
control, and a concatenation of the four encoders, which is the best of them at \num{0.6555}. The non-semantic control is character length, its logarithm, and word
count.

\lead{Fine-tuned arm, the one exception.} A shared encoder with one sigmoid head per model, trained
end to end inside every fold: \texttt{distilroberta-base}, \num{8} epochs, AdamW at learning rate
$2\times10^{-5}$, batch size \num{16}, maximum sequence length \num{128} with dynamic per-batch
padding, fixed seed. \texttt{roberta-base} and \texttt{ModernBERT-base} were also run. One fold is
trained per process: folds sharing a process on Apple Silicon produced non-finite losses from the
second fold onward, which is a hardware interaction rather than an optimisation failure and is why
the artifact spawns a process per fold.

\lead{Encoder scale and what it costs.} The scaling arms are served two ways and the serving
configuration is part of the result, not an implementation detail. We had expected to rule the large
arms out on cost and the arithmetic does not support it. An embedding router reads the request and
does not generate, and requests here are short (mean \num{109} tokens, median \num{21}), so at
provider list price, with tokens estimated at four characters each, the \num{8}B arm costs about
\$$\num{1.1}\times10^{-6}$ per request. Against a
\num{300}-token response that is \SI{1.21}{\percent} of the cheapest model in our pool and
\SI{0.005}{\percent} of the dearest. The break-even price for \SI{1}{\percent} overhead is
\$\num{0.008} per million tokens, and both priced arms sit above it, at \$\num{0.01} (\num{8}B, a
list price that is not the serving provider's) and \$\num{0.02} (\num{4}B), so they cost \SI{1.21}{\percent} and \SI{2.41}{\percent} of the cheapest
call. Cost at that scale does not decide this question. The thin crossing margin, the wide per-model dispersion and the unmatched
precision across sizes do (E42).

\section{The Four Axes in Full}
\label{app:axes}

\begin{figure}[!t]\centering
  \includegraphics[width=.82\linewidth]{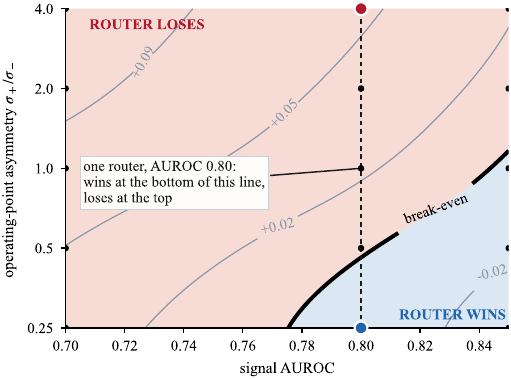}
  \caption{\textbf{Accuracy alone does not decide whether routing helps.}
Excess harm of a full-strength router over an in-sample pin, against signal
accuracy and score asymmetry $\opasym=\sigma_+/\sigma_-$, at pool size
$k=3$. Dots are measured cells and the surface between them is interpolated.
Zero is break-even, and above it the router is worse than pinning one model.
The key reading is at a fixed AUROC of $0.80$, where moving
asymmetry from $0.25$ to $4$ carries the router from winning to losing while
its accuracy never changes.}
  \label{fig:surface3d}
\end{figure}

\subsection{Four Axes, None of Them Accuracy}
\label{app:four-axes}
The threshold a router must clear is a surface over four quantities, each swept in
Appendix~\ref{app:axes}.

\lead{Policy class} costs AUROC on its own, in an amount that depends on the estimator. With
synthetic signal injected at a target AUROC, a full-strength router needs $\astar\approx\num{0.84}$ against
an in-sample pin and $\approx\num{0.69}$ under optimal shrinkage, a gap of \numrange{0.134}{0.172}
over a \num{22}$\times$ range in pool size. Held-out estimators put it at \numrange{0.068}{0.154} against an honest pin and \numrange{0.032}{0.074} against an in-sample one, common
in sign, not in level (Appendix~\ref{app:axes}). On the same injected-signal sweep with a
within-fold rather than a full-sample in-sample pin, the threshold moves with both convention and
pool size. At $k=\num{44}$ under the in-sample convention it does not exist at any accuracy, as
that pin attains zero harm on every evaluation fold (Figure~\ref{fig:baseline}C, Appendix~\ref{app:axes}). Both estimators inject signal at a target
accuracy rather than rebuilding the router on real features.

\lead{Operating point} reverses the verdict at fixed AUROC (Figure~\ref{fig:surface3d}). At
\num{0.80} the same router wins ($-\num{0.0095}$) or loses ($+\num{0.0548}$). A
\num{16}$\times$ change in $\opasym$ moves the deficit \numrange{1.9}{3.1}$\times$ as far as a
\num{0.05} change in AUROC does. This ratio compares two ranges we chose, not a property of the
router. An honest pin shifts a row down by one constant, leaving the spread across $\opasym$
untouched and moving only the sign of individual cells. The induced \emph{leftward} shift of the break-even
contour is not that constant but the constant divided by the local slope of the deficit in accuracy.
On identical folds it is \num{0.11} AUROC at $k=2$, \num{0.15} at $k=3$ and \num{0.19} at $k=5$,
and it is undefined at $k=\num{44}$, where no in-sample crossing exists.
Table~\ref{tab:operating-point} in Appendix~\ref{app:axes} gives the sweep.

\lead{Difficulty} runs the wrong way, and is the one axis we have re-measured under both baseline
conventions. The bands are cut on each request's mean harm across the pool, so they compare
requests of different observed difficulty rather than one deployment at different prevalence, and
the unconditional deficit is not monotone across them. Across the four bands, the tail edge falls monotonically as risk
rises, from $-\num{0.016}$ to $-\num{0.332}$ against an in-sample pin and from $-\num{0.007}$ to
$-\num{0.333}$ against an honest one. The correlation is $r=-\num{0.99}$ and $-\num{0.98}$,
but over $n=\num{4}$ binned points with no null and no interval. The finding is thus the
monotonicity rather than the coefficient, and it survives normalising each edge by its band's base rate
(Appendix~\ref{app:axes}), which matters because the bands are cut on the same labels that enter the edge. Whether some band has the router beating the pin does depend on the baseline (Appendix~\ref{app:axes}).
Table~\ref{tab:base-rate} in Appendix~\ref{app:axes} gives the four bands.

\begin{figure}[t]\centering
  \includegraphics[width=\linewidth]{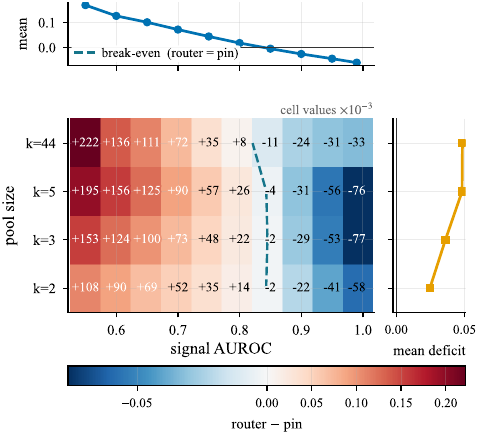}
  \caption{\textbf{For raw scores the break-even contour is nearly vertical.} Deficit over signal accuracy $\times$
  pool size, with marginals. In E4b's raw-score arm against a full-sample in-sample pin, the zero contour runs
  almost parallel to the pool-size axis, so here the accuracy a router needs barely depends on pool size. E4b's rank-calibrated arm needs more, falling with $k$ (Table~\ref{tab:experiments}). Over the swept
  accuracy range the mean deficit does grow with pool size (\num{0.0246} at $k{=}2$,
  \num{0.0360} at $k{=}3$, \num{0.0482} at $k{=}5$, \num{0.0485} at $k{=}44$), which is the right-hand
  marginal. At \emph{fixed} accuracy it is not monotone in $k$. $k{=}44$ sits below $k{=}5$
  from \num{0.60} to \num{0.85}, and above break-even every cell changes sign. The right-hand
  marginal thus averages a deficit below the contour against an advantage above it.}
  \label{fig:heatmap}
\end{figure}

Section~\ref{sec:surface} shows the four-axis plot. The protocols and caveats are detailed here.
Figures~\ref{fig:surface} and~\ref{fig:heatmap} in the body give the four axes as slices and the
break-even contour against pool size. Both are computed against an in-sample pin, which is the
convention that makes them comparable with the published thresholds they answer. Replacing that pin
with an honest one shifts the whole map \emph{down} by the winner's curse of
Section~\ref{sec:baseline}, which is a quantity in harm units and not a displacement along the
accuracy axis. The induced leftward movement of the break-even contour is the curse divided by the
local slope of the deficit in accuracy, and that slope is about \numrange{-0.38}{-0.51} harm per
unit AUROC over the swept range, so the horizontal shift is larger than the curse itself. Measured
within E26 on identical folds at full strength, $\astar$ falls from \num{0.8976} to \num{0.7857} at
$k=2$, from \num{0.9197} to \num{0.7678} at $k=3$ and from \num{0.9305} to \num{0.7389} at $k=5$,
shifts of \num{0.112}, \num{0.152} and \num{0.192} AUROC. They are approximately E24's curses \num{0.0426},
\num{0.0674} and \num{0.0968} divided by that slope, E24 and E26 being different pool draws. We report no ratio of a shift to a curse, the one being an
accuracy and the other a harm. At $k=\num{44}$ the in-sample contour does not exist at all (Axis~1 below), so no shift is defined there.

\subsection{Axis 1: Policy Class (\num{0.147} AUROC)}
Under the synthetic-signal sweep, which injects a classifier at a target AUROC and locates the
crossing, a router needs $\astar=\num{0.8387}$ at full strength and $\num{0.6914}$ under optimal
shrinkage against an in-sample pin. Both levels are means over the ten pool sizes of E9, the dense
version of the E4b and E6 sweeps. The gap between the two policy classes is
\numrange{0.134}{0.172} across that \num{22}$\times$ span in pool size ($k=\numrange{2}{44}$,
mean \num{0.147}), and both curves are flat over it (ranges \num{0.031} and
\num{0.028}), though the full-strength curve carries a weak declining trend inside that range that
the range alone hides (Spearman \num{-0.70} with $k$. \num{0.8492} at $k=2$ against \num{0.8181} at
$k=\num{44}$).

Recomputing on the measured router (E26) gives different levels. At full strength, $\astar$ is
\num{0.8976} at $k=2$, \num{0.9197} at $k=3$ and \num{0.9305} at $k=5$ against an in-sample pin, and
\num{0.7857}, \num{0.7678}, \num{0.7389} and \num{0.6842} at $k=2,3,5,\num{44}$ against an honest pin (Section~\ref{sec:baseline}). One entry of that sweep has to be withdrawn rather
than quoted. At $k=\num{44}$ the within-fold in-sample pin attains \emph{zero} harm on every evaluation
fold, so the deficit is bounded below by zero for any router at any accuracy, including AUROC
\num{1.0}. $\astar$ does not exist under that convention, and its absence is a property of a
degenerate baseline and not a signal requirement.

The shrinkage arm is now usable, which it was not when we last reported it. Its anchor was the
marginal of the \emph{evaluation} columns, which in a synthetically injected arm turns $\lam{=}0$
into an in-sample-pin selector rather than a router and drove every out-of-sample value to \num{0.60},
the left edge of the swept grid. Recomputed with the anchor on training columns only, as E7 already
did, the shrunken threshold is \num{0.6313}, \num{0.6379}, \num{0.6305} and \num{0.6165} at
$k=2,3,5,\num{44}$ against an honest pin, clear of the grid edge at every pool size. The honest policy-class gap is therefore \numrange{0.068}{0.154}, and we withdraw the statement that
we can quote no such gap.

The level of $\astar$ remains estimator-dependent and should not be quoted without one, and the
correction sharpens rather than softens that warning. Three estimators now sit on this axis: the
in-sample sweep of E9 gives a policy-class gap of \numrange{0.134}{0.172}, the held-out sweep
against an honest pin gives \numrange{0.068}{0.154}, and the same held-out sweep against an in-sample pin gives \numrange{0.032}{0.074}. We previously reported that the in-sample-convention gap replicated across estimators. With the corrected anchor it does not, and we withdraw that too. What
survives is the sign, which all three agree on: shrinking the policy class lowers the accuracy a
router needs, by an amount no single number states.

\subsection{Axis 2: Operating Point}
We sweep the operating-point asymmetry \num{16}$\times$ at fixed AUROC.
Table~\ref{tab:operating-point} makes the reversal explicit. At AUROC \num{0.80} the same router wins ($-\num{0.0095}$) or loses ($+\num{0.0548}$)
depending only on the shape of the class-conditional score distributions, and the argmin router has no
threshold.

These deficits are measured against an in-sample pin. The spread across the operating point,
which is the axis, is unaffected by that choice, and provably so, since the pin does not depend on the router's
score-distribution shape, so replacing an in-sample pin with an honest one shifts every cell in a
row by the same constant and leaves their differences identical. What the choice does move is the sign
of individual cells. Against an honest pin each row shifts down by a constant of the order of the
winner's curse at that pool size, so more cells fall on the winning side. We do not import a figure
for that constant from elsewhere in the paper without saying so: this sweep has no train/test split
at all, so no honest pin exists inside its own design, whereas the \num{0.0674} at $k=3$ of
Section~\ref{sec:baseline} is measured on out-of-fold predictions under category-grouped
cross-validation, and the curse depends on the split protocol and not only on $k$. The claim this axis
supports is that the operating point moves the outcome more than accuracy does. The claim that a
particular router at \num{0.80} loses is baseline-dependent and should not be read from this table
alone.

\begin{table}[!t]\centering\footnotesize\setlength{\tabcolsep}{1.5pt}
\caption{At \emph{fixed} AUROC, varying only the asymmetry $\opasym=\sigma_{+}/\sigma_{-}$ of the
class-conditional score distributions over a $16\times$ range, at $k{=}3$. $\opasym$ changes the
\emph{shape} of the ROC curve at fixed area; the router is an $\arg\min$ over models and applies no
threshold, so it does not select an operating point and no row describes a decision rule. The last
column compares two analyst-chosen ranges, a $16\times$ change in $\opasym$ against a $0.05$
change in AUROC at $\opasym{=}1$, and is not scale-free. Deficits are means over \num{150} pools
$\times$ \num{30} repetitions against a full-sample in-sample pin; per-cell dispersion is large
relative to the cells themselves (SD \numrange{0.056}{0.074}) and no interval is reported. The spread
across $\opasym$ is invariant to the baseline choice, since the pin does not depend on $\opasym$; the
sign of individual cells is not.}
\label{tab:operating-point}
\begin{tabular}{lccccc}\toprule
AUROC & $\opasym{=}0.25$ & $\opasym{=}1$ & $\opasym{=}4$ & spread & \shortstack{vs.\ a $0.05$-\\AUROC move} \\ \midrule
0.70 & $+0.0319$ & $+0.0738$ & $+0.1152$ & 0.0833 & 3.13$\times$ \\
0.80 & $-0.0095$ & $+0.0233$ & $+0.0548$ & 0.0643 & 2.41$\times$ \\
0.85 & $-0.0276$ & $-0.0034$ & $+0.0221$ & 0.0497 & 1.87$\times$ \\
\bottomrule\end{tabular}\end{table}

\subsection{Axis 3: Base Rate (monotone degradation over four bands)}
A natural defence of routing is that we measured a benign regime. The opposite holds. Stratifying
scenarios by base rate and recomputing the tail edge within each band gives a monotone negative
relationship (Table~\ref{tab:base-rate}): the conditional edge runs
$-\num{0.007}$, $-\num{0.042}$, $-\num{0.173}$, $-\num{0.333}$ across the four bands under the honest pin, and $-\num{0.016}$, $-\num{0.069}$, $-\num{0.201}$, $-\num{0.332}$ under the in-sample pin.

\lead{The difficulty axis is not a scale artefact.} Dividing each band's edge by that band's base
rate leaves the monotone pattern intact across the four bands, at $-0.192$, $-0.312$, $-0.505$ and
$-0.579$ against the in-sample pin and $-0.086$, $-0.192$, $-0.434$ and $-0.579$ against the honest
pin (E39).

\lead{Baseline dependence.} The monotonicity does not depend on the baseline, but the accompanying
claim does. Against an in-sample pin the router beats the pin in under \SI{1.5}{\percent} of
draws in every band. Against an honest pin it does so in up to \SI{11}{\percent}, and in the \numrange{0.15}{0.30} band its mean
difference is negative ($-\num{0.0021}$), so some band has the router beating the pin. The
monotone degradation, which is the axis, still holds. On this axis alone the honest pin is selected on a random within-band split of scenarios rather than a group-aware
one (Section~\ref{sec:data}), which makes it the weaker of our two honest protocols.

\lead{What the correlation is and is not.} The Pearson $r$ of $-\num{0.98}$ (honest) and $-\num{0.99}$ (in-sample) is computed over the $n=\num{4}$ band summaries, not over scenarios. With four
points and no null it carries no more information than the monotonicity itself, Spearman being
$-\num{1}$ by construction for any four monotone points, so we quote it to two decimals and rest
the axis on the ordering. The bands are also cut on the per-item mean of the same labels that enter
the edge, and the edge lives on the same scale as the harm rate, so part of the growth is mechanical.
Dividing each band's edge by that band's base rate removes the scale (the normalised sequence
above), and that sequence, rather than $r$, is the form of the claim we
would defend.

\lead{The split protocol here differs from the rest of the paper.} This axis was originally measured
against an in-sample pin, the pin selected on the same scenarios it was scored on, which is the error
Section~\ref{sec:baseline} exists to correct and which we did not apply here until late. Re-running
with the pin chosen on disjoint scenarios within each band leaves the relationship intact and the edge
in the top band unchanged at $-\num{0.3325}$. That re-run splits the scenarios inside a band
\emph{at random}, not by category: it is the STATIONARY protocol of Section~\ref{sec:baseline} and not
the group-aware SHIFT protocol used in the core comparisons, a distinction that moves router AUROC
from \num{0.65} to \num{0.86} where we measured it directly. A random split makes the two halves
exchangeable, so this ``honest'' pin is closer to the in-sample pin than the E24 and E26 honest pins are,
and the winner's curse it removes is understated. The top band is the visible symptom: its two
conventions agree to four decimals ($-\num{0.332470}$ in-sample against $-\num{0.332549}$ honest), which
is what one would see if the two pins were the same model in that band. The stored aggregates do not
record the selected pin, so we cannot check that, and we report the coincidence as unverified rather
than as evidence that the relationship is baseline-free.

The re-run does not leave everything intact. We had previously reported
that no band has the router beating the pin. Under the in-sample convention it is rare (under
\SI{1.5}{\percent} of draws in every band) but not absent. Against an honest pin the router beats it in
\SIrange{3.7}{11.0}{\percent} of draws, and in the \numrange{0.15}{0.30} band the mean difference is
negative. The monotone degradation is the finding and it survives. The never-beats claim is withdrawn.

\begin{table*}[!t]\centering\small
\caption{Empirical difficulty is the third axis and it runs the wrong way, under \emph{either} baseline
convention. Bands are cut on each request's mean harm across the pool, its base rate, so they
compare requests of different observed difficulty rather than one deployment at different prevalence. Pool size $k=3$, \num{300} random pool draws per band; $n$ is the number of scenarios in
the band, and each tail edge is a mean over the draws in which the router defers at least once. The
correlation between band base rate and tail edge is $r=-\num{0.99}$ (in-sample pin) and $-\num{0.98}$
(honest pin), but it is a Pearson coefficient over the \num{4} band means, with no null and no
interval, so the monotonicity is the finding and the coefficient is not. The honest pin on this axis
is selected on a \emph{random} half of each band's scenarios, not on a category-grouped split, which
makes it the weaker of our two out-of-sample protocols; in the top band the two conventions agree to
four decimals, and the stored aggregates do not record whether the two pins ever differed there. The
``beats pin'' columns, fractions of the \num{300} draws, are where the conventions part company: the
claim that no band has the router beating the pin holds only against the in-sample pin.}
\label{tab:base-rate}
\begin{tabular}{lcccccc}\toprule
& & & \multicolumn{2}{c}{in-sample pin} & \multicolumn{2}{c}{honest pin} \\
\cmidrule(lr){4-5}\cmidrule(lr){6-7}
base-rate band & $n$ & base rate & tail edge & beats pin & tail edge & beats pin \\ \midrule
0.00--0.15 & 108 & 0.082 & $-0.0159$ & 0.3\% & $-0.0071$ & 11.0\% \\
0.15--0.30 & 67 & 0.221 & $-0.0689$ & 1.3\% & $-0.0425$ & 10.7\% \\
0.30--0.50 & 113 & 0.397 & $-0.2007$ & 1.3\% & $-0.1726$ & 4.7\% \\
0.50--0.70 & 90 & 0.574 & $-0.3325$ & 1.0\% & $-0.3325$ & 3.7\% \\
\bottomrule\end{tabular}\end{table*}

\subsection{\texorpdfstring{Axis 4: Attack Template (\num{10}$\times$ in flag rate)}{Axis 4: Attack Template (10x in flag rate)}}
On a live skill-injection surface, one model flags \numrange{99}{100}\% of attacks under three
templates, \SI{45}{\percent} under a fourth and \SI{6.7}{\percent} under a fifth, which is also the
highest-success template. Its aggregate flag rate of \SI{68}{\percent} overstates its flag rate on the
template an adversary would choose by an order of magnitude, with template and domain confounded
(Table~\ref{tab:flag-suppression}, Section~\ref{sec:mechanisms-die}).

\section{Supporting Measurements}
\label{app:supporting}

Each of these results is summarised in the body. The measurements are reproduced here in full.

\subsection{Moving the Decision Into the Loop}
\label{sec:in-loop}

\begin{figure*}[!t]\centering
  \includegraphics[width=.8593\textwidth]{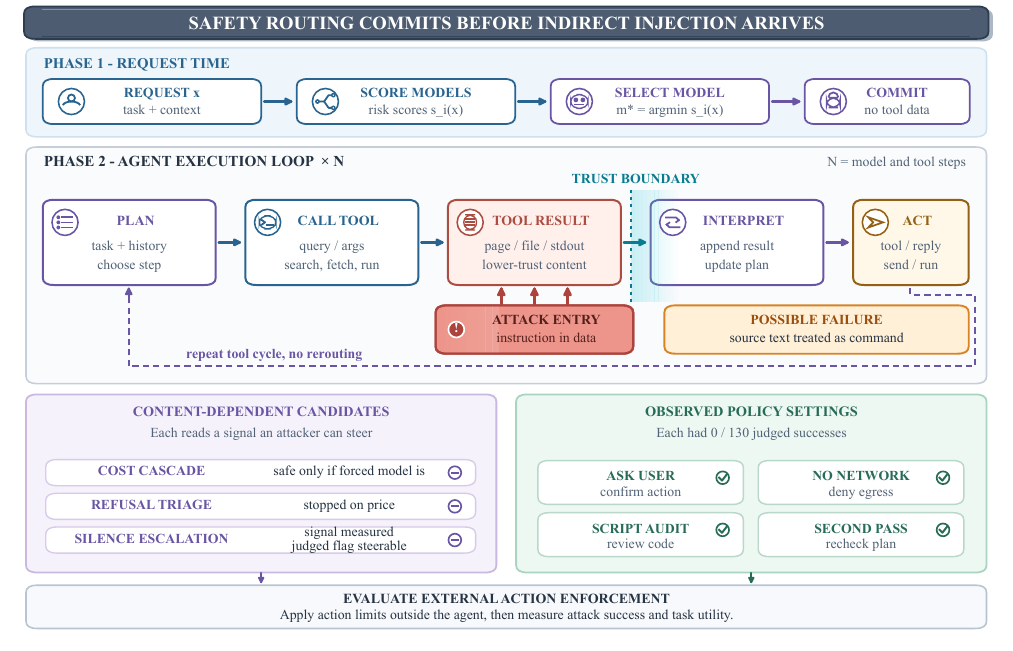}
  \caption{At request time, the router scores $x$ and commits to model $m^*$ before seeing tool
  output. In Phase 2, the agent plans, calls tools, reads lower-trust results, interprets them,
  and acts across $N$ model and tool steps without rerouting. Red arrows show attacker-controlled
  instructions entering a tool result, and teal shading marks where that result enters model context. If
  source text is treated as a command, the next action can change. A1's pre-routing steering (Section~\ref{sec:threat}) is outside this picture. The lower panels show three
  content-dependent candidate defences (one evaluated, one priced, one composed into controllers, E56b) and four policy settings with $0/130$ judged successes
  each under one automated judge. Those results motivate the proposed external action gate (Section~\ref{sec:what-works}).}
  \label{fig:system}
\end{figure*}

Section~\ref{sec:two-settings} showed that a pre-execution router is structurally blind: it commits
before the injection exists. The obvious repair is to stop deciding once. Re-decide at every step, and
the attack becomes visible the moment the agent reads the poisoned content. We measure whether that
visibility is worth anything.

\lead{Two inflating confounds.}
A naive version of this measurement is badly contaminated. Pooling across models, AUROC leaps from
$0.500$ to $0.803$ as soon as the first assistant message enters the prefix, but that is the
classifier recognising which model is speaking, since per-model harm rates span $0.011$--$0.563$.
Prefixes cut at a fixed offset from the injection step leak the label through their length:
$\mathrm{corr}(\text{injection step}, \text{harm}) = -0.259$, and episodes with early injections have \SI{28.7}{\percent} judged harm while those landing at step ten or later cause essentially none. We
therefore compute AUROC within each model and hold the injection step fixed. Removing the confounds
changes the within-population difference from $+0.029$ to $+0.086$ under user-task folds, and a
stricter split, below, removes it.

\lead{What the in-loop monitor reads.}

\begin{table}[t]\centering\small
\caption{What the decision point can see. AUROC within model. The three in-loop rows hold the
injection step fixed ($n={872}$ runs, \num{5} models); the pre-execution row is a different
population, the \num{28}-configuration complete-case grid over \num{478} scenarios, so the
within-population comparison is \num{0.722} against \num{0.808}. These rows measure whether the
signal is \emph{available} at each decision point. No agent was re-executed with a different model
mid-run, which this corpus cannot support. With user tasks and attack families held out together (E18d)
the two steps score \num{0.705} and \num{0.703}, so the signal is available but its rise is not robust.}
\label{tab:in-loop}
\begin{tabular}{lc}\toprule
decision point & AUROC \\ \midrule
pre-execution, prompt only (28-config grid) & 0.647 \\
mid-trajectory, \emph{before} the injection is visible & 0.722 \\
\textbf{the step injected content enters context} & \textbf{0.808} \\
one step later & 0.804 \\ \midrule
tasks and attacks held out, before & 0.705 \\
tasks and attacks held out, at injection & 0.703 \\ \bottomrule
\end{tabular}\end{table}

Table~\ref{tab:in-loop} gives the ladder. Under user-task folds, reading the injected content moves
the monitor from \num{0.722} to \num{0.808}. With attack families also held out, both steps sit at
\num{0.705} and \num{0.703} (E18d), so the signal is available at about \num{0.70} and reading the
injection adds nothing measurable. The $+0.161$ against pre-execution routing is a descriptive
difference across two populations. The $\astar$ thresholds of Section~\ref{sec:surface} are
functionals of a particular pool's marginal-harm heterogeneity, its size and its score-generating
process, and ours were computed on the chat corpus with a synthetically injected signal, whereas
$0.808$ is an AgentDojo measurement. A threshold is not a corpus-free bar, so the comparison that
would place this monitor is an AgentDojo-specific $\astar$.

\lead{Why this does not rescue safety routing.}
First, concurrent work establishes the same point more strongly than we do. \citet{ipiexposure2026}
predict injection exposure from hidden states at \SI{90}{\percent}+ AUROC across eight models,
generalising to unseen attacks, and holding at \numrange{0.956}{1.000} AUROC under two adaptive
attacks on two of those models, neither optimised against the probe. We report
our result as an independent replication with weaker features and not as a contribution.

Second, and more importantly, visibility is not the binding constraint. The defence one would build on
in-loop recognition, escalating when the model fails to flag, reads a signal template selection moves
(Section~\ref{sec:mechanisms-die}). The same model flags \numrange{99}{100}\% of attacks under three
templates, \SI{45}{\percent} under a fourth and \SI{6.7}{\percent} under a fifth, which is also the
most successful attack available. A monitor's input is steerable wherever the attacker can choose whether
the attack registers, as it can here for one model.

\lead{Summary of the in-loop result.}
In-loop visibility gives a usable signal, about \num{0.70} under the strictest split (E18d), and
it is not sufficient, because the recognition a defence reads there is under the attacker's
control, which the template result shows without appeal to any threshold. Both decision points in Figure~\ref{fig:system} remain weak, for different
reasons. The first cannot see the attack, and at the second the model can be kept from noticing it, as we measured on one model of five.

\subsection{Saturation and the Predicted-Marginal Pin}
\label{app:agentic-detail}

Section~\ref{sec:two-settings} reports that the baseline correction is small on AgentDojo, that the
agentic numbers cover one attack family, and that nested tuning on
\texttt{important\_instructions} declines to condition on the request. The supporting arithmetic is
here.

\lead{How little hindsight there is to exploit.}
The winner's curse on the baseline runs \numrange{0.0041}{0.0149} on AgentDojo against
\numrange{0.0426}{0.1127} on HELM, the best fixed configuration already sitting at \num{0.0021}
harm against an oracle of \num{0.0000}, so hindsight has little to exploit. The median difference between router and pin is
\num{0.0000} at every pool size.

\lead{One family, and where its count comes from.}
Section~\ref{sec:data} carries the full family count. The agentic numbers come from one of them.

\lead{Why AUROC \num{0.8839} is severed from the policy's margin.}
At $\lam=0$ a configuration's score is its mean predicted probability over all scenarios, a constant
in the request. The predictor does enter, through that row mean, but AUROC \num{0.8839} is a
within-row \emph{ordering} statistic while the decision is a row \emph{mean}, and permuting each row
across scenarios preserves every number the policy uses while destroying the AUROC, so \num{0.8839}
is severed from the \num{0.0005} margin of Table~\ref{tab:family-router} and is not the strength
of the policy scored there. That margin is itself inside the tie-breaking noise: the pool
resampling is the canonical order plus \num{120} random orderings of one five-configuration set, and the exact tie in one fold's
honest-pin training argmin swings \num{0.0011} of five-fold mean harm on its own,
\SI{201.7}{\percent} of the margin (E48). The \num{605} pool evaluations behind Table~\ref{tab:family-router} are therefore \num{121}
orderings of the same five configurations in each of five folds, not independent replications, and
the marginal pin beats the honest pin in none of them. The comparison is therefore between two pin-selection rules, one ranking
configurations by predicted marginal risk and one by observed training harm. The reading is sharper
than the one we first drew: not that a strong router fails to beat an honest fixed choice by much, but that correct nested tuning, offered a request-conditional signal at AUROC \num{0.8839},
declines to condition on the request.

\lead{The structural-blindness figure is baseline-free.}
The baseline question does not affect the AUROC \num{0.647} a prompt-only router reaches, since it
concerns what the router can observe and not what it is compared against.

\subsection{What the Deferrals Are Worth, and Against What}
\label{app:mechanism}

Section~\ref{sec:baseline} showed that the baseline convention decides the verdict. The conditional
edge is defined relative to the pin, so it inherits the same dependence, and reporting it against one
baseline only produced a claim we now withdraw.

\lead{The edge changes sign with the baseline.}
On the requests the router routes away from the pin, the conditional edge (Eq.~\ref{eq:edge}) is
$\edge = -\num{0.1023}$ against $\pin_{\text{oracle}}$ and $\edge = +\num{0.0770}$ against
$\pin_{\text{honest}}$, on identical held-out folds at $k=\num{44}$. We previously reported the
former, as $-\num{0.0866}$ on the full sample, and drew from it the claim that the router's signal is
anti-correlated with harm on the requests where it acts. That claim does not survive an honest baseline and we retract it. What the negative edge measured was the optimism of a comparator chosen with
hindsight, which here inflated the router's deficit, and it does not make a hindsight-selected fixed model
an upper bound for an input-dependent router, and under a stationary split the router beats it on
average (E25).

\lead{The decomposition is an identity and still holds.}
$\text{deficit} = \defrate \times (-\edge)$ closes exactly in every cell under either baseline, since
off the deferred set the router and the pin issue the same decision. The identity is useful because it
localises the disagreement. Whatever the router gains or loses, it gains or loses only on the fraction
of requests where it deviates, and that fraction is small once $\lam$ is tuned honestly.

\lead{Conditioning matters when reading these numbers.}
Edges are computed on pools in which the router defers at least once. At $k=2$ that is only
\SI{8.5}{\percent} of pools. In the remaining \SI{91.5}{\percent} the tuned router reproduces the
pin and the difference is identically zero. Conditional edges of the magnitude quoted above therefore describe a minority of
pools, and the unconditional mean is an order of magnitude smaller. We report both, because quoting
the conditional figure alone overstates the effect in either direction. The same filter bites hardest
at $k=\num{44}$, where there is only one pool and the unit of resampling is the fold: three of the
five held-out folds have the router choosing the honest pin outright (deferral rate \num{0.000}) and
are dropped by the estimator, so the $+\num{0.0770}$ above and its interval are computed on
\num{146} of \num{393} paired requests, two folds of five. The unconditional edge over all five is
$+\num{0.0308}$, against a router$-$pin difference of $-\num{0.0308}$, and the field names on disk
do not carry the conditioning: any aggregate we label
``mean'' in E26 Part~A is a mean over the folds in which the router deferred.

\lead{Geometry versus signal.}
Shuffling each model's predictions across requests destroys request-level signal while preserving
every marginal. A signal-free router still departs from the pin: deficit $+\num{0.0113}$ and edge
$-\num{0.0202}$ (E9, null arm), against an observed $+\num{0.0509}$ and $-\num{0.0866}$ (E8),
permutation $p < \num{0.001}$. Both pairs are full-sample, in-sample-pin quantities, and they are not on
the same footing as the $-\num{0.1023}$ and $+\num{0.0770}$ above, which come from E26's within-fold
re-analysis. We keep them here only because the null was run against E8's design. The null router
therefore reproduces about a quarter of the observed edge. That is a comparison of a null mean with an
observed value and not an additive decomposition, so it does not license the statement that a quarter
of the edge ``is'' geometry. What it licenses is that a reviewer running the permutation null will
find a substantial $\arg\min$ term over heterogeneous marginals before any request-level signal is
allowed to act. The comparison is baseline-dependent in the same way as the edge itself.

\lead{Shrinkage acts on the deferral rate, and this result strengthens.}
At $k=\num{44}$, moving $\lam$ from \num{1} to \num{0.1} collapses the deferral rate from
\num{0.588} to \num{0.003}. On the single remaining deferral the conditional edge is not estimable,
and at $\lam=\num{0.05}$ the deferred set is empty and the estimator returns \texttt{nan}. The
deficit falls because the rate falls, not because the edge improves. We previously wrote that
shrinkage leaves the edge essentially unchanged, and that is wrong in both directions: at $k=2$ the
deferral rate barely moves (\num{0.072} to \num{0.088}) while the edge moves by \num{0.048}
($-\num{0.1271}$ to $-\num{0.0793}$). 

Under nested cross-validation, with $\lam$ chosen on training groups and applied to held-out groups,
selection drives $\lam$ toward zero rather than always landing on it. In E26 (\num{150} pools scored
on five held-out folds each, \num{750} pool--fold cells), all five folds select $\lam=0$ at
$k=\num{44}$ and, after the nested repair, at every smaller pool size too, $k=2$ included. E24, a
separate committed run of the same procedure over \num{200} pools, selects
$\{\num{0.2},\num{0},\num{0},\num{0},\num{0}\}$ at $k=2$ and zero on all five folds at every
larger pool size. The per-fold value at $k=2$ is not seed-stable, but the endpoint is, and no claim here
rests on a single fold's choice.
The resulting policy issues no deferral at all, being identical to the fixed choice,
in \SI{91.5}{\percent} of pools \emph{at $k=2$}, and in \SI{89.6}{\percent} at $k=3$ and
\SI{80.9}{\percent} at $k=5$. At $k=\num{44}$ that figure is not a fraction of
pools at all, since at $k$ equal to the pool size there is exactly one pool, the unit of resampling is the
fold, and the figure is \SI{60.0}{\percent}, meaning three folds of five. This is the one conclusion the
baseline correction makes stronger. Router collapse, reported as
a pathology by \citet{lai2026routing} and, as convergence to the majority-class model under
artifact-distorted labels, by \citet{garg2026router}, is what correct model selection chooses, and it is what
selection chooses against an honest baseline as well as an in-sample one.

\begin{table}[t]\centering\small
\caption{Encoder ablation and scale sweep on identical category-grouped splits (E23, E41). Sentence
embeddings at small scale are worse than TF-IDF, and the honest-pin bar of \num{0.6842} at full
strength is cleared by only the \num{8}B arm, and thinly. Precision is not matched across the sweep: the local arms are fp32, the \num{4}B
arm's quantisation is not declared by its provider and the \num{8}B arm is fp8, and the single crossing is
thin. $^{\ast}$The bracketed range is the \num{5}th--\num{95}th percentile of the \emph{per-model} AUROC across the \num{44} models: a dispersion across models, not a confidence interval on the mean.}
\label{tab:featurizer}
\begin{tabular}{lcc}\toprule
featuriser & AUROC & [p05, p95]$^{\ast}$ \\ \midrule
length only (non-semantic) & 0.5264 & [0.327, 0.777] \\
MiniLM (\num{22}M) & 0.6060 & [0.500, 0.710] \\
TF-IDF (1--2 gram) & 0.6509 & [0.499, 0.797] \\
Qwen3-Embedding (\num{596}M) & 0.6565 & [0.520, 0.772] \\
Qwen3-Embedding (\num{4}B) & 0.6831 & [0.508, 0.835] \\
Qwen3-Embedding (\num{8}B) & \textbf{0.6901} & [0.529, 0.794] \\
\midrule \multicolumn{3}{l}{honest-pin bar $\astar=0.6842$} \\
\bottomrule\end{tabular}\end{table}

\subsection{Cascade Diagnostics}
\label{app:cascade-detail}

Section~\ref{sec:mechanisms-die} states the cascade, its headline saving, its forced endpoint at
$\advers=1$, and the honest re-run below reverses one of the two claims drawn from it. Every
re-run there selects on training columns or attacker-train reruns and scores on held-out ones,
the discipline of Section~\ref{sec:baseline}. Four supporting points sit here.

\lead{What the cost figures assume.}
Costs are list prices, an assumption.

\lead{How shrinkage acts (E26).} Moving $\lam$ from \num{1} to \num{0.1} at $k=\num{44}$ collapses
the deferral rate from \num{0.588} to \num{0.003}, which is why the effect runs through the rate
and not through the conditional edge.

\lead{The predictability null (E2).} The label-permutation null for per-model harm prediction on
HELM \texttt{harm\_bench} is \num{0.5049} with a \num{95}th percentile of \num{0.5565}, over
\num{30} permutations on randomly drawn models. It is a sanity floor, not a test, and we quote
no $p$-value from it.

\lead{Flag rate by policy setting (E40).} Against \num{0.153} on matched undefended episodes, the
agent's flag rate is \num{0.500} under \texttt{ask\_user}, \num{0.677} under
\texttt{script\_audit}, \num{0.731} under \texttt{no\_network} and \num{0.992} under
\texttt{two\_pass}.

\lead{The best honest cascade (E34).} It escalates \SI{6.9}{\percent} of requests at harm
\num{0.0027} and task utility \num{0.7907}, for \SI{89.1}{\percent} less than the honest pin costs
on the same four folds. Its realised cost, \num{0.0156}, exceeds its hull comparator's \num{0.0138} at
\num{0.0047} lower harm. The comparison fixes the planned budget, not the realised one, and \num{6} of the
\num{13} utility-qualified beaters cost more than their comparator (E34).

\lead{Which folds the cheap-model comparison is over, and why it matters.} The cascade search
evaluates a configuration only in folds where its training cost is below that fold's pin, so a
configuration that \emph{is} the pin in some fold is absent from it. For the cheap model the
correction is not cosmetic. It has \num{6} failures in the whole grid, all \num{6} of them in the
one fold the filter removes, which is also the fold in which it is itself the honest pin. Over the
four folds it qualifies in its harm is exactly \num{0.000}, which is what we previously reported.
Over all five it is \num{0.0128}, against the pin's \num{0.0149} on the same five. The comparison
still favours the cheap model and by a much smaller margin, and it costs \num{0.026} of task
utility (\num{0.7585} against \num{0.7846}). The pin's own harm was likewise averaged over rows
rather than folds, which under-weighted that same fold and put it at \num{0.0079}. On five folds it
is \num{0.0149}. E34 now emits an unfiltered per-fold reference so any claim of this kind is quoted
over a fixed set of folds, and the survivor comparison is made on the four folds the survivor
occupies rather than against a five-fold pin: on those four its task utility is \num{0.7907}
against the pin's \num{0.8132}, a deficit of \num{0.0225} rather than the \num{0.0102} we
previously reported against a pin averaged over a different set of folds.

\lead{Design hypotheses from Section~\ref{sec:what-works}.} These would be tested by the missing
safety--utility--friction experiment rather than confirmed by it. Pin one strong model in preference
to routing on predicted safety. Gate actions, meaning which tools, which destinations, what needs
confirmation and what is audited before execution, in preference to classifying inputs. Assume every
input is adversarially chosen and ask what the worst consequence of an action would be under that
assumption. And if a detector is used at all, report it per attack template and not as a mean.

\lead{Pool-cell detail behind Section~\ref{sec:baseline}.} Nested selection picks
zero over all five folds at $k=2$, so each cell's policy is a fixed model chosen by predicted
marginal harm, and a win over the pin is evidence about choosing that model, not about per-request
dispatch. The router never defers at all in
\SI{91.5}{\percent} of the \num{750} cells at $k=2$, \SI{89.6}{\percent} at $k=3$ and
\SI{80.9}{\percent} at $k=5$. It is ahead of the honest pin in \SI{6.9}{\percent} of
cells and lands on an identical harm outcome in \SI{91.7}{\percent}, two fractions that are close
but not the same object. These come from a seed-\num{26} re-analysis over \num{150} pools, and the
independent run of the same estimand in Table~\ref{tab:pin-protocol} puts the $k=2$ mean at
$-\num{0.0071}$, so the quantity does not flip with the draw. At $k=\num{44}$ there is one pool
rather than a sample, its median is \num{0.00}, and the router is ahead in \num{2} of the
\num{5} held-out folds.

\lead{Why mid-run switching is a separate estimand.} \Citet{replaygap2026} fork live
trajectories and continue under a different model, finding \SIrange{74}{77}{\percent} of early
swaps diverge at the first post-fork action against \SIrange{6}{35}{\percent} for controls. Their
binary task outcome flips far less often, in \num{5} of \num{358} swap branches and none of \num{359} controls, but that
rate is the same order as the entire pre-execution headroom, so a replay approximation would carry
an error the size of the quantity it estimates. Whether that flip rate transfers to a harm label is
itself unestablished.

\lead{Where the encoder sweep crosses the bar.} Of the three scaled arms, only the \num{8}B clears the honest-pin bar of \num{0.6842}. The \num{4}B arm reaches \num{0.6831} and falls \num{0.0011}
short, and the \num{596}M arm reaches \num{0.6565}, so the crossing is a property of the largest arm
alone (Table~\ref{tab:featurizer}). The bar moved from \num{0.6731} to \num{0.6842} when the
shrinkage anchor was corrected, which is enough to take the \num{4}B arm back across it: a crossing
this thin is a statement about the bar as much as about the featuriser.

\lead{How sparse the family coverage is.} Twenty-eight configurations appear somewhere in the
AgentDojo traces, but only \num{3} of the \num{15} attack families are scored by \num{4} or more
configurations on the whole family, and only \texttt{important\_instructions} by \num{5} (E49).
Requiring all \num{28} to be complete therefore retains that family alone.

\lead{Why the agentic grid is \num{477} scenarios in E43 and \num{478} elsewhere.} The two
counts are one scenario apart because E43 drops any run whose AgentDojo \texttt{security} verdict
is JSON null, while the parse every other agentic result is built on reads a null verdict as no
harm. Exactly one attacked run of \num{33119} is affected, on a Llama variant rather than on the
pin, and treating it either way moves the pin's harm by \num{0.000004} and the headroom by the
same (E51). We report the discrepancy rather than reconcile it, since an absent verdict is not a
safe one and neither reading changes any conclusion here.

\lead{How E40 counts policy language.} The \num{86} figure of
Section~\ref{sec:what-works} searches the judge's explanation concatenated with the agent's
output, so it does not separate the agent restating the policy from the judge describing it. The
comparison that carries the argument is against \num{0} of \num{1342} undefended episodes on the
identical field, and the block-marker count is \num{0} of \num{520} defended against \num{2} of \num{1342} undefended.

\lead{Per-family headroom (E31).} Of the three families with enough configuration coverage to
analyse, \texttt{important\_instructions} has headroom \num{0.0158}, against \num{0.0053} for
\texttt{direct} and \num{0.0042} for \texttt{ignore\_previous}. Only the first clears the
\num{0.01} threshold at which the script attempts a router at all.

\lead{Attack-family spread (E46).} Pooled over all configurations, per-family harm runs
\numrange{0.024}{0.461} with sd \num{0.148} against \num{0.0786} across user tasks. Holding the
configuration fixed to the one present in every family, the two spreads are \num{0.1512} and
\num{0.1261}, so the pooled contrast is largely which configurations ran which family.

\lead{The de-saturated agentic pool (E30).} Dropping the dominant configurations raises headroom to
\num{0.0502} at \num{20} configurations, and a router then beats the honest pin by \num{0.0288}
harm on average, winning in \num{3} of the \num{5} held-out folds, at a cost of \num{0.0653} task utility. There is one
pool per fold here, not a population of pools, the pool is constructed with hindsight, and its
shrinkage anchor averages all columns, test columns included, so the figure is not nested and is
likely optimistic
(Section~\ref{sec:two-settings}).

\lead{The adversary parameterisation, and why the sweep is not in the body.}
We let the adversary steer routing, parameterised by the fraction $\advers$ of attacks that can
force cheap-model routing. The sweep over every safety floor and every $\advers$ we ran is
Table~\ref{tab:adversary-full} in Appendix~\ref{app:adversary}. Intermediate $\advers$ measures
cost rather than feasibility, which is why the sweep sits in the appendix rather than in the body.

\lead{Low harm at low utility.}
Three models in the safest quartile score below-median task utility, so their low harm comes with
low capability, a trade-off the harm-only formulation cannot see, and we do not identify why their harm is
low.

\lead{What the $\advers=1$ zero does and does not show.}
At $\advers=1$ no cascade escalates at all, every configuration degenerating to the cheap model, so
the zero count is the definition of a genuine cascade and is independent of the baseline. It is not
an outcome. Evaluated at that endpoint on the same four folds (E47), the twelve Meta-SecAlign
survivors keep harm \num{0} and utility \numrange{0.7805}{0.7933} at cost \numrange{0.0093}{0.0130}, against
\num{0.0156}--\num{0.0992} for their cascades, while Command-R-plus falls from harm \num{0.0027} and
utility \num{0.7135} to \num{0.0594} and \num{0.2967}. The survivors are compared on the four folds
where they qualify, with no prespecified fallback for the fold that removes them, and cost is
pre-execution selection between the cheap model and the pin, without the sunk cost of running the
cheap model before an escalation.

\subsection{Refusal-Provenance Triage in Full}
\label{app:triage}

Section~\ref{sec:mechanisms-die} states the mechanism and the economic count that is cheapest to
check. We stopped it before building it, on three counts. The induced/intrinsic partition is orthogonal to the
false-positive/true-positive partition a router needs. The stability premise is contradicted for a third of prompts in the
published record: \citet{heverin2026rbe} find about a third of initially-refusing prompts admit at
least one ``refusal escape'' under structured perturbation, with flip rates above \SI{20}{\percent}
for some artefact types. The economics also count against it. Triage costs $k$ cold prefills, while the
alternative it must beat, re-sending the request to another model, is priced at cache-read rates,
conventionally a tenth of input. Against a refusal that is itself unbilled, break-even sits at
$k=\num{0.1}$, below a single perturbation. That tenth is a pricing convention we assume rather
than measure, and the conclusion needs only that a cold prefill costs more than a cache read. In
general break-even is the fallback call's input cost divided by one perturbation's. The \num{0.1}
holds only where the fallback is billed at cache-read rates, which needs a warm cache on the
fallback model or a provider fallback credit, and where the refusal itself is unbilled. Output
tokens, latency and cache warming are excluded on both sides.

\lead{Why the superseded saving's honest counterpart is plausibly zero.} The
\SIrange{20.3}{35.2}{\percent} figures of Table~\ref{tab:adversary-full} attach to safety
floors of \num{0.0021} and \num{0.0042}, both tighter than any honest pin reaches. At the honest
pin's harm at $k=\num{28}$, \num{0.0149}, the nearest floors we swept, \num{0.0105} and
\num{0.0200}, give \SI{0}{\percent} saving at every adversary strength. We report this as the
reason the pair is withdrawn rather than as a measured replacement for it, the honest sweep at
that floor not having been run.

\subsection{The Full Adversary Sweep}
\label{app:adversary}

Section~\ref{sec:mechanisms-die} reports the cascade's cost saving in prose at the two adversary
strengths that carry weight, $\advers=0$ and $\advers=1$, and at the floors on which that comparison
is defined. Table~\ref{tab:adversary-full} gives the sweep in full: every floor we ran and every
$\advers$ we ran. The whole sweep lives here rather than in the body because its intermediate columns
are not readable without the explanation that follows, and a body table whose own surrounding
text instructs the reader to discount a column is worse than no table.

\begin{table}[!t]\centering\footnotesize\setlength{\tabcolsep}{3.5pt}
\caption{The full adversary sweep. Cost saving against the cheapest single model that meets the
floor, for every safety floor and every adversary strength $\advers$ we ran. ``no cascade'' means no
cascade meets the floor at any price. The $\advers\in\{\num{0.25},\num{0.5}\}$ columns are reported
here rather than in the body because they are not readable without the artefact explanation in the
surrounding text: the saving they show is a saving conditional on the floor still being met, and the
adversary they model forces routing on a random rather than a chosen subset.}
\label{tab:adversary-full}
\begin{tabular}{lcccc}\toprule
safety floor & $\advers{=}0$ & $\advers{=}0.25$ & $\advers{=}0.5$ & $\advers{=}1$ \\ \midrule
$\leq 0.0021$ & $+20.3\%$ & $+43.3\%$ & no cascade & no cascade \\
$\leq 0.0042$ & $+35.2\%$ & $+43.3\%$ & $+58.7\%$ & no cascade \\
$\leq 0.0105$ & $+0.0\%$ & $+0.0\%$ & $+0.0\%$ & $+0.0\%$ \\
$\leq 0.0200$ & $+0.0\%$ & $+0.0\%$ & $+0.0\%$ & $+0.0\%$ \\
$\leq 0.0500$ & $+0.0\%$ & $+0.0\%$ & $+0.0\%$ & $+0.0\%$ \\
\bottomrule\end{tabular}\end{table}

\lead{Why the saving rises before it vanishes.}
The saving is not monotone in $\advers$: at the tightest floor it rises from \SI{20.3}{\percent} at
$\advers=0$ to \SI{43.3}{\percent} at $\advers=\num{0.25}$ before collapsing to no qualifying cascade
at $\advers=\num{0.5}$. At the next floor down it rises further still, to \SI{58.7}{\percent} at
$\advers=\num{0.5}$, before no qualifying cascade remains at $\advers=1$ (this superseded oracle-floor sweep counts
cascades that meet the floor, and the forced endpoint is evaluated in Appendix~\ref{app:cascade-detail}). That rise is an artefact of what the column measures and not a benefit of
being attacked. The column is a cost saving conditional on the safety and utility floors still being
met. Forcing traffic onto the cheap model is exactly what lowers cost, so while the floor happens to
survive, an adversary who steers a quarter of the attacks makes the cascade look cheaper. The quantity
that matters is not how cheap the cascade is under partial attack but whether the floor can be met at
all, and in this sweep no qualifying cascade remains at $\advers=1$. The forced endpoint's own outcome
is in the E47 paragraph above.

\lead{Two further reasons the intermediate columns understate the attacker.}
First, our $\advers$ parameter forces routing on a \emph{random} subset of attacks rather than on a
subset the attacker chooses, so every intermediate column is a lower bound on what a targeted
adversary achieves. Only the endpoints are faithful to an adversary model we would defend
(Section~\ref{sec:limitations}). Second, the figures in this table select the pin, the cost--harm
frontier and the safety floor on the evaluation data. Re-run with an honest
selection (E34, E47), one of the two claims the table was originally used for is withdrawn. The
$\advers=1$ count is unaffected by both caveats, but it is a definition, and the paper rests on the
forced endpoint's evaluated outcome instead (E47).

\subsection{The Adaptive-Attacker Arms}
\label{app:adaptive}

Section~\ref{sec:mechanisms-die} reports the three attacker arms in prose. Table~\ref{tab:adaptive}
sets them side by side: the static control, the generic adaptive attacker that must pick one template
for the whole pool, and the targeted attacker that knows which model it faces and picks per model.
Templates are selected on attacker-train reruns and scored on held-out ones throughout.

\begin{table*}[!t]\centering\small
\caption{An attacker that adapts on held-out reruns. Templates are selected on half of the
reruns and scored on the rest, pooled over \num{200} splits of \num{130} rerun indices across
\num{5} models. The corpus has no task identifier, so held-out means held-out reruns of the same
configurations, not new tasks. The two adaptive arms differ only in what the attacker
knows. Choosing a single template for the whole pool gains nothing, though the static control is
\texttt{claude\_v35}, the corpus-modal template, which is also the corpus's most successful, so that
arm is scored against an already-optimal control. Choosing per model lowers the flag on
\texttt{gpt-5.4} alone (\texttt{claude-sonnet-4-6} flags at the same rate in both arms, three
models do not flag in either compared arm) and raises success by \num{10.35} points unmatched. Matched on run configuration the flag fall
holds under every weighting and the success gain does not (Appendix~\ref{app:adaptive}), and the
pooled rates are unweighted means over the five models.}
\label{tab:adaptive}
\begin{tabular}{llcc}
\toprule
attacker & what it knows & attack success & flag rate \\
\midrule
    static    & nothing, fixed template & $0.7027$ & $0.0624$ \\
    adaptive  & one template for the pool, chosen on train & $0.7018$ & $0.0631$ \\
    targeted  & \textbf{which model it faces}, chosen per model on train & $\mathbf{0.8062}$ & $\mathbf{0.0225}$ \\
\bottomrule
\end{tabular}
\end{table*}

\lead{The generic arm's prior, corrected.}
Before running we recorded a prior that generic adaptation would fail and attributed it to
\citet{hofer2026stronger}. At the source, that work finds task-universal attacks transfer across
tasks within a model but not across model families, so it gives no clean prior for a pool-wide
template, and the null here is our own measurement. It is also
the arm most likely to be fragile, one template having to serve five models.

\lead{What ``held-out'' means here, and a first matching (E54).} The corpus records no task
identifier. Its undefended episodes rerun six fixed payloads, one per template, over two
domain--harm pairs, with a \texttt{sample} index counting reruns within each run configuration.
\num{1226} of the \num{1342} agent outputs are distinct, so reruns are genuine samples but not new
tasks. The attacker's split is over that index, so its selection is honest across reruns and says
nothing about unseen tasks. For a given model the two arms' held-out episodes also cover run
configurations and sandbox modes unevenly. E54 matches them on the rerun index and mode, which
pairs reruns of different templates by index, an arbitrary pairing we keep only as a first pass:

\begin{center}\footnotesize
\begin{tabular}{lccc}\toprule
aggregation & static & targeted & gain \\ \midrule
published episode means & \num{0.7027} & \num{0.8062} & \num{10.35} pts \\
shared rerun indices & \num{0.7474} & \num{0.7849} & \num{3.75} pts \\
shared index and mode cells & \num{0.7821} & \num{0.7952} & \num{1.31} pts \\
\bottomrule\end{tabular}
\end{center}

\noindent On that pairing the flag falls \num{3.98} points pooled and \num{3.97} on the matched
cells. A cluster bootstrap over rerun indices (\num{300} multinomial reweightings of the \num{130} indices, template
selection rerun on each, the \num{200}-split schedule and observed support held fixed, E54) gives a
\SI{95}{\percent} interval of \numrange{-5.63}{-1.91} points for the matched flag change, below zero
in every replicate, and \numrange{-3.59}{8.24} for the matched success change, below zero in
\SI{37}{\percent}. It is a sensitivity interval conditional on that support, not an estimate for
unobserved templates, modes or models. The flag change
rests on one model: three of the five flag nothing in either arm and a fourth is unchanged because
its static control is already the evading template, so the whole fall is \texttt{gpt-5.4}
(\num{0.2405} to \num{0.0421}), the one model whose judged recognition the abstract reports falling.

\lead{Matched on run configuration (E54b).} The meaningful controls are the run configuration
(\texttt{preset}) and the sandbox mode, since reruns within them are exchangeable. Keeping E54's
split, seed and selection rule and matching the two arms on (preset, domain, harm, sandbox mode)
cells gives, under two weightings, with conditional \SI{95}{\percent} percentile ranges from \num{200} bootstrap replicates that resample reruns within each (model, template, configuration) cell and rerun selection, where empty intersections remove a small share of comparisons:

\begin{center}\scriptsize\setlength{\tabcolsep}{2.5pt}
\begin{tabular}{lcc}\toprule
cell weights & success (pts) & flag (pts) \\ \midrule
equal & \num{15.0} [\num{7.26}, \num{24.66}] & $-\num{3.92}$ [$-$\num{6.11}, $-$\num{1.65}] \\
episode & \num{2.0} [$-$\num{5.26}, \num{11.69}] & $-\num{3.98}$ [$-$\num{5.53}, $-$\num{2.06}] \\
\bottomrule\end{tabular}
\end{center}

\noindent The success change depends on the weighting, because a few small cells with extreme
rates dominate the equal-weight mean, and its per-model sign is not uniform (\texttt{gpt-4o} loses
under both). The flag change does not depend on it. It is below zero in every replicate under both
weightings, and all of it is on \texttt{gpt-5.4} (\numrange{-19.9}{-19.6} points). E54's
\num{1.31}-point success figure is one point in this range, not a corrected effect.

\lead{The drop under a label the judge did not write (E55).} The flag is the corpus judge's
verdict. E55 labels recognition with a fixed lexicon applied to the agent's own output, written and
hashed before the first run and reading no judge field. On \texttt{gpt-5.4} it agrees with the judge
at $\kappa=\num{0.818}$. Comparing \texttt{claude\_v35} with \texttt{claude\_v39} within each run
configuration that contains both, the lexicon rate falls from \num{0.100} to \num{0.000}, \num{0.200}
to \num{0.000} and \num{0.189} to \num{0.022}, against \num{0.233} to \num{0.033}, \num{0.200} to
\num{0.000} and \num{0.244} to \num{0.044} under the judge, with a within-configuration
permutation $p\leq 10^{-4}$ for both, the resolution of \num{10000} permutations. A lexicon over output text
measures expressed recognition, not whether the model registered the attack before acting.

A separate coverage objection does not hold. Three of the six templates are absent for at least one
model, so a per-model argmax could mix choosing well with being offered a longer menu. Restricting
every target to the three templates common to all five models leaves the published gain unchanged,
because across \num{200} splits and five models every one of the \num{1000} picks is already one of
those three.

\lead{Where the targeted attacker's gain comes from.}
The gain is not uniform: two of five models keep the static pick, \texttt{gpt-4o} falls from
\num{0.595} to \num{0.570}, and the rise is carried by \texttt{gpt-5.4-mini} (\num{0.539} to
\num{0.957}) and \texttt{gpt-5.4} (\num{0.504} to \num{0.629}). The flag-rate fall is concentrated
in the same way. Pooled it runs \num{0.0624} to \num{0.0225}, but these are unweighted means over
five models and the entire fall is \texttt{gpt-5.4} (\num{0.2413} to \num{0.0422}). Three models
flag nothing in either arm, and \texttt{claude-sonnet-4-6}, the model Section~\ref{sec:mechanisms-die}'s
flagging argument rests on, is identical under both because its static control is already the
evading template.

\lead{What the adaptive result is a measurement of.} E54 and E54b measure the signal. What the
attacker demonstrably gains is a lower flag rate on one model, under two independent labels, and its
success gain depends on how configurations are weighted. The next subsection composes the signal into
controllers and scores harm.

\subsection{The Flag Through a Controller (E56b)}
\label{app:controller}

\lead{Design.} Section~\ref{sec:mechanisms-die} shows a targeted attacker lowers one model's judged
flag. To ask what that does to harm, we compose the flag into two controllers from the logged
undefended episodes. \emph{Block-on-flag} stops a flagged episode and lets a silent one proceed on
the target. \emph{Escalate-on-silence} lets a flagged episode proceed, since the model is already
declining, and hands a silent one to a fallback model. The fallback's outcome is its own logged
success under the same template, which assumes a re-run by the fallback is a fresh draw from its
episodes on that payload rather than a continuation of the target's run. This is an offline counterfactual composition of
logged episodes, not an executed controller. It also assumes the flag is observed before the target
commits any harmful action, and the released episodes do not record when the flag was raised
relative to the target's actions. The attacker's split,
seed and selection rule are E54's. A target-only attacker picks the template that maximises the
target's own training success, as in E54, and a configuration-aware one picks per target and
fallback to maximise training system harm. Each is compared with the static template on the
target's run-configuration cells that both cover in the held-out reruns, as a paired difference.
The fallback term is taken either pooled or matched on the fallback's own shared cells, and cells
are weighted equally or by episodes, giving four constructions. Intervals come from \num{200}
bootstrap replicates that resample reruns within cells. The hypotheses and their criteria were
recorded before the confirmatory run.

\lead{Result.} The positive target-only contrast under block-on-flag is exploratory. The registered generic-attacker hypothesis covered both controllers, while the directional hypotheses concerned escalation. The
target-only attack raised its estimated harm by \num{14.9} points with matched configurations weighted
equally, but by \num{1.8} with configurations weighted by their episode counts within each model,
models averaged equally in both, and the bootstrap ranges are $[\num{6.9}, \num{23.8}]$ and
$[\num{-6.0}, \num{11.6}]$. The contrast depends on the weighting and is not an increase robust to
both conventions. Under escalate-on-silence,
two of the five models never flag and a third flags \num{2} of \num{420}, so escalation covers
\SIrange{94}{98}{\percent} of tasks and the controlled system is effectively its
fallback:

\begin{center}\scriptsize\setlength{\tabcolsep}{2.5pt}
\begin{tabular}{llccc}\toprule
cells & fallback & knows target & knows fallback too & no \texttt{sonnet} \\ \midrule
cells & pooled & $-\num{5.1}$ [$-\num{8.8}$, $-\num{1.8}$] & \num{8.8} [\num{4.7}, \num{15.0}] & \num{7.3} \\
episodes & pooled & $-\num{4.9}$ [$-\num{8.6}$, $-\num{1.7}$] & \num{8.9} [\num{4.6}, \num{15.0}] & \num{7.4} \\
cells & matched & $-\num{5.4}$ [$-\num{10.2}$, $-\num{0.9}$] & \num{11.9} [\num{5.8}, \num{19.1}] & \num{8.9} \\
episodes & matched & $-\num{10.6}$ [$-\num{14.8}$, $-\num{6.5}$] & \num{2.6} [$-\num{3.3}$, \num{9.8}] & \num{2.3} \\
\bottomrule\end{tabular}
\end{center}

\begin{figure*}[!t]\centering
  \includegraphics[width=\textwidth]{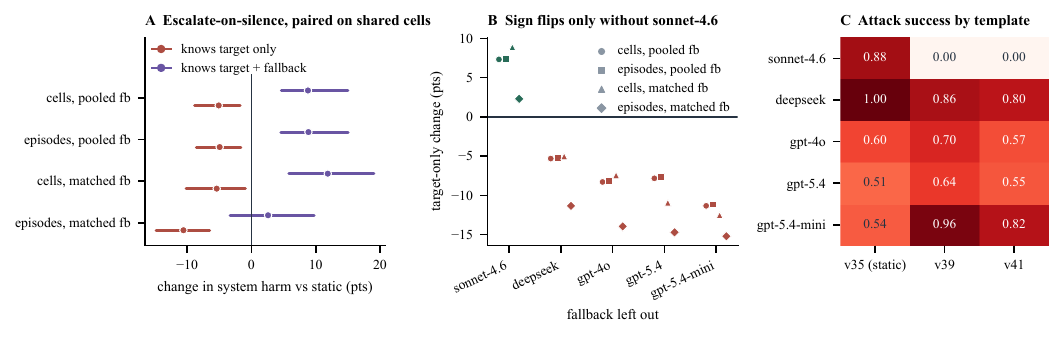}
  \caption{\textbf{Through a flag-gated controller, the fallback decides.} (A) Change in system
  harm against the static template under escalate-on-silence, for an attacker that knows only the
  target and one that also knows the fallback, in four matching constructions, with rerun-bootstrap
  \SI{95}{\percent} intervals (E56b). (B) The target-only change with each fallback left out, one
  marker per construction. Only leaving out \texttt{claude-sonnet-4-6} flips the sign. (C) Attack
  success over all undefended episodes by template, for the three templates every model has.
  \texttt{claude-sonnet-4-6} is vulnerable to the static template alone.}
  \label{fig:controller}
\end{figure*}

\noindent Entries are changes in system harm against the static template, in points, with
\SI{95}{\percent} intervals. The target-only attacker lowers harm in every construction, but the
leave-one-fallback-out check shows why. The effect comes entirely from \texttt{claude-sonnet-4-6}
as fallback, which is vulnerable to the static template and to no other
(Figure~\ref{fig:controller}C). Without it the sign reverses in all four constructions, and removing
any other fallback leaves it negative. The configuration-aware attacker gains in all four constructions, and its conditional range excludes zero in three. Under escalation the generic attacker picks the static
template in every training draw, so its hypothesis cannot be tested here. What survives is a statement about
composition. Through a flag-gated controller, the sign of an attack's effect on harm is set by the
fallback's template vulnerabilities, and a flag-level measurement cannot reveal it.

\subsection{The Four Action-Level Defences}
\label{app:defences}

\begin{figure*}[!t]\centering
  \includegraphics[width=\textwidth]{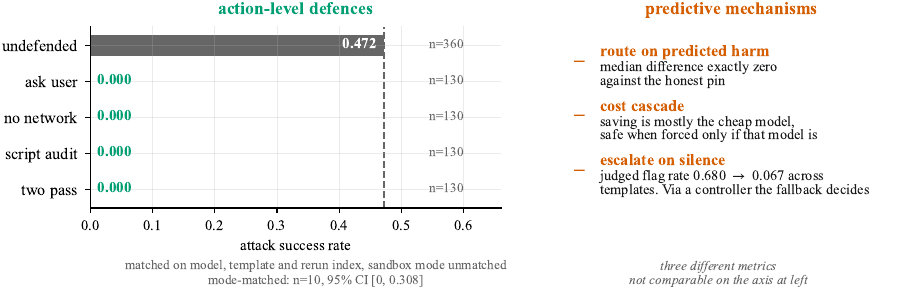}
  \caption{\textbf{Four settings at zero, consistent with recognition.} Each of four action-level settings records zero judged successes on its observed support, on the run cells and models they were
  actually run on, under one automated judge. An external gate can apply trusted policy to untrusted
  arguments while injected instructions cannot change its authorisation, but mechanical enforcement of
  these four could not be verified from the released harness, and the episodes are consistent with
  recognition, so that property is not shown for them.}
  \label{fig:defences}
\end{figure*}

Section~\ref{sec:what-works} reports the four action-level defences and the three denominators
(\num{130}, \num{70} and \num{10}) on which their zeros can be read. Figure~\ref{fig:defences}
states the asymmetry that section argues for, together with the reason it does not cover these four
defences, and Table~\ref{tab:defences-cells} resolves the matched comparator of
Table~\ref{tab:defences} per defended cell.

\begin{table}[t]\centering\footnotesize\setlength{\tabcolsep}{4pt}
\caption{The matched comparator of Table~\ref{tab:defences}, resolved per (model, template) cell,
identically for all four defences. \num{30} defended episodes have no undefended comparator and
\num{30} sit in a cell whose undefended success is exactly \num{0.000}, leaving \num{70} of
\num{130} informative about a reduction. Per-cell model $\times$ template breakdown from
\texttt{e38\_defence\_matched.json:defences.*.per\_cell}.}
\label{tab:defences-cells}
\begin{tabular}{lcc}\toprule
defended cell & defended $n$ & undefended success ($n$) \\
\midrule
\texttt{sonnet-4-6}$\times$\texttt{v32} & 30 & --- (0) \\
\texttt{sonnet-4-6}$\times$\texttt{v35} & 30 & 0.988 (80) \\
\texttt{gpt-5.4}$\times$\texttt{v39} & 10 & 0.633 (30) \\
\texttt{gpt-5.4-mini}$\times$\texttt{v32} & 30 & 0.000 (30) \\
\texttt{gpt-5.4-mini}$\times$\texttt{v35} & 30 & 0.400 (70) \\
\bottomrule\end{tabular}\end{table}

\subsection{Attack Surface Versus Defence}
\label{app:attack-surface}

The released corpora differ greatly across attack surfaces, under different attacks and evaluation designs. In a handcrafted corpus browser-borne
injection did not succeed on two current frontier models, \num{0} of \num{140} multi-step attacks
against Claude Sonnet 4.6 and GPT-5.4 \citep{cua2026handcrafted}, and in AgentDojo tool injection
GCG attacks optimised on open-weights models transfer below \SI{2}{\percent} to three of four frontier
targets (\SI{7.7}{\percent} on the fourth) \citep{hofer2026stronger}, while agent skill files, on the same weights, run in
our corpus at a pooled \SI{50.5}{\percent} undefended, \numrange{0.263}{0.942} across the five
models. A safety layer aimed at hand-crafted browser injection targets a surface where such attacks already
fail on these two models, while the skill-loading path stays open. Attack strength is also not fixed: \citet{autodojo2026} show
black-box adaptive optimisation raising success well above static injections against nearly every
defence they test, which is why we do not read a low static rate as a safety property.

\subsection{The Best Detector Is the Worst Fallback}
\label{app:detection}

\lead{Why the attack factor is not identified, and what pooling would do.} The
\num{5}$\times$\num{6} template-by-model grid is missing \num{9} of its \num{30} cells, so the
attack factor is not identified from this design, and execution mode is unmatched within the grid
as in Section~\ref{sec:what-works}. Separately, the \num{170}$\times$ flagged-versus-unflagged
difference of Section~\ref{sec:mechanisms-die} is computed on undefended episodes alone. Pooling in the defended episodes, which have zero judged success, changes the association to \num{360}$\times$, which measures the defences and not the flag.

\lead{The two nulls behind the variance-share deletion.} The observed attack-factor shares are
\num{0.336} on the \num{5}$\times$\num{6} grid and \num{0.045} on the complete
\num{5}$\times$\num{3} sub-grid. Both sit inside additive zero-interaction resamples
(\numrange{0.292}{0.442} and \numrange{0.042}{0.075}, \num{2000} draws), which we give as a
diagnostic rather than the reason for the deletion, since a null fitted to the observed margins
places its mean beside the observed share by construction. Against a model-only null in which
templates do nothing, the attack factor exceeds the \num{95}th percentile on both grids (against
\numrange{0.122}{0.257} and \numrange{0.000}{0.004}), which is why we keep the effect and not its
size.

\lead{Safety as a detection property.} The model with the highest flag rate (\num{0.680}) has the
highest attack-success rate conditional on failing to flag (\num{0.823}) among the models that flag
at all. That restriction is necessary, not a hedge: two models in the pool never flag anything, so
for them ``conditional on failing to flag'' is the whole distribution and the comparison is vacuous,
and one of them is higher at \num{0.942}. Among models that detect, the best detector has the highest harm when its detection fails. Flag rate also varies \num{5}$\times$ between the two
strongest flaggers, and two models never flag.

\subsection{Comparators and Protocol Robustness}
\label{app:comparators}

\lead{Observation 1 (what an in-sample comparator adds).}
This is the optimizer's curse \citep{smithwinkler2006} combined with the finite-class maximal
inequality \citep{boucheron2013concentration}, stated for our two baselines. Fix a pool of $k$
models and one outer fold with $n$ test requests and $m$ training requests, and assume the test
requests are drawn independently of the training requests from a test distribution. Let
$\hat h_j$ be model $j$'s mean harm on the test requests, $\tilde h_j$ its mean on the training
requests, and $h_j$ its expected harm under the test distribution. The in-sample pin reports
$\min_j \hat h_j$. The honest pin is $\hat\jmath=\arg\min_j \tilde h_j$ and reports $\hat h_{\hat\jmath}$.
Write the selection cost as $C=\hat h_{\hat\jmath}-\min_j \hat h_j$.
\begin{enumerate}
\item $C\geq 0$ on every fold and every pool, deterministically.
\item $\mathbb{E}[C]=R+O$, with the regret of the honest choice
      $R=\mathbb{E}[h_{\hat\jmath}]-\min_j h_j$ and the optimism of the in-sample pin
      $O=\min_j h_j-\mathbb{E}[\min_j \hat h_j]$, both non-negative.
\item If the test requests are independent draws and harm lies in $[0,1]$, the optimism is at most
      $\sqrt{\log k/(2n)}$.
\item If the training requests are independent draws from a distribution on which model $j$'s
      expected harm is $g_j$, the regret is at most $\sqrt{\log k/(2m)}+2\Delta$, where
      $\Delta=\max_j|g_j-h_j|$ is the shift between training and test distributions.
\end{enumerate}

\noindent\emph{Proof.} (1) $\hat h_{\hat\jmath}\geq\min_j\hat h_j$ for any index. (2) Add and subtract
$\min_j h_j$. Because $\hat\jmath$ depends only on the training requests, which are independent of the
test requests, $\mathbb{E}[\hat h_{\hat\jmath}\mid\hat\jmath]=h_{\hat\jmath}$. The regret is non-negative
because $h_{\hat\jmath}\geq\min_j h_j$, and the optimism is non-negative because
$\mathbb{E}[\min_j\hat h_j]\leq\min_j\mathbb{E}[\hat h_j]=\min_j h_j$. (3)
$\min_j h_j-\min_j\hat h_j\leq\max_j(h_j-\hat h_j)$, and each $h_j-\hat h_j$ is a mean of $n$
independent centred variables in an interval of length one, hence sub-Gaussian with variance proxy
$1/(4n)$ by Hoeffding's lemma. The maximal inequality, which needs no independence across models,
gives $\mathbb{E}\max_j(h_j-\hat h_j)\leq\sqrt{2\cdot\tfrac{1}{4n}\log k}=\sqrt{\log k/(2n)}$. (4) Let
$j^\star=\arg\min_j h_j$, a fixed index. Using $\tilde h_{\hat\jmath}\leq\tilde h_{j^\star}$,
$h_{\hat\jmath}-h_{j^\star}\leq (h_{\hat\jmath}-g_{\hat\jmath})+(g_{\hat\jmath}-\tilde h_{\hat\jmath})
+(\tilde h_{j^\star}-g_{j^\star})+(g_{j^\star}-h_{j^\star})$. The first and last terms are at most
$\Delta$ each. The third has expectation zero because $j^\star$ is fixed, and the second is at most
$\max_j(g_j-\tilde h_j)$, whose expectation is at most $\sqrt{\log k/(2m)}$ by the same argument. $\square$

\noindent\emph{Reading.} Part 1 is why the sign of the selection cost carries no evidence: an
in-sample comparator can only flatter a fixed model, so what matters is the size. Parts 3 and 4
locate that size. The optimism bound grows as $\sqrt{\log k}$ and shrinks with the test fold, and
the regret bound carries a shift term that is zero for a stationary split and need not be for a
category-held-out one. The data are consistent with this. The measured cost rises with $k$
(\num{0.043}, \num{0.067}, \num{0.097}, \num{0.113} at $k=2,3,5,44$ under the union label, E24) and
stays below the optimism bound at every $k$ (\num{0.155} at $k=44$). Two caveats bound what this
shows. An upper bound that grows with $k$ does not imply that the cost does, and under
category-held-out folds the independent-draw premise is an assumption about the population a
category is sampled from, not a property of the split. We use the observation to explain why the
cost is positive and where it can come from, not to predict its size.

\lead{The benchmark's own full-sample convention (E25f).} Choosing the in-sample comparator once
on all evaluation data, as the benchmark does, rather than within each fold, the HELM cost is
\numrange{0.001}{0.018} under random splits and \numrange{0.022}{0.084} with categories held out.

\lead{The same contrast on AgentDojo and on LLMRouterBench (E25a, E57).} On AgentDojo, on E27's
complete-case grid of \num{28} configurations and \num{478} scenarios, with the user-task arm asserted
equal to E27, the selection cost in harm under three splits of increasing shift is:

\begin{center}\scriptsize\setlength{\tabcolsep}{3pt}
\begin{tabular}{lccc}\toprule
$k$ & random & user task held out & suite held out \\ \midrule
2 & \num{0.003} & \num{0.004} & \num{0.028} \\
3 & \num{0.004} & \num{0.006} & \num{0.027} \\
5 & \num{0.004} & \num{0.008} & \num{0.027} \\
28 & \num{0.002} & \num{0.015} & \num{0.017} \\
\bottomrule\end{tabular}
\end{center}

\noindent Only three suites survive the complete-case filter, so the suite arm has three folds. On
LLMRouterBench \citep{commercial2026} we use the official release, whose per-query results cover more
datasets and models than the paper's own tables. We drop ArenaHard's sub-splits, which repeat its
queries, keep per dataset the record set most models share and then the models complete on every
kept dataset, leaving \num{20} models on \num{18} datasets, a wider grid than the benchmark's
own \num{15}-dataset setting. Null scores are kept as missing. The benchmark's Best Single maximises
the mean of per-dataset accuracies and is chosen once over all datasets. We also score a comparator
chosen per dataset, under the random 70/30 split and with the dataset held out and a matched
\SI{30}{\percent} test, since comparing a per-dataset choice under shift with a global one under a
random split would confound shift with granularity. In accuracy points:

\begin{center}\scriptsize\setlength{\tabcolsep}{2.5pt}
\begin{tabular}{lcccc}\toprule
$k$ & Best Single & per dataset & per dataset & \\
 & random & random & held out & ratio \\ \midrule
2 & \num{0.03} & \num{0.32} & \num{2.58} & \num{8.16} \\
3 & \num{0.03} & \num{0.48} & \num{3.37} & \num{6.95} \\
5 & \num{0.04} & \num{0.73} & \num{3.50} & \num{4.80} \\
20 & \num{0.00} & \num{2.16} & \num{3.10} & \num{1.43} \\
\bottomrule\end{tabular}
\end{center}

\noindent The benchmark's own comparator costs nothing at its full pool, and its selection under a
held-out dataset is unchanged there too. The per-dataset shift effect is large in small pools and
small at the full pool, so the fivefold criterion we registered before the run fails at $k=5$ and
$k=\num{20}$. No router output is used, so these figures size the comparator, not the benchmark's
routers.

\lead{Every eligible corpus (E58).} The rule, fixed before any cost was computed, admits every corpus
on disk that scores at least five models on shared items and carries a grouping of at least four
values, the attack method where one exists and otherwise the corpus's own harm category or domain.
Harm is the corpus's unsafe-compliance label. HarmBench keeps the \num{19} models run on all
\num{16} attacks and the \num{2210} attack and behaviour pairs all of them have. SORRY-Bench \citep{xie2025sorrybench} items are
its \num{44} category fulfilment rates, each over ten prompts, grouped by the paper's four domains.
XSTest keeps its \num{200} unsafe contrast prompts with human labels, harm being full compliance.
AgentDojo has three suites after the complete-case filter and is reported above, outside the count.
Two HELM scenarios, \texttt{simple\_safety\_tests} \citep{vidgen2023simplesafetytests} and \texttt{xstest}, were missed by our first
eligibility check and added after an independent verifier found them, and the same verifier found
that resampled copies of one item could fall on both sides of a random split, which narrows the
random arm. Folds are now assigned over original items, and E25b was corrected the same way. At
$k=2$, with \num{200} bootstrap reps over items within groups and \num{200} group-label permutations,
Table~\ref{tab:e58} gives every corpus.

\noindent The corpora overlap. HELM \texttt{harm\_bench} and HarmBench share behaviours, the three
HELM scenarios share their \num{44} models, and the two XSTest rows share prompts, so they are fewer
than six independent tests. The last three rows are the corpora where some models never produce
harm (\SIrange{16}{40}{\percent} of models), a pattern we note after the fact. The permutation
$p$ for HELM \texttt{xstest} and \texttt{simple\_safety\_tests} moves with the pool draw (the
verifier's draw gives \num{0.05} for both), and HarmBench's permutation result holds with all
\num{29} models on the eight attacks they share. The registered prediction that the cost grows with
ranking instability, one minus the mean Kendall $\tau$ between groups, failed ($\rho=\num{-0.03}$).

\lead{Is the cost large against what routing could gain (E59)?} On the same folds and pools, with
the selection costs asserted equal to E58, a perfect router that serves each held-out item the pool
model with the lowest harm on it bounds what any router could gain over the honest pin. We
registered that the in-sample comparator can change a verdict where its cost is at least a tenth of
that bound at $k=2$. As a ratio of means the share is \num{0.57} on HELM, \num{0.40} on SORRY-Bench and
\num{0.03} on HarmBench with groups held out, against \num{0.04}, \num{0.21} and \num{0.01} on random
splits, so on all three corpora with room to route it rises under shift. The mean of per-pool ratios
is lower, \num{0.23}, \num{0.16} and \num{0.014}, with the same verdicts. The per-item bound is the
loosest ceiling and credits a router with item-level label noise, so a stricter one raises $R$. On
HarmBench, a router that may use the attack but not the item gives \num{0.19}. That ceiling is not
comparable across corpora, since where a held-out fold holds one group it coincides with the in-sample
pin and $R$ is one by construction. SORRY-Bench items are category means, so its ceiling is coarser
and its $R$ overstated. On the three saturated corpora headroom is at most \num{0.011} of harm and $R$
is a ratio of near-zeros. The rule we registered, which set no minimum headroom, is met on all but
HarmBench, three of those five being these. We report no interval for $R$, since the bootstrap median
exceeded the point estimate in every cell. The HarmBench result is not an instance of stable
rankings. Its overall safest model is not the safest in \num{14} of \num{16} attack groups, against one
of four domains on SORRY-Bench, so we do not claim a mechanism for the split.

\lead{AIR-Bench 2024, registered separately (E61, E62).} After E58 we registered a seventh corpus
before downloading any score, HELM's AIR-Bench 2024 runs \citep{zeng2024airbench}, \num{87} models on
\num{5694} prompts, harm being one minus the judge's refusal score, grouped by its \num{16} level-2 risk
categories and, as a secondary grouping, its four level-1 domains. It shares \num{40} of HELM
\texttt{harm\_bench}'s \num{44} models. The interval test and the permutation null are met under both
groupings, and $R$ is \num{0.12} with the categories and \num{0.10} with the domains. Our registered
prediction that the held-out cost would reach \num{0.02}, HELM \texttt{harm\_bench}'s order, failed. It
is \num{0.0053}, against \num{0.0002} under random splits. These values enumerate every pair of models.
Enumerating the other corpora the same way moves their points by at most \num{0.0019} and changes
no verdict, but AIR-Bench's sampled draw, \num{0.0068}, sat well above its exact value.

\lead{How often and how far the choice moves (E60).} At $k=2$ the cost is the share of pools in which
the model chosen on training items is not the safest held-out one, times the harm gap when it is
not. Holding HELM categories out raises both, \num{0.21} times \num{0.20} against \num{0.06} times
\num{0.04} under random splits (E62). This is an identity, and across corpora its factors move with
model count, grader and base rate, so we do not read it as the cause of the differences.

\lead{Which groups share a fold (E64).} Every held-out figure uses one GroupKFold assignment of
groups to folds. Over \num{200} random assignments the exact $k=2$ held-out cost is \num{0.037}
$[\num{0.026}, \num{0.045}]$ on HELM \texttt{harm\_bench}, where our assignment sits at the 89th
percentile, \num{0.0036} $[\num{0.0021}, \num{0.0056}]$ on HarmBench, where it sits below every random
one, and \num{0.0043} $[\num{0.0032}, \num{0.0054}]$ on AIR-Bench. Each lower end exceeds the
random-split cost, \num{0.0026}, \num{0.0006} and \num{0.0002}. SORRY-Bench and
\texttt{simple\_safety\_tests} hold one group per fold and so have no spread. The direction does not
depend on the assignment, and the size moves by a factor of up to \num{2.4}. Holding out one attack
at a time raises HarmBench's cost further (E63). On AIR-Bench the cost rises with finer categories,
but in the finest ones we could not separate that from near-duplicate prompts, so granularity does
not yet explain the differences between corpora either.

\lead{Each judge alone (E65).} The union label is derived from the two published judges, so this is
two independent labels and their union, not three. Held-out minus random has a \SI{95}{\percent}
bootstrap interval above zero at $k=2$, $3$ and $5$ under each, for E25's folds and for
leave-one-category-out, which was the registered criterion. At $k=2$ the held-out cost is
\num{0.0454}, \num{0.0289} and \num{0.0434} (union, GPT, Llama) in E25's folds and \num{0.0374},
\num{0.0285} and \num{0.0379} leaving one category out, against \num{0.0026}, \num{0.0021} and
\num{0.0023} under random splits. Over \num{200} random assignments of categories to folds the exact
$k=2$ cost is \num{0.027} $[\num{0.019}, \num{0.034}]$ under the GPT judge and \num{0.038}
$[\num{0.030}, \num{0.046}]$ under the Llama judge. At the full pool the cost is \num{0.1127},
\num{0.0120} and \num{0.0921}, so only the direction, not the size, survives the choice of judge there.

\lead{The same comparison under each judge and each held-out fold (E24j).} E24 routes on the
union of HELM's two published judges. Rerun on its folds and pool draws with each judge's own label
as the target, the union rows reproduce E24 exactly, and the others give the following (a negative
router-minus-pin favours the router):

\begin{center}\scriptsize\setlength{\tabcolsep}{2.2pt}
\begin{tabular}{lcccccc}\toprule
& \multicolumn{3}{c}{selection cost} & \multicolumn{3}{c}{router $-$ honest pin} \\
\cmidrule(lr){2-4}\cmidrule(lr){5-7}
$k$ & union & GPT & Llama & union & GPT & Llama \\ \midrule
2 & \num{0.043} & \num{0.030} & \num{0.043} & $-\num{0.007}$ & $-\num{0.006}$ & $-\num{0.009}$ \\
3 & \num{0.067} & \num{0.038} & \num{0.068} & $-\num{0.014}$ & $-\num{0.008}$ & $-\num{0.012}$ \\
5 & \num{0.097} & \num{0.043} & \num{0.088} & $-\num{0.024}$ & $-\num{0.006}$ & $-\num{0.019}$ \\
44 & \num{0.113} & \num{0.012} & \num{0.092} & $-\num{0.031}$ & \num{0.000} & $-\num{0.010}$ \\
\bottomrule\end{tabular}
\end{center}

\noindent Nested selection picks $\lam=0$ in \num{57} of the \num{60} judge, fold and pool-size fits. That
count rests partly on the tie-break, since \num{20} fits have exact ties on the training deficit and
breaking them toward the largest $\lam$ leaves \num{37} at zero, while the held-out router-minus-pin
is unchanged in \num{11} of \num{12} judge and pool-size settings and moves by under \num{0.001} in
the other, because tied values make the same choices (E24t). The
collapse onto a fixed choice therefore holds under every label as behaviour, not as a $\lam$ count, while the size of the selection cost
depends on the label, most at $k=44$. Per held-out fold, the cost for two-model pools is
positive in all five folds under all three labels, and the router is at least level with the
honest pin on average in all fifteen fold-label cells. At $k=44$ there is one pool per fold, and the GPT
judge's cost is zero in three of the five folds.

\lead{Why a fixed mixture is not the comparator.}
A \emph{mixing} baseline, serving model $i$ with fixed probability $w_i$ regardless of $x$,
degenerates to the pin under one objective: expected harm is linear in the weights, so the hull
minimum sits at a vertex. Conditioning on $x$ \emph{enlarges the feasible set} rather than
guaranteeing a policy outside the hull, and two conditions are worth stating because our own results
meet the second. A router that ignores $x$ is a fixed mixture, so it is in the hull by definition.
An input-dependent router whose induced mixture happens to coincide with a hull point is also in
it. At $\lam=0$ the shrinkage estimator selects on a row mean and issues one model for every
request, which is exactly that case (Section~\ref{sec:baseline}), so the policy this paper measures
at full pool size lies \emph{inside} the hull. A second objective changes the comparator, not the
router: the best fixed mixture can then beat the best fixed model under a constraint, though not
always, since a model minimising both harm and cost is already the constrained optimum and no
mixture improves on it. What is true without qualification is only that a mixture is not the
interesting comparator here.

\lead{Where the deferral identity closes.}
Equation~(\ref{eq:decomp}) is exact under either baseline convention, but it closes per fold and per pool, not after averaging, since cell means
satisfy $\overline{\text{deficit}} = \overline{\defrate}\cdot\overline{(-\edge)} +
\widehat{\mathrm{Cov}}(\defrate,-\edge)$, the covariance being a sampling term
rather than a Jensen gap.

\lead{Stationarity is a second axis, not a nuisance.}
Selecting the pin on categories disjoint from those it is evaluated on models a deployer meeting a
novel attack type. A stationary attack mix is a different position. Under a random split instead of a
category-grouped one the router's own signal also improves, AUROC \num{0.6509} to \num{0.8561}
(\num{0.6418} to \num{0.8549} with the vocabulary fitted on training requests only),
because category identity becomes usable, which is legitimate information in a genuinely stationary
deployment rather than leakage. Both protocols leave the mean favouring the router over the honest pin, but only the stationary
one has it beating the in-sample pin as well, so the verdict against a hindsight comparator depends
on the protocol (E25).

\section{Extended Related Work}
\label{app:related-extended}

The body states the results the argument leans on directly. The fuller accounting sits here.

\lead{Prior results we build on.}
Established already: the winner's curse in selection over noisy estimates
\citep{smithwinkler2006, bastani2025winners}, routers that fail signal-free baselines
\citep{routerbench2024, commercial2026}, and deferral from a single best model to a second with the
threshold set by gain on the deferred set (their RS-I rule, RS-D uses overall validation
accuracy) \citep{lassance2026dcs}, so Eq.~\ref{eq:decomp} is not
novel. The two-axis (harm, over-refusal)
objective predates the routing literature: paired harmful/benign refusal testing
\citep{rottger2024xstest}, the trade-off curve \citep{bianchi2024safetytuned}, dedicated
over-refusal benchmarks \citep{cui2024orbench, li2024mossbench}, and the safety-shift /
harmfulness-discrimination decomposition we use \citep{jailbreakensemble2025}.
Perturbation-neighbourhood refusal instability is the premise of SmoothLLM
\citep{robey2023smoothllm}, with an entropy-metric treatment in \citet{heverin2026rbe}. Hazard rates
over attacker effort date to \citet{littlewood1993} and reach LLM jailbreaks in
\citet{topol2026survival}. MTDeep \citep{mtdeep2018} defends a pool by randomised selection at a
two-objective Stackelberg equilibrium, drawing a network per input from a mixed strategy rather than conditioning on it: the
equilibrium version of our mixing baseline (Section~\ref{sec:setup}), not of a signal-driven
router. Hidden-state probes predicting injection
exposure, with a named knowledge--action gap, are due to \citet{ipiexposure2026}.

\lead{Concurrent results.}
\Citet{dekoninck2024router} find routing survives injected noise, but against a cost$\times$quality
Pareto hull their own optimal-routing characterisation places inside routing's feasible set: it
lower-bounds the \emph{optimal} strategy rather than a fitted one, and on their own SWE-Bench
numbers their router loses to the hull with ten models and beats it with five. With a single
objective the hull optimum is a vertex, so our best-fixed pin is a harder comparator.
\Citet{chen2026gap} attribute \SIrange{12}{36}{\percent} of the reported router-to-oracle gap to
single-draw label noise no single-commit router can close. \Citet{chen2026cofailure} bound
combination gains by a co-failure ceiling, under which accuracy cannot exceed $1-\beta$ for $\beta$
the rate at which every model fails the same query, which they note is classical (Kuncheva's oracle combiner, alongside the
majority-vote limits of \citet{kuncheva2003limits}), concluding that on their pool and verifiable
tasks combining rarely beats the single best model without a strong query-level signal. Our result is complementary, since they bound the headroom while we show the measured router
deficit is substantially inflated by how the single best model was chosen, which leaves the oracle
headroom, if anything, larger under an honest pin.

\lead{Deployed safety routing.} \citet{vllmsemanticrouter2026} classify each request and select an
eligible model from a configured pool of heterogeneous models, running jailbreak and PII
classifiers inline and then blocking or steering the request by configured policy before dispatch,
which is the construct Section~\ref{sec:intro} describes.
Routing has also been applied to safety in a different sense, \citet{mo2026triaging} triaging a
conversation to specialised guardrail models rather than to the model that answers, so their router
selects a detector and not the model whose output can become an action.

\lead{Supporting numbers relocated from Section~\ref{sec:related}.}
\Citet{lassance2026dcs} cut the other way, their two dynamic-selection rules beating the single
best model on \num{9} and \num{12} of \num{15} datasets under label corruption, so deferral does win
where the competence signal is not adversarial, with single-best and threshold chosen on a
validation split, which is the convention we advocate. \Citet{routingplateau2026} find many of \num{21}
routing methods across five benchmarks converging to a narrow accuracy plateau, which they
attribute to routers learning global model-performance trends rather than query-specific signal,
and \citet{lai2026routing} show the collapse persists when train and test are identical, so it is
not a generalisation artefact.

In AI control, \citet{adaptive2024untrusted}'s Claude-3-Haiku sensitivity arm (AUROC \num{0.67})
attains \SI{17.9}{\percent} of the untrusted model's usefulness at \SI{99.1}{\percent} safety,
while pinning GPT-3.5, their main trusted model, yields pass@1 \num{23.9} against the untrusted
model's \num{47.4}, about half its usefulness, at \SI{100}{\percent} safety (their Table~8). Their conclusion that strong trusted models are needed is the same
observation, under a different, scheming-model threat model.

\Citet{manczak2024primeguard} report attack success falling from \SI{100}{\percent} to
\SI{8}{\percent} with helpfulness maintained. Their only cross-model comparator is a different base
model rather than a pin chosen over a pool.

\Citet{ipiexposure2026}'s probes hold at AUROC \numrange{0.992}{0.999} under a search-based rewrite
attack and \numrange{0.956}{1.000} under an explicit instruction to suppress the signal, over two
models in each case (their Sec.~3.4). Neither attack optimises against the probe, their own
limitations section recording that ``neither attack directly optimizes against the probe or
AGRI''. \Citet{benchmarkless2026} report
that under target-to-target deltas the judge is ``the least consequential of the three factors'' and
that ``ordering and critical-status are preserved even when absolute levels are not''. On safety
data with a \num{44}-model pool the ordering is not preserved. Their instrument is a different
measurement object, so we agree a score must not be collapsed to a ranking but locate the
instability in the judge where they locate it in scenario category and risk measure. That grader
choice moves measured attack success at all is the standing result of \citet{strongreject2024}, with
\citet{chouldechova2026comparison} and \citet{cyclicjudge2026} on judge-induced ranking instability.

\lead{The empty cell, in detail.} Mid-trajectory model switching has been studied twice, by
\citet{replaygap2026} on SWE-bench, which measures trajectory divergence and replay validity, and
by \citet{handofftax2026} on coding agents, which measures task success and cost. Neither measures
an attack outcome. \Citet{agentdrift2026} benchmark injection-hijacked trajectories with
step-level labels, but generate them from a single model, so they carry no cross-model comparison.
Where routing is applied to safety it selects a detector rather than the model that answers
\citep{saferoute2025, mo2026triaging}. Our recall is bounded by what a web and arXiv search
surfaced and does not prove absence. For re-scoring we looked for a published safety-routing
evaluation with a best-single comparator and released per-query outcomes, searching arXiv titles and
abstracts from 2024 to 2026, GitHub and Hugging Face, and the safety subsets of routing benchmarks,
and found none. SafeRoute \citep{saferoute2025} routes between two guard models against a per-instance
oracle and releases no outputs, \citet{kassem2025routerrobust} measure jailbreak routing to the weaker
model without a released matrix, and CASCADE \citep{luo2026cascade} selects defences and releases code
only. Semantic Scholar, Google Scholar, OpenReview, Zenodo and workshop proceedings were not searched.

\section{Defences in Detail}
\label{app:defences-detail}

This appendix gives the full evidence behind Section~\ref{sec:mechanisms-die}.

\subsection{Cascade, Triage and the Recognition Signal}
\label{app:mech-detail}

Section~\ref{sec:baseline} scored a baseline on something no deployer has. Three candidate defences
of our own fail the same way, each looking sound on a signal and not on harm through the system that
reads it: an evaluated cascade, a pricing argument, and a held-out attack on the flag a controller
reads.

\lead{Cost-saving cascade.}
Treating safety as a constraint ($\min \mathbb{E}[\text{cost}]$ s.t.\ $h \leq h_f$, $u \geq u_f$),
a cascade sends each request to a cheap model or, on predicted risk, to the pin. With pin, hull mixture and budget fixed on
training data, \num{16} of the \num{96} genuine cascades ($0<\escfrac<1$) have
lower held-out harm than the honest hull at the same planned budget, at $\advers=0$, and \num{13} hold utility $\geq\num{0.70}$. Against the oracle
hull, \num{3} beat it and none held that utility. The oracle frontier had hidden these configurations. Margins are at most \num{0.47} points, and \num{6} of the \num{16} cost more than the hull at realised cost (E34).

The thirteen are \num{12} thresholds on one checkpoint under two prompt wrappers plus one other
model. Their harm equals the pin's in all \num{52} fold-rows. What escalation adds, however, is measured
against the cheap model alone on the same folds. There it lowers harm by \num{5.68} points for the other
model (\num{0.0027} against \num{0.0594}) and raises it by \num{0.27} points for the twelve, whose cheap
model alone has harm zero (E47). The saving is mostly the cheap model, and on matched folds the cheapest
survivor is less useful than the pin (\num{0.7907} against \num{0.8132}).
Cost is pre-execution selection, paying the cheap model or the pin, not a sequential cascade paying
both.

Suppose the adversary can force cheap-model routing on a fraction $\advers$ of attacks, an assumed
steering ability rather than a search we ran. At $\advers=1$ no configuration escalates, so by
definition none counts as a cascade (Table~\ref{tab:adversary-full}). At that forced
endpoint twelve survivors keep harm zero and utility $\geq\num{0.78}$, while the other falls to
harm \num{0.0594} and utility \num{0.30}, so steering costs what the forced destination costs
(Appendix~\ref{app:cascade-detail}).

\lead{The second defence was stopped on arithmetic.} Refusal-provenance triage, distinguishing an
adversarially induced refusal from an intrinsic one, was not built because of three feasibility
counts. The cheapest to check is that $k$ perturbations cost $k$ cold prefills against an unbilled refusal and a
cache-read-priced switch, so break-even is $k=\num{0.1}$ where the fallback is billed at cache-read rates (Appendix~\ref{app:triage}).

\lead{Escalation on template-dependent recognition.}
This one looked strong. On the skill surface flagging is near-perfectly protective. Harm
occurs in \num{0.0037} of undefended episodes where the agent flags the content and
\num{0.6309} where it does not, a \num{170}$\times$ difference (E44). The flag is the judge's verdict, written in the same call as the success label, so the rubric may account for part of this gap. Pooling the defended
episodes would inflate that to \num{360}$\times$ (Appendix~\ref{app:detection}).

\begin{table}[!t]\centering\footnotesize\setlength{\tabcolsep}{4pt}
\caption{Flagging and attack success for
\texttt{claude-sonnet-4-6} across five templates (no \texttt{v32} runs).
Templates are not scenario matched, so template and domain are confounded;
\texttt{v35} has both the lowest flag rate and the highest success.}
\label{tab:flag-suppression}
\begin{tabular}{lccc}\toprule
attack template & $n$ & flag rate & attack success \\ \midrule
\texttt{claude\_v41} & 70 & 1.000 & 0.000 \\
\texttt{claude\_v53} & 20 & 1.000 & 0.000 \\
\texttt{claude\_v39} & 100 & 0.990 & 0.000 \\
\texttt{claude\_v54} & 20 & 0.450 & 0.000 \\
\texttt{claude\_v35} & 90 & 0.067 & 0.878 \\
\bottomrule\end{tabular}\end{table}

Flagging varies with template, confounded with domain (Table~\ref{tab:flag-suppression}). One model flags
\numrange{99}{100}\% of attacks under three templates, \SI{45}{\percent} under a fourth and
\SI{6.7}{\percent} under a fifth, and the fifth is also the highest-success template. On both grids the template effect survives a null in which templates do nothing, but this design
does not identify its size as a variance share (Appendix~\ref{app:detection}). That model's conditional harm rate when it fails
to flag is \num{0.823}, so under the evasion template the
``safest'' model is the worst choice among those that flag at all (Appendix~\ref{app:detection}).

\lead{From selection to adaptation, on held-out reruns.} That template is selected
post hoc from the five this model was run on (the corpus has six), which licenses no adaptivity
claim. We therefore let the attacker choose its template on half of the reruns and scored it on the
rest (Table~\ref{tab:adaptive}, Appendix~\ref{app:adaptive}). The corpus records no task
identifier, and its undefended episodes rerun six fixed payloads over two domain--harm pairs, so
held-out means held-out reruns of the same configurations, not new tasks. One choosing a single
template for the whole pool gains nothing (\num{0.7018} against a static \num{0.7027}), though that
control, the corpus-modal \texttt{claude\_v35}, is already the evasion template. One that knows
which model it faces reaches \num{0.8062}, but the two arms cover run configurations and sandbox
modes unevenly. Matched on both, its success gain depends on how cells are weighted, \num{15.0}
points with each cell counted once and \num{2.0}, with an interval spanning zero, weighted by
episodes (E54b). The flag does not. It falls from \num{0.0624} to \num{0.0225} unmatched and by
\numrange{3.9}{4.0} points matched under either weighting (rerun-bootstrap \SI{95}{\percent}
intervals \numrange{-6.11}{-1.65} and \numrange{-5.53}{-2.06}, below zero in every replicate). All
of it is on \texttt{gpt-5.4}, whereas the model this section's flagging argument rests on is
identical under both arms. For \texttt{gpt-5.4} the matched judged flag falls from \num{0.226} to
\num{0.030}, \num{19.6} points (\num{19.9} episode-weighted), and its template moves from
\texttt{claude\_v35} to \texttt{claude\_v39} in \num{186} of \num{200} splits, chosen for training success with the flag only breaking ties. The
drop does not depend on the judge. A fixed lexicon applied to the agent's own output, reading no
judge field, agrees with the judge on \texttt{gpt-5.4} ($\kappa=\num{0.82}$) and reproduces the
drop within each of the three run configurations that contain both templates (permutation
$p\leq 10^{-4}$, the resolution of \num{10000} permutations, E55).

\lead{Through a controller, the fallback decides.} Composing the flag into controllers from the logged
reruns (E56b), block-on-flag adds almost nothing on its own, since flagged episodes already fail (\num{1} of
\num{269}). Escalate-on-silence covers \SIrange{94}{98}{\percent} of tasks (E56), because two models never
flag and a third flags \num{2} of \num{420}, so the system is effectively its fallback. The flag-lowering attacker then cuts system
harm by \numrange{4.9}{10.6} points, but only because one fallback, \texttt{claude-sonnet-4-6},
resists every template but the static one. Without it the sign reverses (\numrange{2.3}{8.9}). The
fallback's vulnerabilities, not the flag, set the outcome (Appendix~\ref{app:controller}).

The reduction appeared only in the arm that knew the target model, which in a routed deployment the router determines.

\lead{The common pattern.}
\begin{quote}
\emph{Signals derived from attacker-influenced content, or from the model's own recognition of it,
can be adversarially steered. Signals grounded in authenticated provenance, tool permissions,
sandbox state or deterministic policy are not covered by our experiments and may well be sound.}
\end{quote}
Cost saving is exposed to steerable routing wherever the forced model is unsafe, refusal provenance
did not survive its pricing and flagging is steered by template selection. This generalises beyond routing, since any defence gated on a model recognising
something must
report per-template rates and not a mean. That extends to probe-gated defences only against an adversary that optimises against the
probe, which the reported attacks of \citet{ipiexposure2026} do not (Section~\ref{sec:related}).

\subsection{Action-Level Policy Settings}
\label{app:what-works-detail}

Section~\ref{sec:mechanisms-die}'s candidate defences read signals an attacker can steer, as measured for the flag and assumed for the cascade. On the same live surface, four
action-level policy settings record zero judged successes on the same \num{130} episodes,
drawn from \num{30} rerun indices, so they are one replicated test rather than four independent ones, namely user confirmation, denied network egress, code audit before execution
and a second pass over the plan (Table~\ref{tab:defences}, Appendix~\ref{app:defences}).

\begin{table}[t]\centering\footnotesize\setlength{\tabcolsep}{4pt}
\caption{Skill-injection attack success by policy setting.
Each defended setting has \num{130} episodes on three of five models. The
\num{360}-run undefended comparator matches scenario and model coverage,
but not execution mode. Matching execution mode as well leaves \num{10}
informative episodes and a Clopper--Pearson upper bound of \num{0.308} on residual judged success. Zeros
depend on one automated judge, and mechanical enforcement was not verified.}
\label{tab:defences}
\begin{tabular}{lcc}\toprule
defence & $n$ & attack success \\ \midrule
\texttt{normal} (all models) & 1342 & 0.5052 \\
\texttt{normal} (matched to defended runs) & 360 & 0.4722 \\
\texttt{ask\_user} & 130 & 0.0000 \\
\texttt{no\_network} & 130 & 0.0000 \\
\texttt{script\_audit} & 130 & 0.0000 \\
\texttt{two\_pass} & 130 & 0.0000 \\
\bottomrule\end{tabular}\end{table}

The comparator has to be matched. The corpus-wide undefended rate of \num{0.5052} is not the right
baseline, because the defended episodes cover a subset of scenarios and three of the five models.
Restricted to those, it is \num{0.4722} ($170/360$, \SI{95}{\percent} interval
\numrange{0.4197}{0.5252}). Four zeros are
not zero risk. The exact Clopper--Pearson interval on $0/130$ is
$[0, \num{0.028}]$, and this bound on residual success counts all \num{130} episodes as independent, which \num{30} shared rerun indices over two domain--harm pairs do not justify. That is
not a bound on the \emph{reduction}, because \num{30} defended episodes have no undefended
comparator and \num{30} sit in a cell whose undefended success is exactly \num{0.000}
(Table~\ref{tab:defences-cells}, Appendix~\ref{app:defences}), so only \num{70} are informative
there. The Clopper--Pearson interval on $0/70$ is $[0, \num{0.051}]$.

\lead{The matching is still incomplete, and it matters.} Matching on model, template and rerun index
leaves the corpus's \texttt{sandbox\_mode} field unmatched, and undefended success depends on it
more than on anything we do match: \num{0.702} in \texttt{multi\_turn} ($n=\num{822}$),
\num{0.232} in \texttt{single\_turn} ($n=\num{250}$) and \num{0.159} where the field is absent
($n=\num{270}$). The defended episodes are \num{60} \texttt{single\_turn}, \num{60} field-absent
and \num{10} \texttt{multi\_turn}, so the matched comparator draws much of its harm from a mode the
defended episodes were mostly not run in. Matching on mode as well, two of the three contributing
cells stop being informative. All \num{28} successes in
\texttt{gpt-5.4-mini}$\times$\texttt{v35}'s \num{0.400} comparator fall in \texttt{multi\_turn},
and its \num{30} mode-matched \texttt{single\_turn} episodes have success $0/\num{30}$.
The comparator for \texttt{claude-sonnet-4-6}$\times$\texttt{v35}, \num{0.9875}, shares no mode with
its defended episodes at all. What survives is \num{10} informative episodes and a Clopper--Pearson upper bound of \num{0.308} on residual judged success (E40). We report all three denominators (\num{130},
\num{70} and \num{10}) because the headline reduction is an artefact of the coarsest of them.

Three qualifications belong with the zeros: the defences were run on one corpus and on five
(model, template) cells spanning three of the five models and three of the six templates, not on
the full nine such cells (Table~\ref{tab:defences-cells}). None was evaluated against an adaptive attacker, which is also true of the cascade, whose steering is
assumed, and of refusal-provenance triage, stopped on cost
before it was built. No utility, refusal or friction
cost was measured, so the comparison is on harm alone. Every zero is also read through one automated
judge, \texttt{claude-sonnet-4-6}, which is itself the target model in \num{60} of the \num{130}
defended episodes per defence (Section~\ref{sec:limitations}).

The asymmetry we argue for is one of authority. An enforced policy may inspect attacker-influenced
arguments and provenance, but injected instructions cannot rewrite it, whereas a prediction's output
is what the attacker moves. The release does not establish mechanical enforcement. Every configuration sets
\texttt{docker\_sandbox: false}, and the released tree carries the \texttt{safety\_policy} field but
not a harness that enforces it. The episodes are consistent with recognition. None of the \num{520} defended episodes contains a mechanical-block marker, whether a
permission error, an unreachable host or a refused connection, while \num{86} of them carry policy language in the
agent's output or the judge's explanation of it, against \num{0} of \num{1342} undefended episodes
(E40). A network namespace cannot stop an agent that never issues the request, and the agent's flag
rate rises with the same pattern across the four settings, from \num{0.153} on matched undefended
episodes to \num{0.992} at the highest (Appendix~\ref{app:defences}).

\lead{Consistent with recognition.} The episodes are consistent with these four settings stopping attacks the way the undefended
models sometimes do, by \emph{recognition}, the agent reading the policy and declining. Recognition
has an input the adversary chooses, and Section~\ref{sec:mechanisms-die} shows that an attacker's
template choice can move it. The observations are consistent with recognition but do not distinguish mechanisms, and the design they suggest, untested here, places policy authority outside the agent's context.

The routing controller already sits at the decision point, so it should enforce policy over actions
rather than predict intent (design hypotheses in Appendix~\ref{app:defences}).

\section{Experiment Register and Reproduction}
\label{sec:experiments}
\label{app:experiments}

Every experiment, its status, and the script that reproduces it. Status \textsc{superseded} means a
corrected rerun exists. \textsc{void} means a design flaw makes the output uninterpretable. Both are
listed and not silently dropped.

Table~\ref{tab:experiments} lists every run. A register is only useful if a reader can check it
against the paper, so we also mark, with a dagger, every \textsc{kept} run whose headline we do not
quote anywhere in the paper, and give the reason for each in the table's note. Six runs carry that
mark. The intended reading is that nothing in this register is both load-bearing and absent from the
text: a run is either reported, or superseded, or void, or daggered with a reason.

\begin{table*}[!t]\centering\footnotesize\renewcommand{\arraystretch}{0.88}
\caption{Experiment register. Superseded and void runs are listed rather than dropped.}
\label{tab:experiments}
\begin{tabular}{llll}\toprule
id & question & headline & status \\ \midrule
E0 & headroom at full pool & vacuous at $M{=}44$ & \textsc{superseded} \\
E0b & deployable-$k$ vs 1-D null & 6/6 cells no excess & \textsc{kept} \\
E1 & grader swap & safest model changes on 2/4 scenarios & \textsc{kept} \\
E2 & Gate 0: predictability + routing & AUROC 0.6509; in-sample-pin comparison & \textsc{superseded} \\
E3 & interaction vs additive null & real on one corpus, 84\% noise on another & \textsc{kept} \\
E4b & threshold $\astar$, raw arm & 0.843/0.845/0.843/0.821 (calibrated 0.946--0.867) & \textsc{kept}$^{\dagger}$ \\
E5 & label-reliability curve & confounded by difficulty; features plateau & \textsc{kept} \\
E6 & shrinkage family & $\astar$ falls to $\approx0.70$; hull arm is sampler slack & \textsc{kept} \\
E7 & real shrunk router & deficit 10--25$\times$ better, pin still wins & \textsc{kept}$^{\dagger}$ \\
E8 & conditional tail edge & $-0.0866$ against an in-sample pin & \textsc{superseded} \\
E9 & CIs, dense $\astar(M)$, permutation & signal-free router already loses & \textsc{kept} \\
E10 & is AUROC sufficient? & no: operating point 1.9--3.1$\times$ & \textsc{kept} \\
E11 & tune the operating point & edge worsens on the confident tail & \textsc{kept}$^{\dagger}$ \\
E12 & parse agentic corpus & 33{,}119 runs; two data traps caught & \textsc{kept} \\
E14 & agentic router, nested & beats signal-free, not the pin & \textsc{kept}$^{\dagger}$ \\
E16 & adversary + safety floor & $\advers{=}1$: none escalates, by definition & \textsc{kept} \\
E17 & + utility floor & saving 0\% even at $\advers{=}0$ & \textsc{superseded} \\
E18c & in-loop signal, controlled & 0.722 $\rightarrow$ 0.808 at the injection step & \textsc{kept} \\
E18d & E18c, tasks and attacks held out & 0.705 $\rightarrow$ 0.703: level holds, rise does not & \textsc{kept} \\
E19 & counterfactual escalation & headline withheld: judge unreliable & \textsc{kept}$^{\dagger}$ \\
E20 & routing on the skill surface & AUROC exactly 0.5000; 38 scenarios & \textsc{void} \\
E24 & honest vs in-sample pin & winner's curse 0.0426$\to$0.1127 & \textsc{kept} \\
E24j & E24 per judge and category & cost 0.012--0.113 at $k{=}44$; collapse under every judge & \textsc{kept} \\
E25b & bootstrap of the shift contrast & difference $>0$ at every $k$ & \textsc{kept} \\
E25a & random vs shifted cost, AgentDojo & suite held out 0.027--0.028 vs random 0.003--0.004 & \textsc{kept} \\
E25f & full-sample pin convention & random 0.001--0.018; shift 0.022--0.084 & \textsc{kept} \\
E24t & $\lam$ tie-break sensitivity & 37/60 at $\lam{=}0$ if ties go high; outcomes unchanged & \textsc{kept} \\
E25 & shift vs stationary protocol & router AUROC 0.651 vs 0.856 & \textsc{kept} \\
E26 & re-analysis, honest pin & edge $+0.0770$ at $k{=}44$; median difference zero & \textsc{kept} \\
E27 & agentic re-run, honest pin & no deferral at $k{=}28$, edge undefined & \textsc{kept} \\
E28 & six frozen featurisers & plateau 0.606--0.656 & \textsc{kept} \\
E29 & held-out adaptive attacker & generic 0.7018; targeted 0.8062 & \textsc{kept} \\
E30 & non-saturated pool & headroom 0.0502 once dominants drop & \textsc{kept} \\
E31 & per-family rebuild & $\lam{=}0$; equals the honest pin, margin within tie swing & \textsc{kept} \\
E32 & item ceiling + fine-tune & 0.7993 item diagnostic; fine-tune 0.6674 & \textsc{kept} \\
E33 & over-refusal axis & harm-pin ranks 37/44 on over-refusal & \textsc{kept} \\
E34 & cost frontier, honest selection & 3$\to$16 beat the hull, 13 at $u{\geq}0.70$ & \textsc{kept} \\
E35 & can E31's rebuild rescue E20? & no family has ASR and $\geq$3 models & \textsc{kept}$^{\dagger}$ \\
E36 & per-model flag rates & flag 0.680, not 0.822 & \textsc{kept} \\
E37 & rank inversion vs its null & observed 43.8\% below null 99.2\% & \textsc{kept} \\
E38 & defence comparator, matched & undefended 0.4722, not 0.5052 & \textsc{kept} \\
E39 & base-rate axis, honest pin & $r=-0.9806$; never-beats claim withdrawn & \textsc{kept} \\
E40 & defence comparator, mode-matched & 10 informative episodes; $[0,0.308]$ & \textsc{kept} \\
E41 & encoder scale sweep, 22M--8B & clears $\astar$ only at 8B; 4B crossing withdrawn & \textsc{kept} \\
E42 & router cost against pool price & 8B adds 1.2\% to the cheapest call & \textsc{kept} \\
E43 & pre-execution routing ceiling & oracle headroom $\leq$0.0105 on every subgrid & \textsc{kept} \\
E44 & flag protectiveness, undefended & 170$\times$, not the 360$\times$ of the pooled set & \textsc{kept} \\
E45 & AgentDojo defences, matched & every defence lowers harm; inversion withdrawn & \textsc{kept} \\
E46 & attack-vs-task spread, composition & 1.89$\times$ pooled, 1.20$\times$ model-fixed & \textsc{kept} \\
E47 & cascade decomposition & vs cheap alone: $-5.68$ pts for 1, $+0.27$ for 12 & \textsc{kept} \\
E48 & family-router tie-break & 1 of 5 folds ties; the swing is twice the margin & \textsc{kept} \\
E49 & AgentDojo grid reconciliation & 28-config complete set is one family, \num{477} scenarios & \textsc{kept} \\
E50 & attack-key parse check & path and JSON attack labels agree on all runs & \textsc{kept} \\
E51 & null \texttt{security} verdicts & 1 of \num{33119} attacked runs; pin harm shifts \num{0.000004} & \textsc{kept} \\
E21 & base-rate axis & in-sample pin; $r=-0.9896$ & \textsc{superseded} \\
E22 & flag suppression & one template drops flagging to 0.067 & \textsc{kept} \\
E23 & featuriser ablation & MiniLM $-$ TF-IDF $=-0.0449$ & \textsc{kept} \\
E13 & router analysis in the agentic setting & 28-config AgentDojo grid; superseded by E14 & \textsc{superseded} \\
E15 & cost--safety trade-off, oracle frontier & superseded by E34's honest frontier & \textsc{superseded} \\
E18 & router inside the agent loop & superseded by E18c's length-controlled ladder & \textsc{superseded} \\
E52 & predicted vs realised marginals & Spearman \num{0.9591}; the two pins coincide & \textsc{kept} \\
E53 & per-surface harm floor & safest model nonzero on all three surfaces & \textsc{kept} \\
E54 & adaptive gain, rerun-index matching & flag $-3.97$ pts, success $+1.31$ on shared cells & \textsc{superseded} \\
E54b & adaptive gain, configuration matched & flag $-3.9$ to $-4.0$ pts; success weighting-dependent & \textsc{kept} \\
E56 & flag-gated controllers, unmatched & exploratory; superseded by E56b & \textsc{superseded} \\
E57 & Best Single on LLMRouterBench & harmless at its pool; shift test failed at $k\geq5$ & \textsc{kept} \\
E58 & every eligible safety corpus, fixed rule & interval test met on 2 of 6; null beaten on 3 & \textsc{kept} \\
E59 & selection cost against perfect-router headroom & HELM $0.57$, SORRY $0.40$, HarmBench $0.03$ & \textsc{kept} \\
E60 & flip $\times$ gap decomposition & an identity, HELM held out raises both factors & \textsc{kept} \\
E61 & AIR-Bench 2024, pre-registered & clear shift effect, small size, prediction failed & \textsc{kept} \\
E62 & k$=$2 over all model pairs & no E58 or E59 verdict changes, AIR-Bench $0.0068\to0.0053$ & \textsc{kept} \\
E63 & mechanism, one factor at a time & pooling matters on HarmBench, granularity not separable & \textsc{kept} \\
E64 & spread over fold assignments & direction robust, size moves up to $2.4\times$ & \textsc{kept} \\
E65 & each judge alone & direction robust at $k\leq5$, size judge-dependent at $k{=}44$ & \textsc{kept} \\
E66 & fixed choice versus dispatch & held out it picks a fixed model, random it routes & \textsc{kept} \\
E67 & group-level view of the headline & positive dropping any one group & \textsc{kept} \\
E56b & flag-gated controllers, matched & fallback sets the sign; flips without one model & \textsc{kept} \\
E55 & recognition under an independent label & $\kappa{=}0.82$; within-configuration drop, $p{=}0.0001$ & \textsc{kept} \\
\bottomrule
\multicolumn{4}{@{}p{0.97\textwidth}@{}}{\footnotesize $^{\dagger}$Kept, but its
headline is not quoted anywhere in this paper, and deliberately so. E4b: the paper's $\astar$ headline comes from E9's denser sweep. E4b's values appear only in this row, and its sweep geometry in Figure~\ref{fig:heatmap}.
E7: the conclusion (the pin still wins under shrinkage) is reported throughout
Sections~\ref{sec:surface} and~\ref{sec:baseline}; the \numrange{10}{25}$\times$ improvement factor is
not, because it is a ratio of two small deficits. E11: a refuted hypothesis, since tuning the
operating point does not make the router beat the pin, and the operating-point axis is reported
from E10 instead. E14: superseded in substance by E27, which corrects E13 and E14 for having
selected the pin on the evaluation scenarios; the status stays \textsc{kept} because its own
comparison is internally consistent, and what E27 replaces is the baseline convention. E19: the headline is withheld because the run replays
a poisoned prefix rather than re-running the environment, so it cannot score the corpus's own \texttt{security}
flag, and the reliability of the judge we substituted for it is not established. E35: a negative
result reported only through its consequence, that E20 stays \textsc{void} as a measured statement
rather than as an artefact of one filtering choice. Separately, E6's mixing-baseline arm is not
daggered but is withdrawn: it minimises over Dirichlet$(1,\dots,1)$ draws, which place no mass on the
vertices of the simplex, so its ``hull versus pin'' gap (\num{+0.0011}, \num{+0.0156}, \num{+0.0538}
at $k=2,3,5$) is Monte-Carlo slack that grows with $k$ and verifies nothing. That a fixed mixture
cannot beat the best fixed model is a one-line consequence of linearity, and we rest it on that.}
\end{tabular}\end{table*}

Routing on the
live skill surface (E20, Table~\ref{tab:experiments}) is inconclusive for a different reason, since its
complete-case grid of \num{38} scenarios gives every model AUROC \num{0.5000}, so a router fed it
must lose for arithmetic rather than evidential reasons, and the same filtering raises one model's success rate
from \num{0.263} corpus-wide to \num{0.605} (\texttt{per\_model\_asr\_complete\_case} in E20).

\lead{Reproduction.}
All experiment scripts are in the artifact's \texttt{experiments/} and write JSON to
\texttt{results/}. Tables in \texttt{tables/} are generated by \texttt{scripts/make\_tables.py} from those JSON files. Tables written inline in the section files, among them Table~\ref{tab:corpora} and Table~\ref{tab:e58}, are checked against the same files by the claim map. Corpora and retrieval steps are recorded in \texttt{DATA.md} and fetched by
\texttt{scripts/fetch\_data.py}, and large downloads are
excluded from version control, and experiments that read a raw corpus rather than the shipped derived
tables are marked as such in the registry.

\lead{Standing method rules.}
Group-aware cross-validation by default, never random splits over templates, with the exceptions
Section~\ref{sec:data} lists. AUROC computed within
fold and then averaged, never pooled. A matched null or permutation test on every claim that a signal
exists. A non-semantic baseline on identical input. Over-refusal or utility reported alongside the core harm comparisons. An adaptive attack where one was run (the held-out template selectors, A3 and A4). The cascade's steering is
assumed, and the four policy settings were not attacked adaptively. And a statistic that cannot
be re-run from data in the repository does not exist.

The departures from those rules are listed in Section~\ref{sec:data}, and two are restated here. The base-rate axis of Appendix~\ref{app:axes} selects its honest pin on a random split of scenarios within each band rather than a group-aware one, which makes its
two halves exchangeable and understates the winner's curse it is meant to remove. And the four
policy settings of Section~\ref{sec:what-works}, like the cascade, are evaluated without an adaptive attacker, and the argument offered in their favour is an argument about where the
adversary's degrees of freedom sit, not a measurement.

Two references remain unread in full text, and no claim here depends on either.
\citet{kuncheva2002switching}, whose restated finding (``our choice of one individual classifier
over the rest will depend on how much we trust the estimate'') is the closest prior framing we are
aware of, is closed access. We have its metadata and restatements only. \citet{didaci2005dcs} is
likewise closed access: its abstract and six independent restatements establish it as a
ceiling-characterisation result with a positive conclusion and no single-best threshold, but its
body text is unread. More broadly, no citation-graph search was available and one venue's
submissions were behind a bot wall throughout, so where a novelty claim rests on the absence of
prior work we say so and give the query vocabulary, having twice found that a zero-result phrase
query was a terminology artefact.

\end{document}